\documentclass[aps,prd,groupedaddress]{revtex4}

\usepackage{epsfig} 
\usepackage{amsmath}
\usepackage{amsbsy,color}
\usepackage{array}
\newcolumntype{L}[1]{>{\raggedright\let\newline\\\arraybackslash\hspace{0pt}}m{#1}}
\newcolumntype{C}[1]{>{\centering\let\newline\\\arraybackslash\hspace{0pt}}m{#1}}
\newcolumntype{R}[1]{>{\raggedleft\let\newline\\\arraybackslash\hspace{0pt}}m{#1}}
\usepackage{dcolumn}
\usepackage{bm}
\usepackage{graphicx}
\usepackage{amssymb,latexsym,mathrsfs}
\usepackage[breaklinks, colorlinks, citecolor=blue, colorlinks=true,linkcolor=blue]{hyperref}%,draft
\newcommand{\beq}{\begin{equation}}
\newcommand{\eeq}{\end{equation}}
\newcommand{\beqa}{\begin{eqnarray}}
\newcommand{\eeqa}{\end{eqnarray}}

\definecolor{grey}{rgb}{0.5,0.5,0.5} 
\definecolor{darkred}{RGB}{175,0,0}

\def\ben{\begin{enumerate}}
\def\een{\end{enumerate}}
\def\bi{\begin{itemize}}
\def\ei{\end{itemize}}
\def\be{\begin{equation}}
\def\ee{\end{equation}}
\def\bea{\begin{eqnarray}}
\def\eea{\end{eqnarray}}

\def \sn2{\left(S/N\right)^2}

\def \arcs{^{\prime\prime}}
\def \farcs{\overset{\prime\prime}{.}}
\def\fnl{$f_{\rm NL}^{\rm loc}$}
\def\fnlequil{$f_{\rm NL}^{\rm eq}$}
\def\fnlloc{$f_{\rm NL}^{\rm loc}$\ }

\def\spherex{SPHEREx\ }

\graphicspath{{Paper_plots/}}

\begin{document}
%============================

\title{Cosmology with the SPHEREX All-Sky Spectral Survey}
\author{Olivier Dor\'e$^{1,2}$, Jamie Bock$^{2,1}$, Matthew Ashby$^9$,  Peter Capak$^3$, Asantha
 Cooray$^8$, Roland de Putter$^{1,2}$, Tim Eifler$^{1}$, Nicolas Flagey$^{11}$, Yan Gong$^8$,
 Salman Habib$^{10}$,  Katrin  Heitmann$^{10}$, Chris Hirata$^6$, Woong-Seob Jeong$^7$, Raj Katti$^2$, Phil Korngut$^{2}$, Elisabeth
 Krause$^4$, Dae-Hee Lee$^7$, Daniel  Masters$^3$, Phil
 Mauskopf$^{12}$, Gary Melnick$^9$, Bertrand Mennesson$^{2}$, Hien
 Nguyen$^1$, Karin \"Oberg$^9$, Anthony Pullen$^{13}$, Alvise
 Raccanelli$^5$, Roger Smith$^2$, Yong-Seon 
 Song$^7$, Volker Tolls$^9$, Steve Unwin$^1$, Tejaswi Venumadhav$^2$,
 Marco Viero$^4$, Mike Werner$^1$, Mike Zemcov$^{2}$} 

\affiliation{\vspace{0.5cm}\\
$^{1}$Jet Propulsion Laboratory, California Institute of Technology, Pasadena, CA 91109, USA\\
$^2$ California Institute of Technology, Pasadena, CA 91125, USA\\
$^3$ IPAC, 1200 E. California Blvd., Pasadena, CA 91125, USA\\
$^4$ Stanford University, Palo Alto, CA 94305, USA\\
$^5$ Department of Physics \& Astronomy, Johns Hopkins University, Baltimore, MD 21218, USA\\
$^6$ Ohio State University, Columbus, OH 43210, USA\\
$^7$ Korea Astronomy and Space Science Institute, Daejeon 305-348,
Korea\\
$^8$ Department of Physics and Astronomy, University of California, Irvine, CA 92697, USA\\
$^9$ Harvard-Smithsonian Center for Astrophysics, Cambridge, MA 02138, USA\\
$^{10}$ Argonne National Laboratory, Lemont, IL 60439, USA\\
$^{11}$ Institute for Astronomy, Hilo, HI 96720, USA\\
$^{12}$ Arizona State University, Tempe, AZ 85287, USA\\
$^{13}$ Carnegie Mellon University, Pittsburgh, PA 15213, USA
}

\date{\today}

\begin{abstract}
\vspace{0.5cm}

SPHEREx (Spectro-Photometer for the History of the Universe, Epoch of
Reionization, and Ices Explorer) \href{http://spherex.caltech.edu}{[Website]} is a proposed all-sky spectroscopic survey satellite
designed to address all three science goals in
NASA's Astrophysics Division: probe the origin and destiny of our Universe; explore 
whether planets around other stars could harbor life; and explore the
origin and evolution of galaxies. SPHEREx will scan a series of Linear
Variable Filters systematically across the entire sky. The SPHEREx
data set will contain R=40 spectra fir 0.75$<\lambda<$4.1$\mu$m and
R=150 spectra for 4.1$<\lambda<$4.8$\mu$m for every 6.2 arcsecond pixel
over the entire-sky.  In this paper, we detail the extra-galactic and
cosmological studies SPHEREx will enable and present detailed
systematic effect evaluations. We also outline the Ice and Galaxy
Evolution Investigations.
%We focus in particular on the inflation studies enabled by SPHEREx.
\end{abstract} 

\maketitle

%\section{Introduction and motivation for the mission concept}

\section{\spherex mission overview}
\label{sec:intro}

SPHEREx (Spectro-Photometer for the History of the Universe, Epoch of
Reionization, and Ices Explorer; PI: J. Bock) is a proposed all-sky survey satellite
designed to address all three science goals in NASA's Astrophysics Division:
probe the origin and destiny of our Universe; explore 
whether planets around other stars could harbor life; and explore the
origin and evolution of galaxies. All of these exciting science themes
are addressed by a single survey, with a single instrument, providing
the first near-infrared spectroscopy of the complete sky. In this
  paper, we will focus on the cosmological science
  enabled by SPHEREx and outline the Galactic Ices and the Epoch of
  Reionization (EOR) scientific investigations.

SPHEREx will probe the origin of the Universe by constraining the
physics of inflation, the superluminal expansion of the Universe that
took place some $10^{-32}$ s after the Big Bang. SPHEREx will study its
imprints in the three-dimensional large-scale distribution of matter
by measuring galaxy redshifts over a large cosmological volume at low
redshifts, complementing high-redshift surveys optimized to constrain
dark energy.

SPHEREx will investigate the origin of water and biogenic molecules in
all phases of planetary system formation - from molecular clouds to
young stellar systems with protoplanetary disks - by measuring
absorption spectra to determine the abundance and composition of ices
toward $>2\times10^4$ Galactic targets.  Interstellar ices are the likely
source for water and organic molecules, the chemical basis of life on
Earth, and knowledge of their abundance is key to understanding the
formation of young planetary systems as well as the prospects for life
on other planets. 

SPHEREx will chart the origin and history of galaxy formation through
a deep survey mapping large-scale structure. This technique measures
the total light produced by all galaxy populations, complementing
studies based on deep galaxy counts, to trace the history of galactic
light production from the present day to the first galaxies that ended
the cosmic dark ages. 

SPHEREx will be the first all-sky near-infrared spectral survey,
creating a legacy archive of spectra (0.75 $\leq\lambda\leq$ 4.8 $\mu$m with
$\lambda$/$\Delta\lambda$ = 41.5,
and a narrower filter width $\lambda/\Delta\lambda$ = 150 in the range $\lambda = 4.12 - 4.83 \mu$m) with the high sensitivity
obtained using a cooled telescope with large spectral mapping
speed. Space-borne measurements are essential, because the opacity of
the Earth's atmosphere makes these observations functionally
impossible across the relevant spectral range. The SPHEREx archive
applies to numerous exciting and diverse astronomy investigations,
some of which we can only imagine today. All-sky surveys such as the
IRAS, COBE and WISE Explorer missions have played a major role in
modern astrophysics, with a proven legacy that lasts for decades. 

\begin{figure}[!t]
\centering
\includegraphics[width=0.9\textwidth]{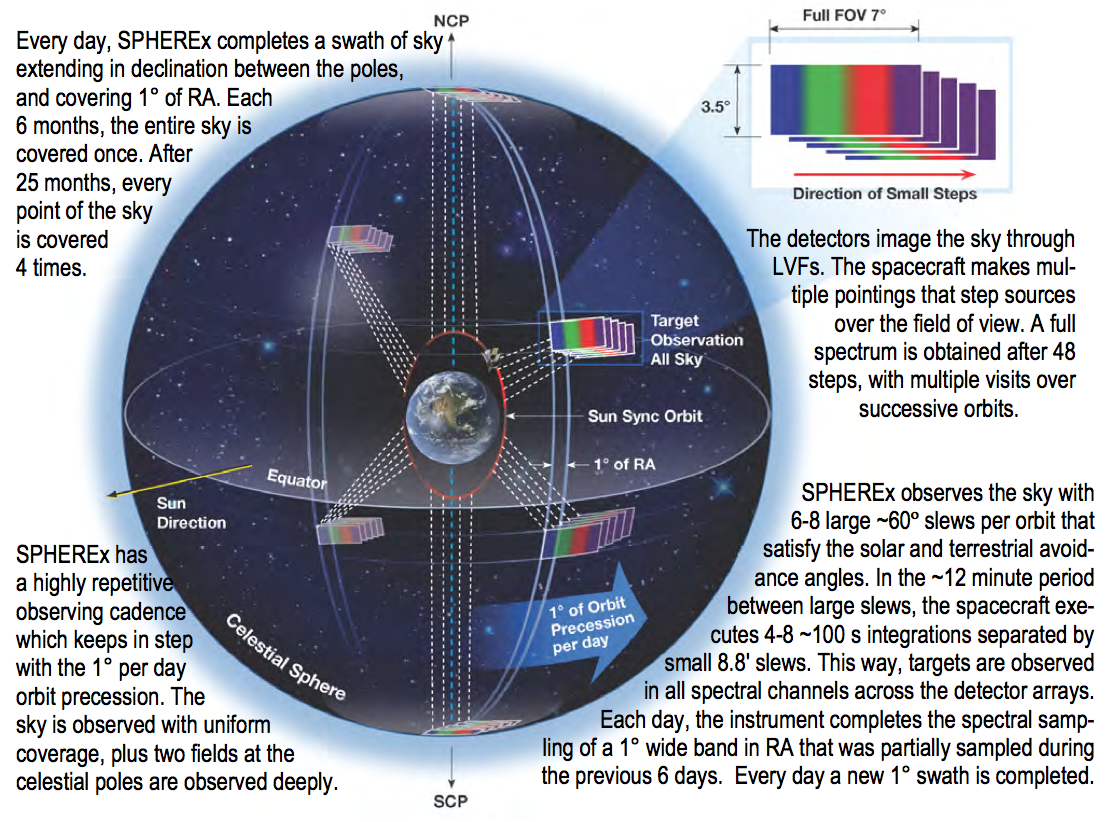}
\caption[]{\spherex completes an All-Sky map in a step and repeat
  fashion, one targeted observation at a time, separated by spacecraft
  slew maneuvers.}
\label{fig:scanning}
\end{figure}

The SPHEREx Mission will implement an extremely simple and robust
design that maximizes spectral throughput and efficiency. The
instrument is based on a 20 cm all-aluminum telescope with a wide
$3.5^\circ\times 7^\circ$ field of view, imaged onto four 2k x 2k
HgCdTe detector arrays arranged in pairs separated by a
dichroic. These H2RG arrays have been fully qualified for space
observations by JWST, and are built upon a long successful history of
space instruments using arrays in smaller formats. Spectra are
produced by four space-demonstrated linear-variable filters. The
spectrum of each source is obtained by moving the telescope across the
dispersion direction of the filter in a series of discrete steps (see
Figure~\ref{fig:scanning}). This simple method was demonstrated by
LEISA on New Horizons to obtain remarkable spectral images of
Jupiter. Using thermal methods demonstrated by Planck, Spitzer and
WISE, the telescope is radiatively cooled to $\leq$ 80 K and two of
the detectors are cooled to $\leq$ 55 K with 450 \% margin on the
total heat load. SPHEREx has no moving parts except for one-time
deployments of the sunshields and aperture  cover.

\begin{table}[!b]
\centering
%\begin{tabular}{|p{3.0cm}|C{3.0cm}|C{3.0cm}|C{3.3cm}|C{3.3cm}|C{2.5cm}|}
 %       \hline & {LSST} & DESI & Euclid & SPHEREx & CHIME \\
  %  \hline
\resizebox{0.5\textwidth}{!}{
\begin{tabular}{|L{5.0cm}|C{4.0cm}|}
 \hline
\hline
Parameter & Value \\
\hline
Telescope Effective Aperture & 20 cm\\
\hline
Pixel Size & 6.2'' $\times$ 6.2''\\
\hline
Field of View & 2 $\times$ (3.5$^\circ\times 7.0^\circ$); dichroic\\
\hline
Spectrometer
Resolving Power and & R=41.5;
$\lambda$=0.75-4.1$\mu$m\\
Wavelength Coverage & R=150; $\lambda$=4.1-4.8 $\mu$m\\
\hline
Arrays & 2 x Hawaii-2RG 2.5 $\mu$m\\
& 2 x Hawaii-2RG 5.3 $\mu$m\\
\hline
Point Source Sensitivity & 18.5 AB mag (5$\sigma$) on average per frequency
element with 300\% margin \\
\hline
Cooling & All-Passive\\
\hline
2.5 $\mu$m Array and Optics Temperature & 80K with 700\% margin on total
heat load\\
\hline
5.3 $\mu$m Array Temperature & 55K with 450\% margin on total heat load\\
\hline
Payload Mass & 68.1 kg (current best estimate + 31\% contingency)\\
\hline
\hline
\end{tabular}}
\label{tab:instrument}
\caption{SPHEREx Key Instrument Parameters.}
\end{table}

SPHEREx will observe from a sun-synchronous terminator low-earth
orbit, scanning repeatedly to cover the entire sky in a manner similar
to IRAS, COBE and WISE. The feasability study of such a sky strategy has
been presented in \cite{Spangelo:2014}. During its two-year nominal mission, SPHEREx
produces four complete all-sky spectral maps for constraining the
physics of inflation. These same maps contain hundreds of thousands of
high signal-to-noise absorption spectra to study water and biogenic
ices. The orbit naturally covers two deep, highly redundant regions at
the celestial poles, which we use to make a deep map, ideal for
studying galaxy evolution.  All aspects of the SPHEREx instrument and
spacecraft have high heritage. SPHEREx requires no new technologies
and carries large technical and resource margins on every aspect of
the design. The projected instrument sensitivity, based on
conservative performance estimates, carries 300 \% margin on the
driving point source sensitivity requirement.

In this paper, we present an overview of the all-sky survey focusing
on its cosmological implications. We explain briefly the principle of
our measurement in Sec.~\ref{sec:spectro_princile} and motivate it briefly in
Sec.~\ref{sec:motivation} before giving an overview of the galaxy
redshift survey in Sec.~\ref{sec:overview}. In Sec.~\ref{sec:sims} we
describe our simulation pipeline before quantifying the
cosmological implications in Sec.~\ref{sec:cosmo}. In
Sec.~\ref{sec:sys} we discuss potential systematics and the associated
mitigation strategies enabled by SPHEREx. In Sec.~\ref{sec:legacy} we
illustrate briefly the legacy value of such a data-set, before
detailing the scope of the SPHEREx Ice Investigation in
Sec.~\ref{sec:ice_investigation} and the Epoch of Reionization
Investigation in Sec.~\ref{sec:eor_investigation}.

\section{spectroscopy without a spectrometer}
\label{sec:spectro_princile}

SPHEREx performs its all-sky spectroscopic survey using a compact
wide-field telescope with no moving parts that images the sky directly
onto the focal plane through Linear Variable Filters (LVFs). Spectra
are built up by stepping the instrument field across the sky with
small motions of the spacecraft (see Figure~\ref{fig:scanning}).

\begin{figure}[!t]
\centering
\includegraphics[width=0.65\textwidth]{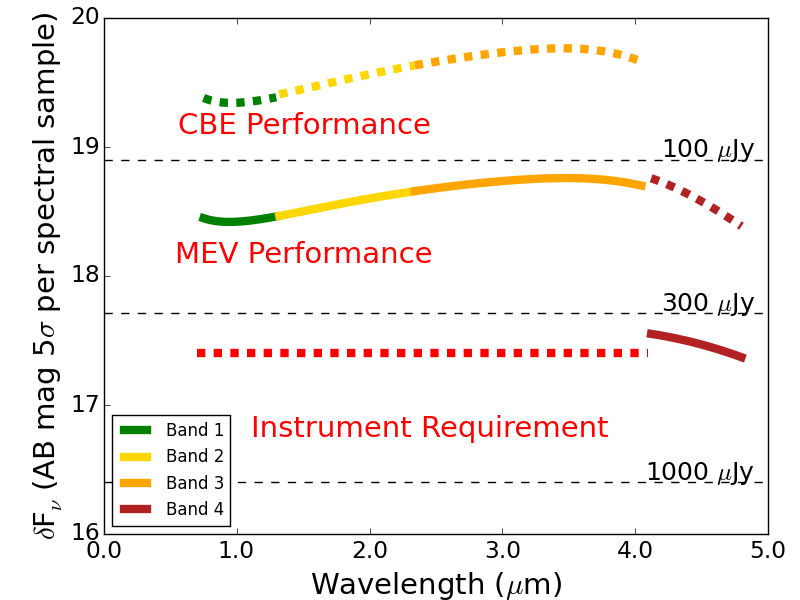}
\caption[]{SPHEREx scientifically required point source
  sensitivity has large margin over the estimated instrument
  performance. We show above the instrument requirement (red dashed
  line) and the MEV (solid colored curve) and CBE (dashed current
  curve) performance. The Maximum Expected Value (MEV) performance
  refers to  the instrument performance based on specifications that
  each sub-system can meet with contingency. The Current Best
  Estimated (CBE) performance is the instrument performance based on
  currently estimated sub-system performance. The scientific margin is
  defined as the difference between MEV Performance and the Instrument
  Requirement. The Band 4 sensitivity easily meets the 12
  AB mag (5σ) requirement, and will return hundreds of thousands of
  high-quality ice absorption spectra (see Sec.~\ref{sec:ice_investigation}).
}
\label{fig:pt_src_sensitiviy}
\end{figure}

SPHEREx is designed to maximize the spectral surveying power needed to
map the entire sky (see Table~\ref{tab:instrument}). In spite of its small aperture, SPHEREx achieves
high spectral speed and overall power (see
Figure~\ref{fig:pt_src_sensitiviy}) by the use of high-throughput,
efficient optics coupled to large-format low-noise focal plane
arrays. The instrument uses a small telescope that provides excellent
image quality over a wide field, and LVFs that eliminate the reimaging
optical elements and mass associated with conventional spectrometers. 

LVFs specifically fabricated for the required wavelength range and
resolving power, are mounted above the near-IR detector arrays -- one
for each array. Every column of pixels has a unique spectral
response. Complete spectra are obtained by stepping these columns
across the sky. The LVF wavelength ranges and resolving powers are
chosen to allow a uniform step size survey while realizing higher
spectral resolution in the longest wavelength band (see Figure~\ref{fig:scanning}).

%SPHEREx has four 4 spectral bands, all observed simultaneously. A
%dichroic beam splitter centered at 2.34 $\mu$m reflects Bands 1  and 2 onto the short wavelength FPA (Fig. E.1-3) and transmits Bands 3
%and 4 to the longer wavelength FPA. 

% \begin{figure}[!t]
% \centering
% \includegraphics[width=0.85\textwidth]{plot_lam_spectroawesome_v12.png}
% \caption[]{SPHEREx (colored in green, yellow, orange and red) achieves high spectral mapping
%   speed compared with other spectroscopic survey instruments. Spectral
%   mapping speed is defined to be the inverse of the on-sky time
%   required to map the full sky to a line flux of 10$^{-18}$ W/m$^2$ (1$\sigma$) at
%   each wavelength. Spectral power, summarized in the legend, is
%   defined as the integral $\int {\rm Speed}\ d\lambda/\lambda$ over the spectral range of each
%   instrument. A key to obtaining high spectral power is observing from
%   space where the foreground sky brightness is minimized, most
%   prominently at 2-5 $\mu$m. 
% }
% \label{fig:spectral_speed}
% \end{figure}

\section{\spherex all-sky survey motivation}
\label{sec:motivation}

\begin{figure}[!t]
\centering
\includegraphics[width=0.85\textwidth]{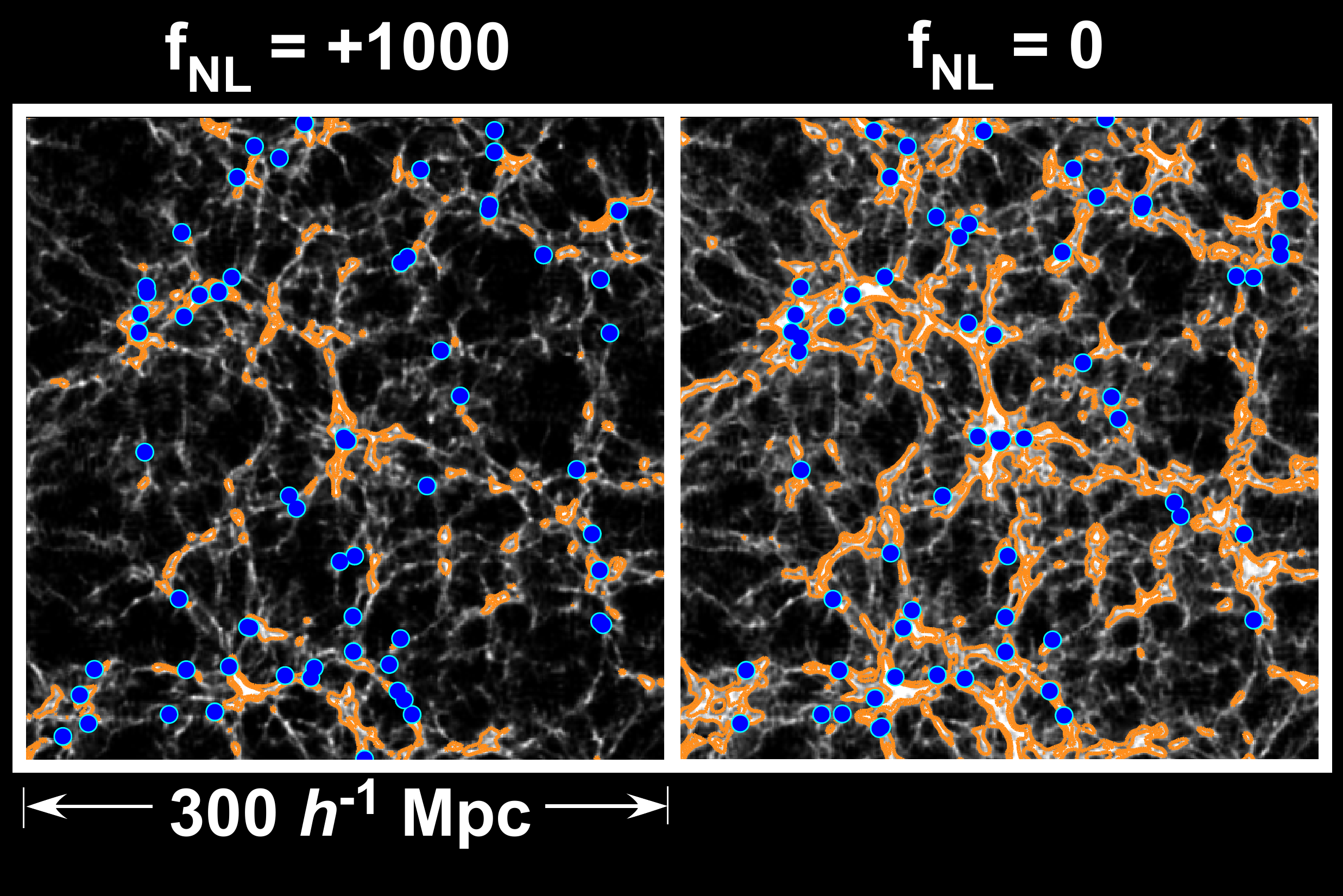}
\caption[]{By precisely measuring the spatial clustering of
galaxies, SPHEREx will measure primordial non-Gaussianity described
conveniently by the parameter \fnlloc. This plot shows two simulated
distributions of matter (color scales) and galaxies (blue dots) for
different values of \fnlloc. By precisely measuring the
clustering of galaxies over a volume more than $10^5$ times larger than
depicted here, SPHEREx will measure \fnlloc to high accuracy.}
\label{fig:fnl_sims}
\end{figure}

The physics of inflation can be probed by measuring the imprint of
inflationary ripples in the large-scale structure (LSS) distribution
of galaxies (see Figure~\ref{fig:fnl_sims}). Precise measurements of this distribution constrain the statistics of the
inflationary ripples and provide a method for concrete discrimination
between various models of the inflationary field(s)
\cite{Huterer:2010en,Bartolo:2004if,dalaletal08}. In particular, LSS
of the low redshift Universe provides a powerful tracer of primordial
non-Gaussianity, and therefore contains great potential to improve
constraints on the physics of inflation \cite{Abazajian:2013vfg,Alvarez14}.  

% Detecting non-Gaussianity would elucidate the nature of the scalar
% field(s) driving inflation. Fundamental scalar fields are a central
% theory in particle physics, the recent discovery of the Higgs particle
% representing experimental confirmation of this concept.  Just as the
% Higgs particle is generated by a phase transition in the Higgs field,
% inflation is associated with a phase transition in a scalar field(s)
% in the early Universe. Understanding inflationary fields would thus
% impact particle physics broadly.  The vigorous scientific interest in
% inflation is reflected in 1000s of papers published on non-Gaussianity
% in the past decade \cite{Bartolo:2004if,Huterer:2010en}. 

The design of the SPHEREx all-sky survey has been optimized to measure the large-scale
distribution of galaxies, precisely quantifying the statistical
distribution of the inflationary ripples. We can characterize the
non-Gaussian departure in the distribution of the
inflationary fluctuations from a Gaussian bell curve by the \fnlloc
parameter \cite{Salopek:1990jq,Komatsu:2003iq}. Inflation models fall into broad
categories, and measuring \fnlloc provides a unique test. Inflation models
driven by multiple fields generally predict a high-level of
non-Gaussianity ($|$\fnlloc$|>$ 1) while simpler models with a single field
generally predict low levels of non-Gaussianity ($|$\fnlloc$|<$ 1)
\cite{Alvarez14}. 

% The physics of inflation may be probed further through two classes of
% measurements, both necessary and complementary
% \cite{Komatsu:2009kd,Abazajian:2013vfg}. The first approach probes the
% energy scale of inflation by searching for space-time ripples from
% inflation, a background of gravitational waves. These gravitational waves produce a
% characteristic â€˜B-modeâ€™ polarization pattern in the CMB. Increasingly
% precise CMB polarization measurements (BICEP2 2014; Planck 2013 XVI)
% probe the amplitude of this pattern, and in turn the inflation energy
% scale.

% Detecting non-Gaussianity would elucidate the nature of the scalar
% field(s) driving inflation. The concept of a fundamental scalar field
% is a pillar of particle physics, of which the recent discovery of the
% Higgs particle at CERN (ATLAS; CMS 2012) was the experimental
% confirmation. Just as the Higgs particle is generated by a phase
% transition in the Higgs field, inflation is associated with a phase
% transition in a scalar field(s) in the early Universe. Advancing our
% understanding of inflationary fields would thus be a major next step
% in our understanding of fundamental scalar fields and would impact
% physics broadly. This is reflected by the vigorous research throughout
% the cosmology community, with 1000s of papers published on
% non-Gaussianity in the past decade (Bartolo et al. 2004; Huterer et
% al. 2010). 

The Planck satellite team has led a vigorous search for
non-Gaussianity in the statistical properties of the CMB; however this
approach is now near fundamental limits. Experimental advances in CMB
technology produced a large improvement in \fnlloc measurements from WMAP
to Planck \cite{plancknongauss14,Bennett:2012zja}. The most recent
results constrain \fnlloc to be $\leq$ 14.3 (2$\sigma$), i.e., the energy fluctuations
in the primordial Universe followed a Gaussian distribution to at
least one part in 100,000. However, this constraint is not strong
enough to rule out multi-field models. Unfortunately, the CMB is now
largely exhausted for improving \fnlloc constraints. The ultimate CMB
measurement, limited by fundamental cosmic variance, could only
improve on the Planck errors by 60\% \cite{Baumann:2008aq}.

Improved \fnlloc measurements thus require a new approach. Because an LSS
survey maps the Universe in 3-D, it accesses many more modes than are
available from the more limited 2-D CMB surface of last
scattering. This allows LSS surveys to set much tighter constraints on
\fnlloc. SPHEREx will access the unprecedented large cosmic volume
available and improve on Planck's \fnlloc measurement by a factor of 10.
It will achieve the sensitivity of $|$\fnlloc$|<$ 1 (2$\sigma$) required to distinguish
between the single-field and multi-field scenarios.

\section{\spherex all-sky galaxy redshift overview}
\label{sec:overview}

\begin{figure}[!t]
\centering
\includegraphics[width=0.60\textwidth]{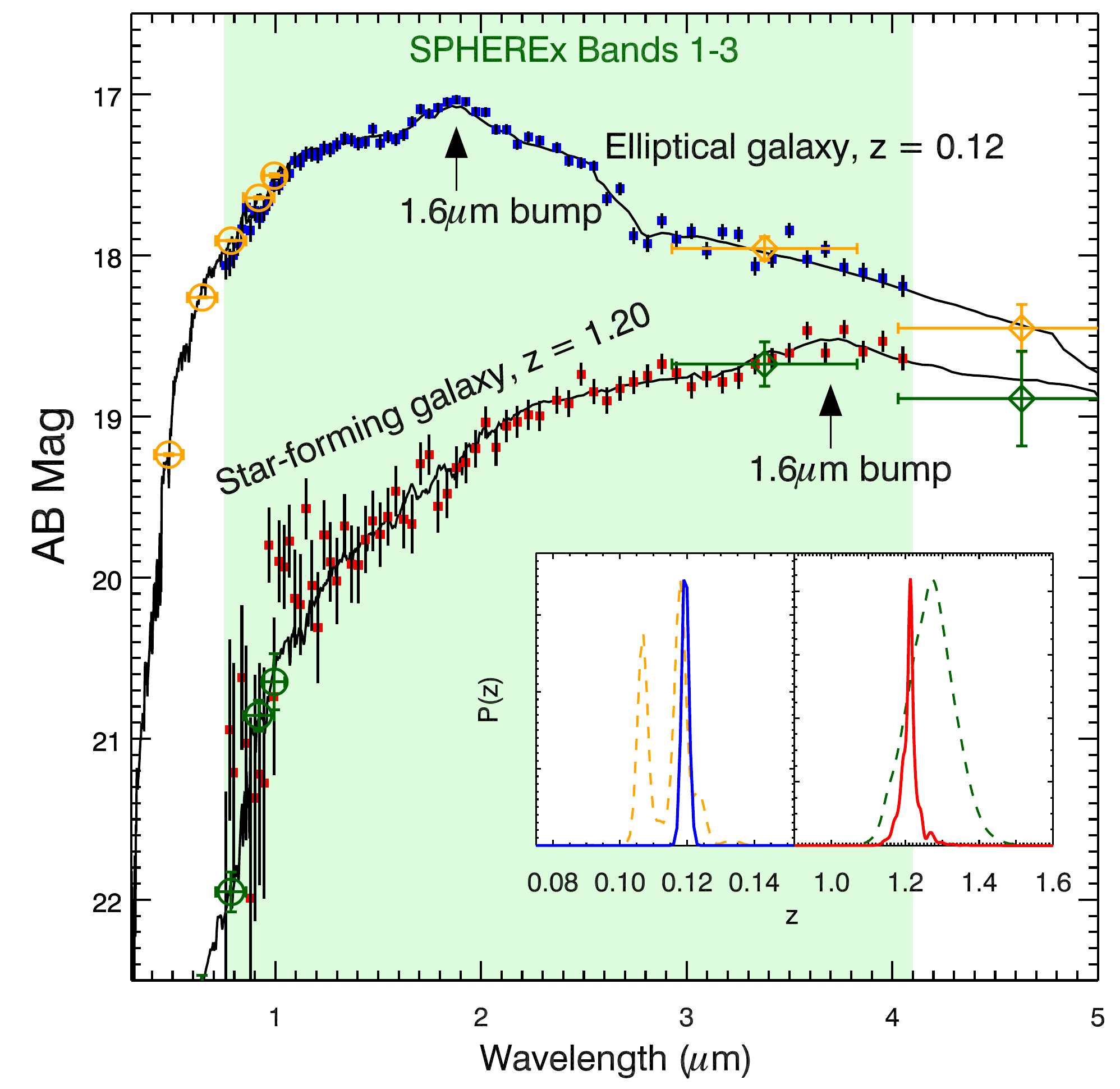}
\caption[]{\spherex determines the redshifts of WISE (diamonds) and
  Pan-STARSS/DES (circles) galaxies by fitting their measured
  spectra. Each redshift is assigned an error, a process we have
  extensively simulated from the COSMOS galaxy catalog. The
  determination is strongly driven by the 1.6 $\mu$m bump, so the
  target  redshift range is well-matched to the 0.75$\leq\lambda\leq$4.1
  $\mu$m wavelength  coverage. We require a minimum redshift error
  $\Delta z/(1+z)$ = 0.5\% to access the finest useful physical scales, which requires a spectral resolution of 35.}
\label{fig:spectra}
\end{figure}

SPHEREx measures 3-D LSS using a small telescope with an innovative
spectroscopic capability, focusing on the low-redshift Universe to
build a unique survey of unprecedented volume (see
Figure~\ref{fig:fnl_effect}). %SPHEREx measures the entire sky spectroscopically at
%moderate $\lambda/\Delta\lambda$ = 41.5 resolving power (between 0.75
%and 4.12 $\mu$m), chosen for sufficient redshift accuracy to access
%the finest useful spatial scales where clustering is in the linear regime.
SPHEREx will determine redshifts for hundreds of millions galaxies
from the all-sky catalog created from WISE \cite{Wright:2010qw},
Pan-STARRS \cite{Metcalfe:2013qea} and DES \cite{Abbott:2005bi}. The
redshift of these galaxies, pre-selected to avoid source blending, is
obtained by fitting the measured spectra to a library of galaxy
templates \cite{Ilbert:2008hz}. By observing at infrared wavelengths,
SPHEREx leverages the well-known rest-frame 1.6 $\mu$m bump that is nearly
universal in galaxy spectra and is a powerful redshift indicator
(Figure~\ref{fig:spectra}, \cite{Simpson1999,Sawicki:2002bi}). Since Pan-STARRS
  and DES are subsantially deeper than SPHEREx, using their galaxy
  catalog we can reliably estimate the confusion noise. Furthermore,
  using the above catalog, we will know before-hand these  galaxies
  and will simply ignore the associated pixels when measuring 
  the clustering of galaxies over the all-sky (both of these effects
  are included in the simulations described before).

The template fit produces a galaxy type, an expected redshift, and a
redshift uncertainty. We built a simulation and redshift fitting
pipeline based on the tools used to produce the official photometric
forecasts for the Euclid and WFIRST missions \cite{Jouvel:2009mh}
including realistic source confusion, clustering, colors, and both
random and systematic photometric errors. The power of low-resolution
(R$\simeq$20-100) spectra such as those obtained by SPHEREx for
cosmological studies has been successfully demonstrated by the
ground-based PRIMUS \cite{Cool:2013vva}, COSMOS \cite{Ilbert:2008hz},
and NMBS \cite{vanDokkum:2009xj} surveys, and our pipeline reproduces
the observed redshift accuracy in these surveys (it also motivates
planned survey such as J-PASS and PAU
\cite{Benitez:2014ibt,Marti:2014nha}). The redshift error is strongly 
dependent on the brightness of the galaxy -- brighter galaxies 
produce smaller errors. Therefore, we classify galaxies according to
their redshift accuracy, obtaining $\simeq$300 million with accuracy $\sigma(z)/(1+z)
\leq 10$\% and $\simeq$9 million with 0.3\% accuracy. As shown in
Figure~\ref{fig:nbar_vs_z} the distribution of galaxies extends out to
moderate redshifts, covering an very large effective volume (Figure~\ref{fig:fnl_effect}).

\section{SPHEREx Full-sky Survey Simulation Pipeline}
\label{sec:sims}

In this section, we present an overview of our
data-analysis/simulation pipeline. While most of it follows standard procedure in the era of
CCD astronomy, we emphasis some \spherex peculiarities.

\subsection{Overview}

\begin{figure}[!t] 
\centering
\includegraphics[width=0.9\textwidth]{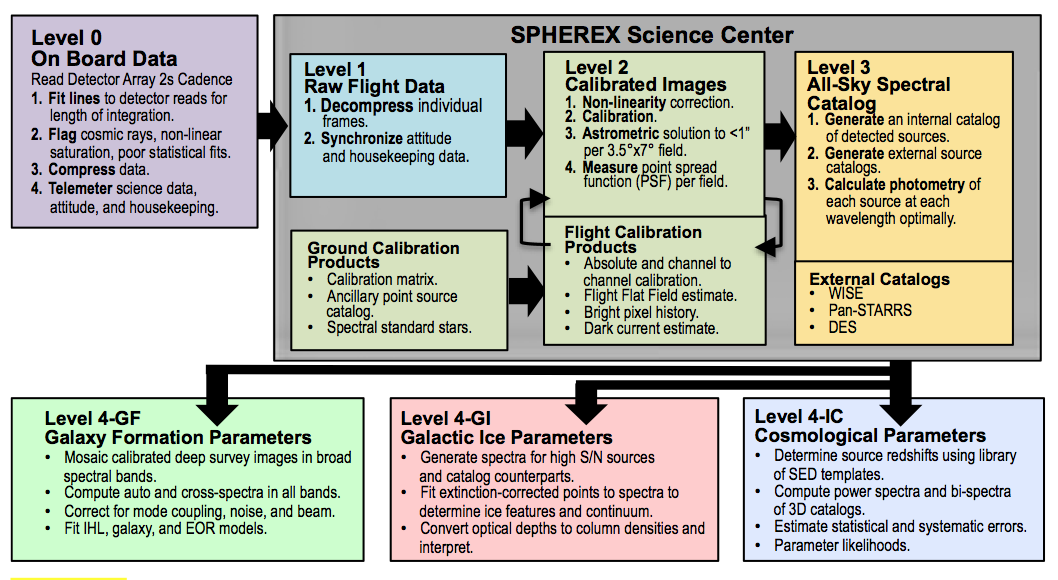}
\caption[]{Overview of the planned \spherex analysis pipeline, going
  from raw telemetered data to calibrated multi-wavelenth images and
  various science studies.}
\label{fig:pipeline_flow}
\end{figure}

An overview of our pipeline is described in the flow diagram depicted
in Figure\ref{fig:pipeline_flow}. It highlights the interplay between
pre-flight, on-board and ground processing. It also shows the role of
external full-sky catalogs we will rely on for the astrometry, the 
photometric calibration and the optimal photometry solution. We
describe each of these steps in the coming sub-sections.

% \subsection{On-board processing}
% The various steps that will be performed on-board are illustrated in
% Figure\ref{fig:on-board_flow}. 

% \begin{figure}[!t]
% \centering
% \includegraphics[width=0.9\textwidth]{on-board_data_analysis.pdf}
% \caption[]{Overview of the on-board data processing steps.}
% \label{fig:on-board_flow}
% \end{figure}

%%%%%%%%%%%%%%%%%%%%%%%%%%%%%%%
\subsection{Reduction and calibration}

% \begin{figure}[!t]
% \centering
% \includegraphics[width=0.9\textwidth]{red_cal_flow.pdf}
% \caption[]{Going from raw telemetered data to calibrated
%   multi-wavelength images}
% \label{fig:red_cal_flow}
% \end{figure}

The first step of the low level data analysis is to read raw
telemetered data and produce calibrated, flat-fielded and
astrometrically aligned images for each individual exposure. The data
reduction pipeline to final maps is outlined in
Figure\ref{fig:pipeline_flow}, and we summarize the essential elements
below. The accuracy to which we estimate controlling the various
contributions of systematic error of each step in the pipeline will be
discussed later in section~\ref{sec:sys}. 

\bi
\item {\bf Raw Telemetered Data to Sky Images}: The SPHEREx focal
  planes are populated with a total of four H2RG HgCdTe charge
  integrating detectors, manufactured by Teledyne systems. They are
  imaged to the sky with an instantaneous FOV of  3.52$^\circ$$\times$3.52$^\circ$ sampled by 6.2 arcsecond
  pixels. Due to bandwidth constraints, the raw 2048 $\times$ 2048
  $\times$ 4 array frames sampled at 1.5$\simeq$s intervals can not be
  telemetered directly to Earth.  Instead, we will implement on-board
  software to fit lines to the charge integration for each pixel to
  produce photocurrent images of each exposure.  The slope fit images
  in units of e$^{-}/s$ are then downlinked to Earth. A series of
  binary flags which identify events such as cosmic rays, saturation
  and non-linearity accompany the images.  The final data product
  returned for each exposure is a sub-sample of 8$\times$32 pixels at
  the full frame rate used to monitor the noise and verify the
  on-board software is performing adequately. 
   
\item {\bf Dark Current Correction}. The detectors have a zero-signal
  response termed dark current (DC), for H2RGs, the mean value is
  typically $\sim$0.01 e$^{-}/s$. A DC template for the specific
  arrays will be constructed from extensive laboratory
  characterization prior to launch and will be subtracted from each
  image at the first level of ground based processing. Additionally,
  sections of each array will be masked in flight to monitor drifts in
  the DC throughout the mission.  
  
\item {\bf Flat Fielding}. Due to inhomogeneities in fabrication, each
  pixel has a unique responsivity to incoming photons. A measurement
  of the  instrument's response to a spatially flat input illumination (a flat
  field) is used to correct the sky images for these variations. We will
  estimate the flat field response both from laboratory measurements using
  a field-filling, uniform radiance source with a solar-like
  spectrum before launch, as well as the diffuse sky brightness itself
  in flight. The redundant and dithered scan strategy at the celestial
  poles offer an excellent dataset to apply this self calibration
  technique.  

\item {\bf Astrometric Registration}. Generation of the absolute
  pointing solution for each field will begin with the spacecraft
  attitude information, accurate to 15 arcseconds.  We will then apply
  a demonstrated astrometry identifying algorithm \citet{Lang:2009nb} to
  the SPHEREx data itself, referenced to a USNO-B +2MASS stellar
  catalog, producing a solution accurate to the sub-arcsecond level.  

\item {\bf Point Spread Function Generation}.  Flux extraction of
  unresolved sources using optimal photometry requires an accurate
  understanding of the PSF. For each wavelength in each exposure, we
  will generate a PSF template by stacking all stellar sources in the
  2MASS catalog.  Each spatial/spectral sample in SPHEREx spans a
  solid angle of 0.53 deg$^2$, for moderately high Galactic latitude
  fields, this will contain ~350 stars with J-band magnitudes brighter
  than 14th.  

\item {\bf Flux Calibration}. SPHEREx will rely on a combination of
  laboratory measurements and in-flight characterization for absolute
  flux calibration.  In the laboratory, injecting a broad-band light
  source into a uniformly illuminating integrating sphere monitored by
  NIST-calibrated reference photodetectors has been demonstrated to
  produce percent level accuracy of absolute calibration
  \citep{Korngut13,Tsumura13,Zemcov13}.  Additionally, as SPHEREx is
  an all sky survey, we will observe all of the 97 spectral standard
  stars available from the STSCI database.   The resulting calibration
  factors will be applied to the data at this stage. 

\ei

After the aforementioned processing, the data for each exposure will
consist of 4 calibrated images with wavelength varying across the FOV.
These will be used as the input to the source flux extraction
described below. 

\begin{figure}[!t]
\centering
\includegraphics[width=0.7\textwidth,height=0.5\textwidth]{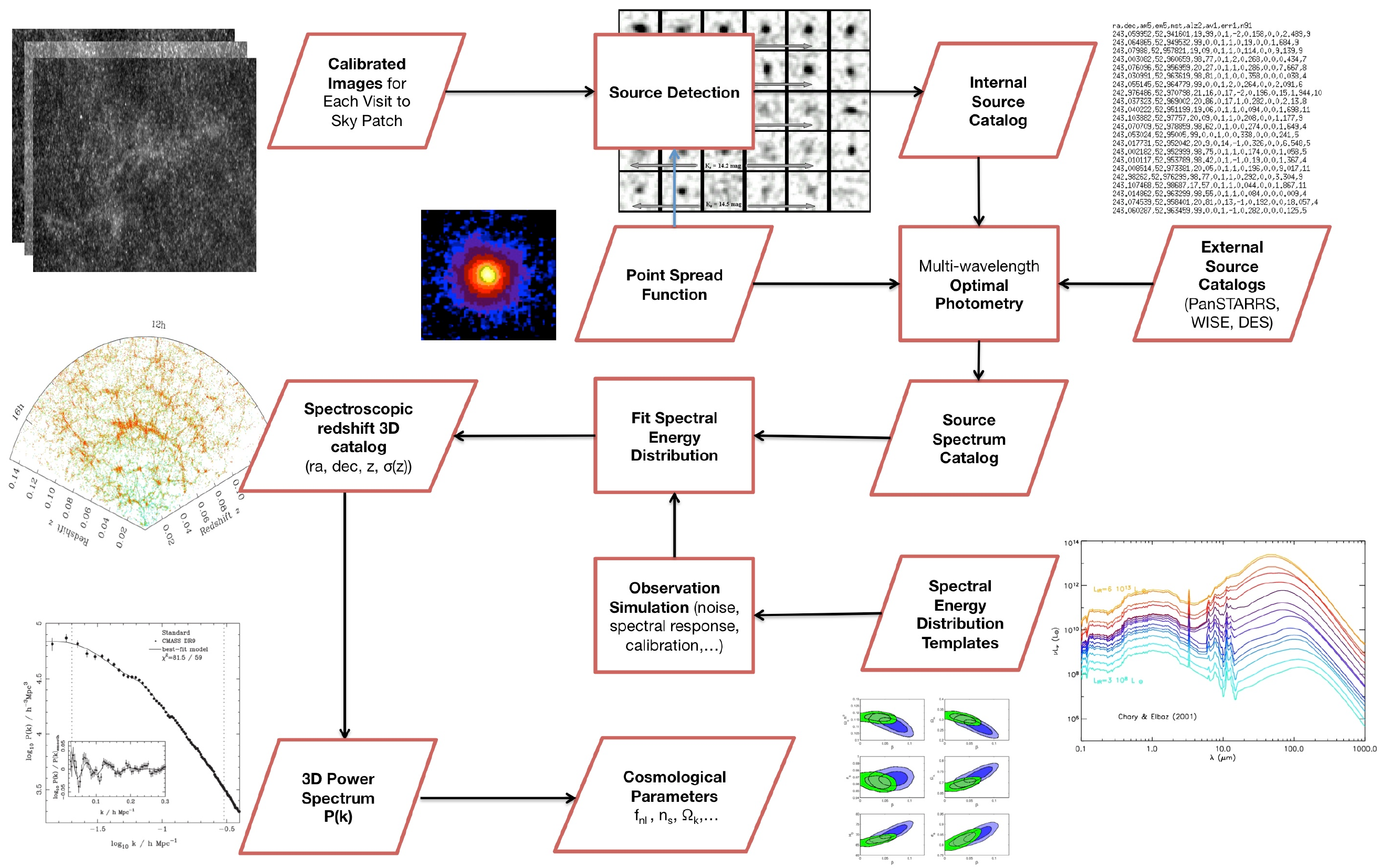}
\caption[]{Flow diagram illustrating the process to go from calibrated
  multi-wavelength images to  cosmological constraints.}
\label{fig:gal_catalog_flow}
\end{figure}

%\subsection{Full sky survey}

%\section{Astrophysical contaminant}

% \begin{figure}[!t]
% \centering
% \includegraphics[width=0.9\textwidth]{dgl_sim.pdf}
% \caption[]{Caption}
% \label{fig:dgl_sim}
% \end{figure}

% \begin{figure}[!t]
% \centering
% \includegraphics[width=0.9\textwidth]{zodi_sim.pdf}
% \caption[]{Caption}
% \label{fig:zodi_sim}
% \end{figure}

%%% 

\section{Building a galaxy catalog}
\label{sec:gal_catalog}

In this section, we describe how we go from a collection of
calibrated images at multiple wavelengths to a galaxy catalog
characterized by its position and its redshift. We outline the method
we employ, demonstrated using a state of the art simulation
pipeline. In particular, this section describes the path from
calibrated images to the galaxy catalog as outlined in the top half of
Figure\ref{fig:gal_catalog_flow}.

\subsection{Simulation pipeline principles}

We use real galaxy data from the Cosmic Evolution Survey (COSMOS,
Scoville et al. 2007) to create realistic simulated SPHEREx
observations. Extracted photometry from objects in these simulated
observations is passed to a high-performance redshift template-fitting
code. The quality of the derived redshifts is then analyzed in detail
to understand the ability of SPHEREx to constrain cosmological
parameters. Our goal with the simulation pipeline is to produce the 
most realistic forecast possible, based on our empirical knowledge of
the galaxy population.  We note that this pipeline is used jointly by
Capak et al. to produce official forecasts for the Euclid consortium,
the WFIRST SDT as well as the Hyper-Suprime-Cam (HSC) Subaru survey.

\subsection{Simulation pipeline overview}

There are five main steps to the SPHEREx simulation pipeline:
\begin{enumerate}
\item{Create the input galaxy catalog based on real data from the multi-wavelength COSMOS survey.}
\item{Generate simulated SPHEREx images at each wavelength step of the LVF based on the simulated galaxy spectral energy distributions (SEDs).}
\item{Optimally extract the R$\sim$40 photometry from the simulated SPHEREx images.}
\item{Derive the photometric redshifts for all detected objects using a highly-optimized template fitting code based on \emph{Le Phare} (Arnout \& Ilbert 2011).}
\item{Compare the derived photo-z's with the input galaxy catalog to test for
    redshift quality and derive the galaxy number counts $N(z)$ at different
    levels of redshift precision.}
\end{enumerate}
In the following sections we describe these steps in more detail.

\subsection{Generating the Input Galaxy Catalog}

The COSMOS survey comprises broad, medium, and narrow-band imaging
over 2 deg$^{2}$ in 30 photometric bands spanning the near-UV to the
infrared. Extensive SED fitting analysis has been done on galaxies in
the COSMOS field, resulting in estimates for each source of (1)
photometric redshift, (2) SED type (from the 31 galaxy templates used
by Ilbert et al. 2009), (3) selective dust attenuation E(B-V), and (4)
reddening law. Together with the magnitudes of the objects, these
parameters let us generate realistic models of the SEDs of the COSMOS
sources over the entire wavelength range spanned by the SPHEREx observations. The input catalog thus consists of a set of
known galaxy coordinates, magnitudes, redshifts, and SEDs. Because the
simulated galaxies are based on real observations, they closely
resemble the actual distribution of galaxies both in physical
coordinates and parameter space. 

\subsection{SPHEREx Image Simulation}

\begin{figure}[!t]
\centering
\includegraphics[width=0.6\textwidth]{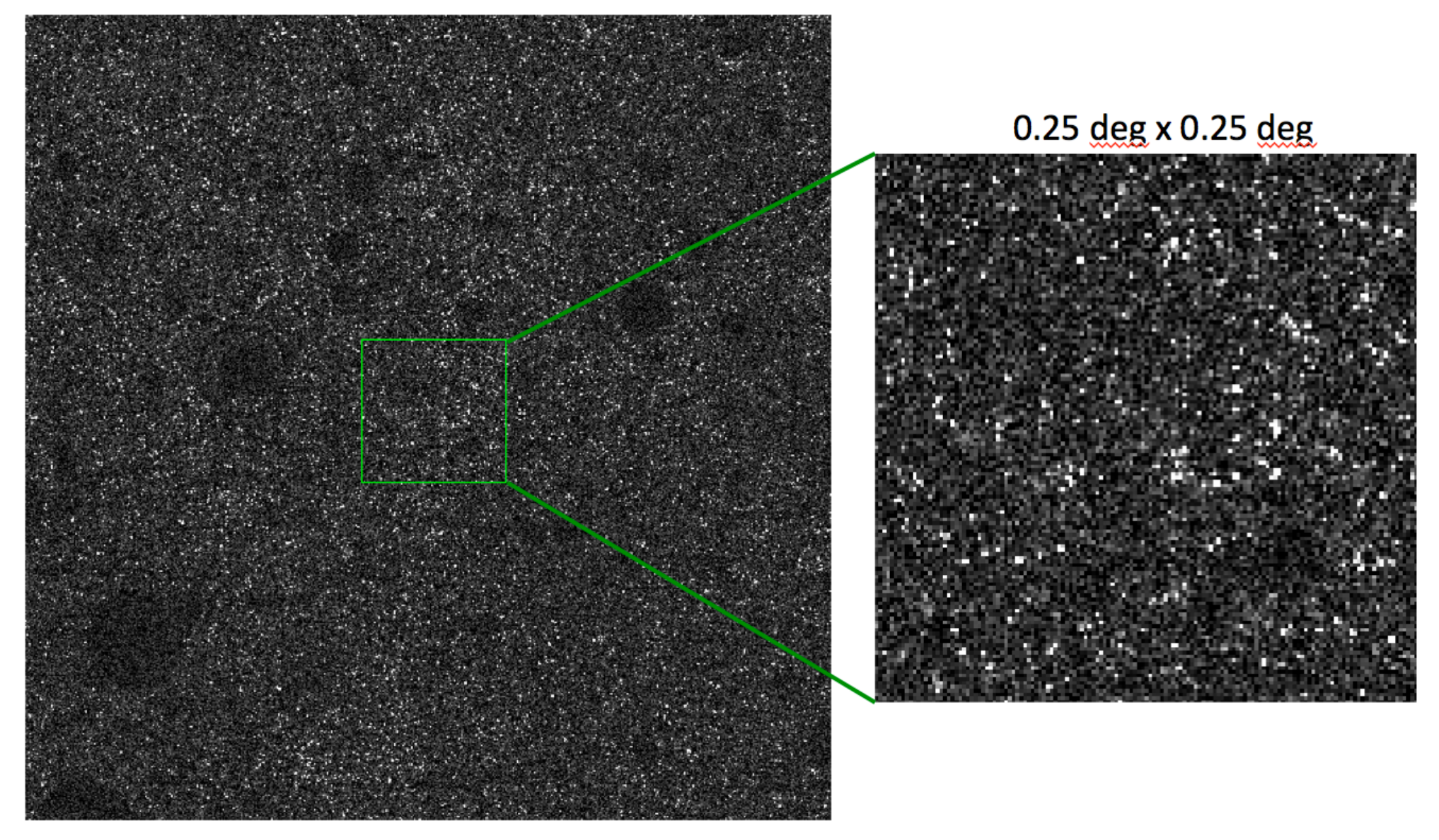}
\caption[]{Simulated data from SPHEREx in a spectral element of width
  $\Delta\lambda/\lambda$=40 centered at 1.93$\mu$ on the COSMOS field.}
\label{fig:simulated_image}
\end{figure} 

\label{subsection:imsim}
We assume a single set of LVF filters matching the design
specifications. For each wavelength step in the LVF, the predicted flux of each object is
computed by integrating its modeled SED against the filter profile at
that wavelength. This flux is then added to a simulated image
oversampled by a factor of 8 relative to the SPHEREx detector (i.e.,
to a grid with pixel scale 6$\farcs$2/8). This is repeated for all
simulated objects. Once the fluxes of all simulated objects have been
added to the oversampled images, the images are smoothed with a
Gaussian with FWHM 1.22$\lambda$/D to represent the SPHEREx PSF at a
given wavelength. Another Gaussian smoothing is performed to take into
account the effect  of geometric aberrations that dominate over
diffraction (by design) over much of the wavelenght coverage. A last
Gaussian smoothing is performed to simulate the effect of pointing
smear. Finally the images are rebinned to the  SPHEREx pixel size of 6$\farcs$2 and Gaussian noise is added to
reflect the depth of the SPHEREx observations. Because the simulated
images are based on real galaxies and include sources far 
below the SPHEREx sensitivity, the effects of source confusion and
noise due to faint objects are realistically captured in the simulated
images. 

%\bi
%\item Single set of LVF filters at R~40.
%\item Predicted fluxes in each filter.
%\item Grid of 1/8 pixel scale, i.e. 6$\farcs$2/8.
%\item Smooth with a Gaussian for optical aberration, 1/2 a pixel.
%\item Smooth with a Gaussian $\lambda/D$ FWHM
%\item Need to add pointing smear 2â€ Gaussian.
%\item Need also alignment tolerances.
%\ei

In addition to the simulated SPHEREx images, we use the modeled COSMOS
galaxy SEDs to generate simulated Pan-STARRS \cite{Metcalfe:2013qea}
ground-based photometry in \emph{ugriz} and WISE photometry in the
3.4~$\mu$m (W1) and 4.6~$\mu$m (W2) bands
\cite{Wright:2010qw}. Appropriate noise is added to these simulated
observations, which represent ancillary data that will be available to
the SPHEREx mission.  

\subsection{Source extraction}

The large instantaneous field-of-view required for a survey such as
SPHEREx, combined with the constraints placed by maintaining a
reasonable total number of detectors, led to the optimized design
plate scale of 6$\farcs$2/pixel. The point spread functions of SPHEREx
will have a range of full width at half maxima (FWHM) of 5$\arcs$ -
7$\arcs$ depending on the wavelength.  This combination of parameters
results in an under-sampled PSF, in which the relative alignment of a
point source on the sky and the coarse SPHEREx detector grid alters
the shape of the observed PSF.  

\begin{figure}[!t]
\centering
\includegraphics[width=0.48\textwidth,height=0.42\textwidth]{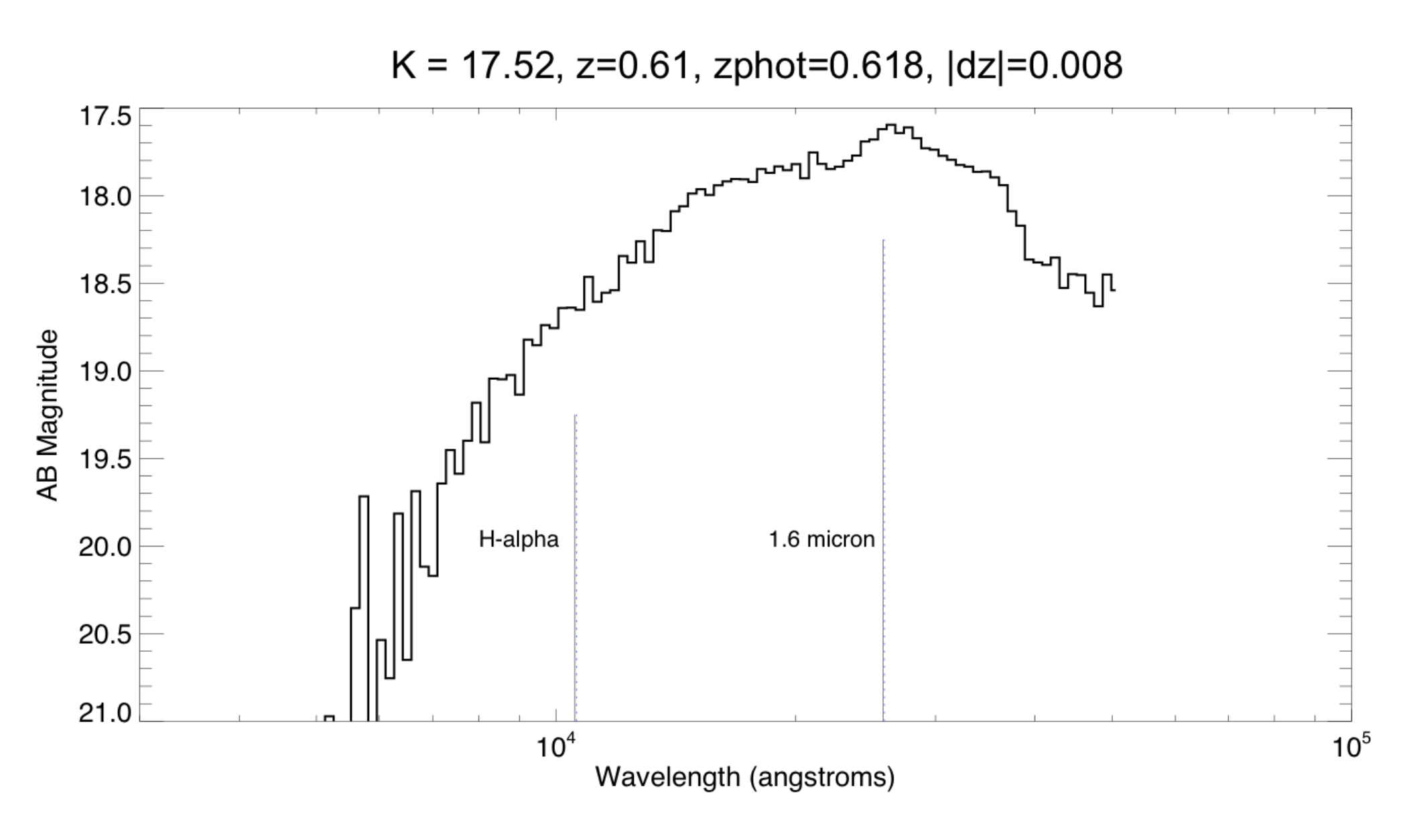}
\includegraphics[width=0.48\textwidth,height=0.4\textwidth]{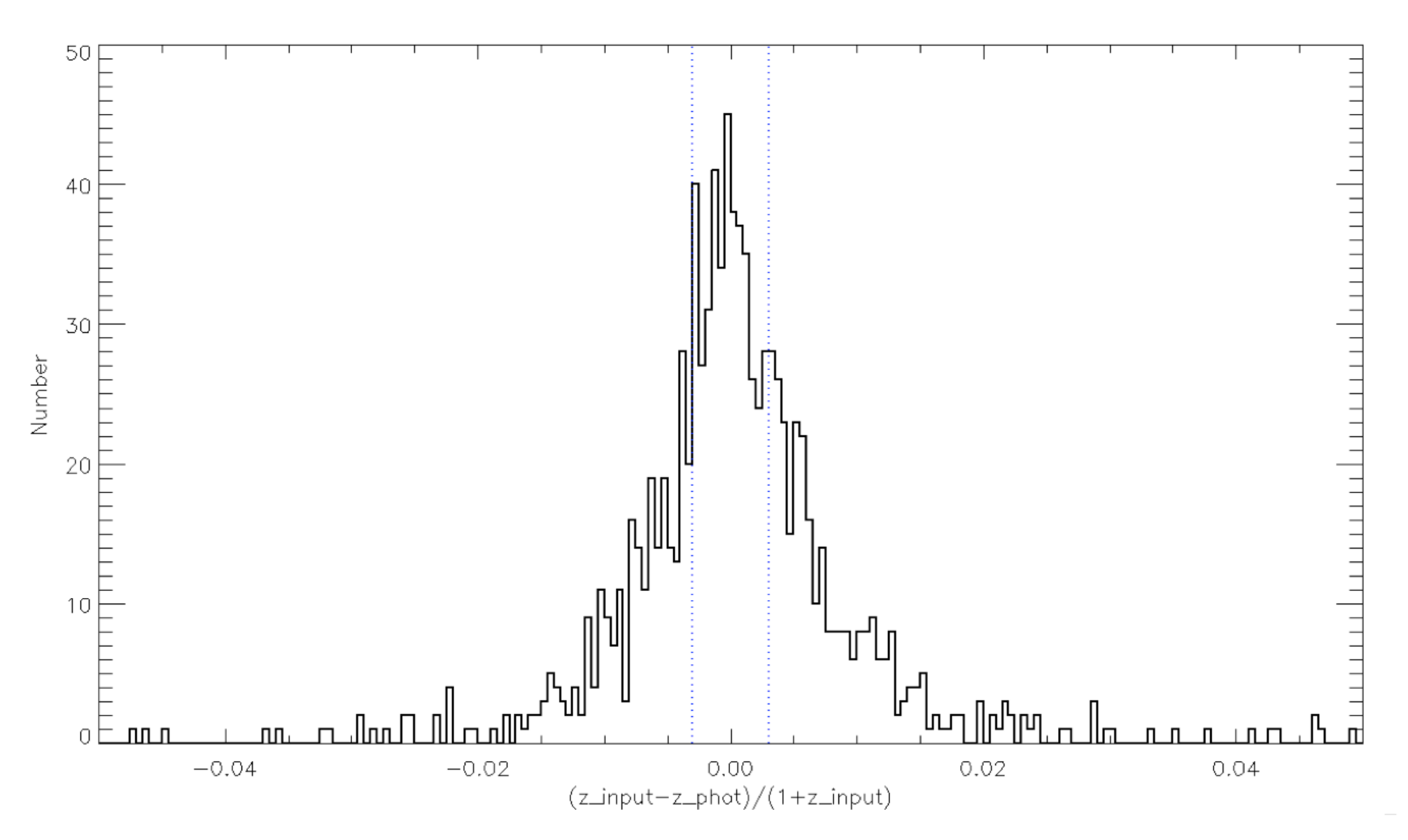}
\caption[]{Left :Example extracted SED for a
  bright object in the SPHEREx simulation pipeline. Right: The actual
  error distributions of objects with
  $\sigma(z)/(1+z)\leq$0.003. The errors are computed as described in
  Sec.~\ref{sec:sigz_error}.}\label{fig:input_output_flux}
\end{figure}

% \begin{figure}[!t]
% \centering
% \includegraphics[width=0.32\textwidth,height=0.35\textwidth]{extracted_vs_input_phot.pdf}
% \includegraphics[width=0.32\textwidth,height=0.385\textwidth]{spherex_extracted_sed.pdf}
% \includegraphics[width=0.32\textwidth,height=0.35\textwidth]{check_zphot_errors.pdf}
% \caption[]{Left: The optimally extracted flux of objects in the F55 filter
%   versus the input flux ($\mu$Jy). The 1-to-1 line is shown in
%   blue. Note that contamination leads to elevated flux estimates for
%   intrinsically fainter sources. Middle:Example extracted SED for a
%   bright object in the SPHEREx simulation pipeline. Right: The actual
%   error distributions of objects with $\sigma(z)/(1+z)\leq$0.003.} 
% \label{fig:input_output_flux}
% \end{figure}

For an unresolved source in a survey with a well known PSF $P$, the
optimal photometric extraction of the total flux $F$ is obtained by
the weighted sum 
\begin{equation}
F = \sum\limits_{i,j}w_{i,j}D_{i,j},
\end{equation}
where $D_{i,j}$ is the sky-subtracted flux measured in pixel $(i,j)$ and the weight function $w$ is constructed as
\begin{equation}
w_{i,j} = \frac{P_{i,j}}{\sum\limits_{i,j}P^{2}_{i,j}}.
\end{equation}
Here $P_{i,j}$ is the fraction of the PSF falling in pixel $(i,j)$.
In the all-sky survey proposed for SPHEREx, the noise budget is
dominated by photon noise from zodiacal light, which is diffuse and
nearly uniform across the FOV.  To quantify the signal-to-noise of a
given detection and how it varies with the relative alignment of the
detector grid, it is useful to define the parameter 
\begin{equation}
N_{eff} = \sum\limits_{i,j}\frac{1}{P_{i,j}^2},
\end{equation}
which represents the effective number of pixels spanned by the PSF. As
$N_{eff}$ increases, the signal from the source is spread over more
pixels and the noise increases by the square root of the number of
pixels.  Therefore, in optimal photometry, the uncertainty in
recovered source flux is related to $N_{eff}$ by 
\begin{equation}
\delta F= \sqrt{N_{eff}} \left[ \delta F(1)\right]
\end{equation}
where $\delta F(1)$ is the uncertainty in flux achieved for a PSF
which is a perfect single pixel square step function. 

\begin{figure}[!t]
\centering
\includegraphics[width=0.4\textwidth]{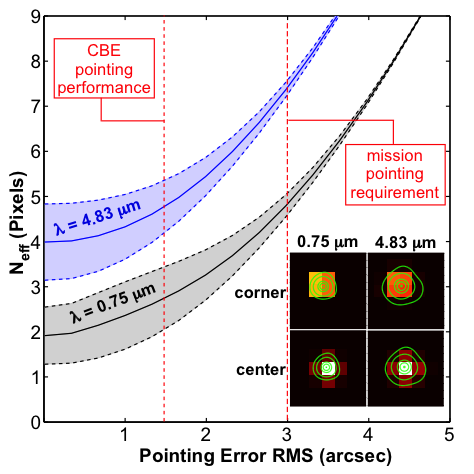}
\caption[]{SPHEREx measures the flux of known sources by combining
  pixels with optimal weights. The effective number of pixels,
  $N_{eff}$ = 1/$\Sigma p_j^2$ affects the flux sensitivity as
  $N_{eff}^{1/2}$, where $p_j$ is the PSF of the $j^{th}$ pixel. The
  plot shows $N_{eff}$ at two wavelengths, calculated from the optical
  PSF with geometric and diffractive PSFs combined with optical
  tolerance errors, as a function of system pointing. $N_{eff}$ depends on
  source position relative to pixel edges (shaded range).}
\label{fig:optphot}
\end{figure}

For SPHEREx, $N_{eff}$ will increase systematically with wavelength
because of the growth of the diffraction limited PSF. Additionally,
for a single wavelength, there is a significant spread in $N_{eff}$
caused by the random alignment of sources with the detector grid. If a
source falls closer to the corner of a pixel, its flux will spill over
to neighboring pixels, while a source landing in the center will
deposit most of its energy in a single pixel. Figure \ref{fig:optphot}
shows the expected range of $N_{eff}$ as a function of wavelength. 

We extract sources in the simulated SPHEREx images optimally, with the
true position of the objects on the detector known from
the higher-resolution ground-based Pan-STARRS/DES data. This is close to
how extraction will be performed in practice. With the position of the
object known, the absolute pointing reconstructed to $< 1"$, and the
PSF accurately characterized in every image,  the values of $P_{i,j}$ that go into the weight function
for optimal extraction can be computed. To do this, the normalized PSF
is resampled onto a fine grid centered at the known position of the object on the detector, and
the value of $P_{i,j}$ in the nearby pixels is found by summing the
portion of $P$ falling within pixel $(i,j)$. In general, the optimally
extracted object fluxes match the input fluxes well.

\subsection{Redshift Determination}

Once the fluxes of the objects have been extracted from the simulated
SPHEREx images, we compute redshift estimates by fitting
template SEDs to the observations. Because of the large number of
filters and the high degree of precision with which we wish to measure
the redshift, off-the-shelf photo-z codes cannot be used. Instead, we
developed a version of the  \emph{Le Phare} code written in highly
optimized C. As with  \emph{Le Phare}, template fitting against the 31
basis SEDs from COSMOS is performed against each extracted object to
determine the redshift probability density function (PDF) $P(z)$. For
each model SED, we sample redshift from 0-2 in steps $dz=0.0005$,
E(B-V) values from 0-1 in steps of $d$E(B-V)=0.05, and four reddening
laws, for 10,418,604 total model grid points. The expected flux in the
SPHEREx bands of each model in the grid is pre-computed by integrating
the modeled SED against the LVF filter profiles and stored in a file
prior to the template fitting. 

The fitting of the photometry from the observed SPHEREx objects to the
model grid is parallelized for efficiency. For each object, the
$\chi^{2}$ offset between the object's photometry and each point in
the model grid is computed after normalizing the model to the observed
photometry. The value of $P(z)$ at each redshift step is computed by
summing the likelihood $e^{-\chi^{2}/2}$ of all models (SED, E(B-V),
reddening law) at that redshift step, thus marginalizing out those
parameters. In this way we derive finely sample $P(z)$ estimates for
the simulated SPHEREx sources.  

\subsection{Error characterization}
\label{sec:sigz_error}

From the likelihood function $P(z)$ we find the best redshift estimate $\bar{z}$ for each object through
the expectation value,
\begin{equation}
\bar{z} = \frac{\sum\limits_{i}z_{i} P(z_{i})}{\sum\limits_{i}P(z_{i})},
\end{equation}
where $i$ runs over all redshift steps. The error in this estimate is extracted from the PDF through
\begin{equation}
\sigma_{\bar{z}} = \sqrt{\frac{\sum\limits_{i}(z_{i}-\bar{z})^{2} P(z_{i})}{\sum\limits_{i}P(z_{i})}}.
\end{equation}

These estimates can then be compared with the known input redshifts to test the performance. 

%\bi
%\item Error validation and catastrophic failures.
%\item Analytical predictions.
%\item PDF(z) for each source\item z,sig(z) from the pdf
%\item Get the solid sig(z)< 0.003/(1+z), etc..
%\item Mention validation of the pipeline on the PRIMUS data.
%\item Mention runs with WISE/DES/PanStarr only.
%\ei

\subsection{Final catalogs and characteristics}

\begin{figure}[!t]
\centering
\includegraphics[width=0.48\textwidth]{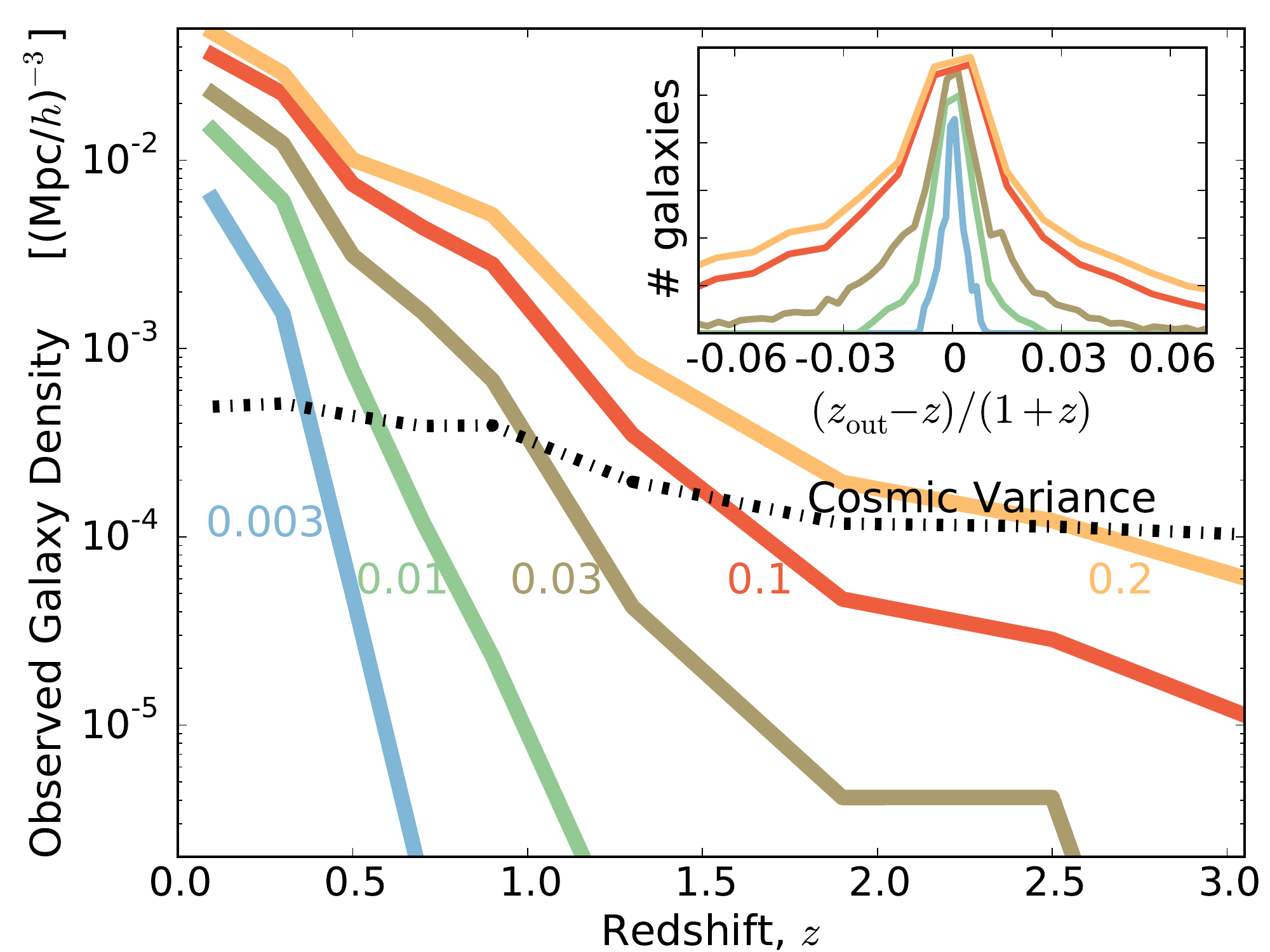}
\includegraphics[width=0.48\textwidth]{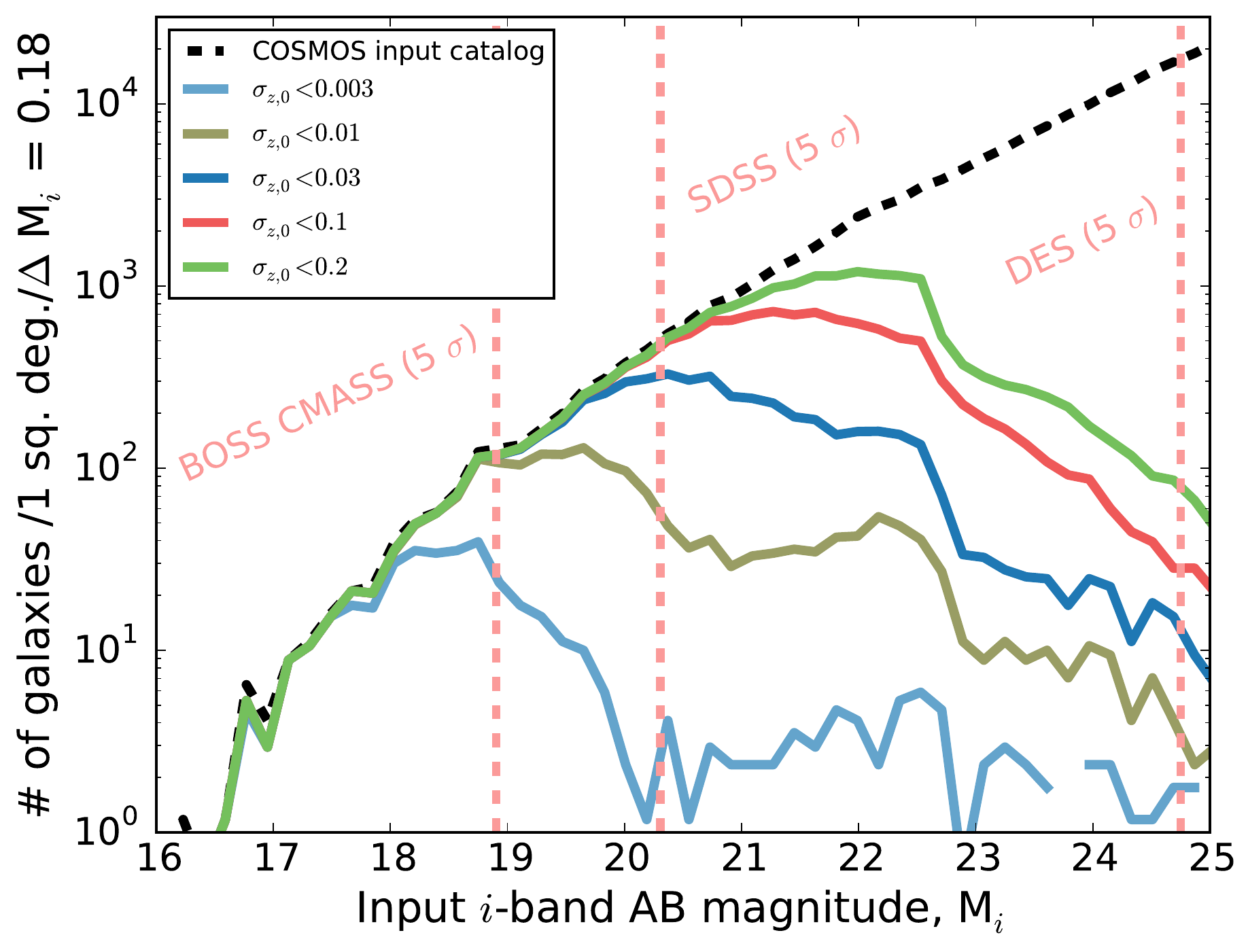}
\caption[]{Left: SPHEREx measures the redshifts of over 500 million
  galaxies, an unprecedented dataset. Accuracy bins are labeled by
  their maximum $\tilde{\sigma}_z \equiv \sigma(z)/(1+z)$. The large sample of low-accuracy
  (0.1) redshifts constrains \fnlloc in the power spectrum. The sample of
  high-accuracy (0.003) redshifts allows clustering measurements on smaller scales and constrains power spectrum parameters
  and the bispectrum. The low-accuracy sample ($\tilde{\sigma}_z < 0.1$) achieves the cosmic
  variance limit (black dashed curve) out to redshift $z$ = 1.6. The
  unnormalized histograms of the simulated redshift errors are shown
  in the inset. Right: Illustration of the $i$-band magnitude distribution of
  SPHEREx galaxies as a function of redshift accuracy.
  The dashed black line shows the distribution of the input COSMOS catalog.
  Also shown are approximate completeness limits for existing surveys (vertical dashed lines).}
\label{fig:nbar_vs_z}
\end{figure}

\begin{figure}[!t]
\centering
\includegraphics[width=0.48\textwidth]{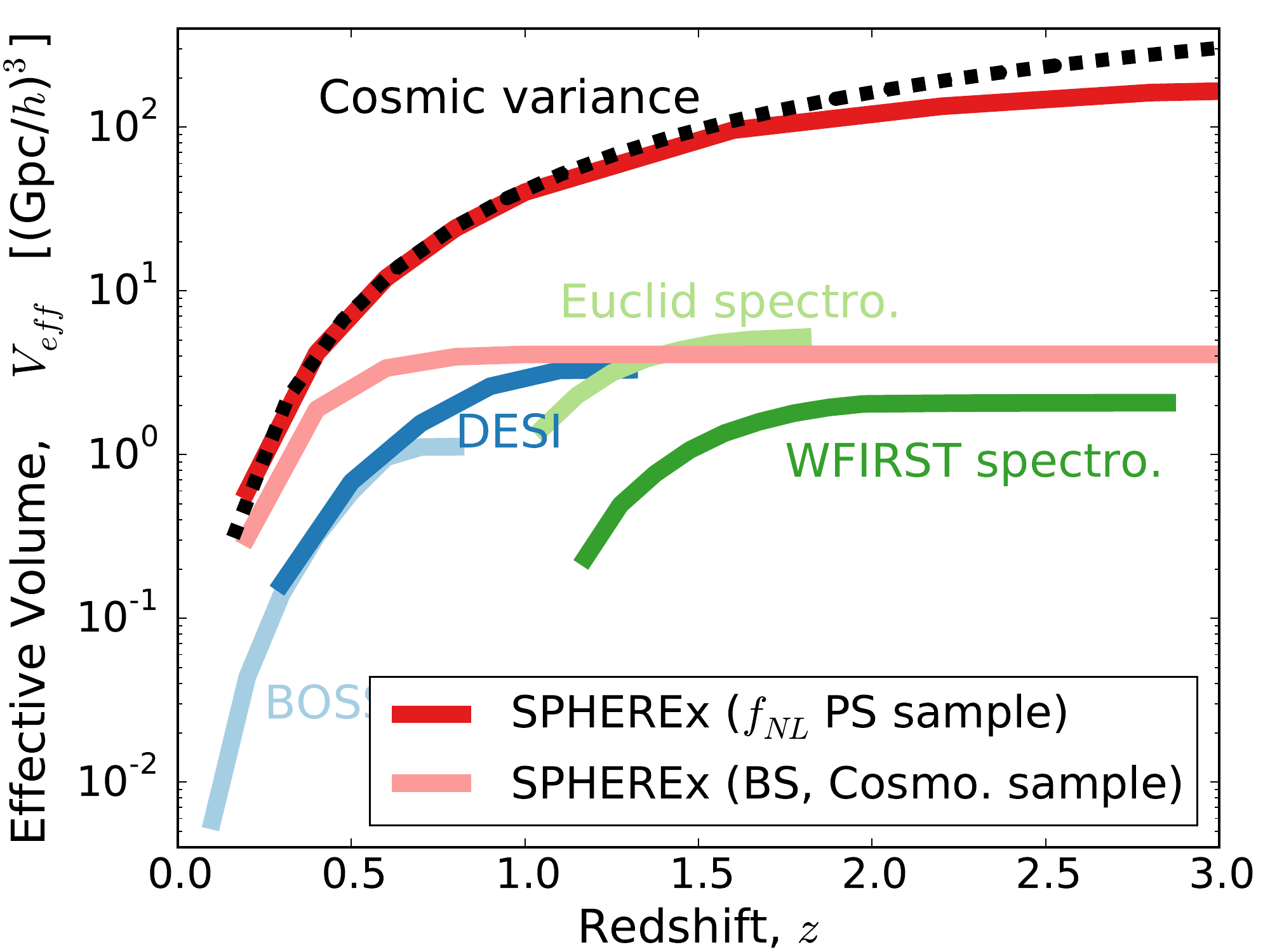}
\caption[]{Effective volume mapped by SPHEREx. The effective volume is
  the physical volume mapped by a given survey, corrected for the
  sampling noise of a finite number of galaxies ($V_{eff}(z) = V_{survey}(z)\sqrt{P_{gg}(k_0,z)/(P_{gg}(k_0,z)+1/\bar{n})}$). For a well-sampled
  survey, the effective volume equals the physical volume. The
  cosmological information content of a given survey is directly
  proportional to the number of independent spatial modes, and
  directly proportional to the effective volume. The SPHEREx \fnl
  power spectrum (PS) sample (red curve) extracts all the cosmological 
  information up to z$\simeq$1.5 (black dashed curve). The SPHEREx
  bispectrum  (BS) and cosmological parameters sample (pink curve),
  based on a  smaller sample of high redshift accuracy galaxies, is uniquely
  powerful at z $<$ 1 compared with other planned surveys. The effective volume is calculated based on
  the galaxy power spectrum at $k_0 = 0.001 h/$Mpc, which is approximately the scale that delivers most of the information
  on $f_{\rm NL}^{\rm loc}$.}
\label{fig:fnl_effect}
\end{figure}

Applying the previous pipeline to our synthetic data-set leads to a 3-D galaxy
catalog with well characterized properties. As we will discuss
below it is beneficial to separate the catalog into multiple
populations with different measured redshift accuracies as they will
correspond to galaxy populations with different bias, and will thus
lead to the potent use of multi-tracer techniques proposed in
\citet{McDonald:2008sh}  and demonstrated observationally with the
GAMA survey \cite{Blake:2013kq}. Besides, different cosmological
parameters, for example \fnlloc or $\alpha_s$ have different redshift accuracy
requirements. In Figure\ref{fig:nbar_vs_z} we illustrate the
reconstructed comoving number density of our sample for various
redshift accuracy bins and the associated $i$-band AB magnitude
distribution. In Figure\ref{fig:fnl_effect} we combine this comoving
  densities with SPHEREx all-sky capabilities to compute the survey effective
  volume and compare it to other current or planned cosmological
  surveys. We note that for SPHEREx we use only 75\% of the sky to
  account for the required galaxy masking. Similarly, bright regions
  will be masked when performing the cosmological analysis in an
  analogous manner to what is performed e.g., for the WMAP or Planck
  analysis.

%Comparing the input and output catalog allow us to illustrate in Figure\ref{lam_mag} \spherex
%selection function for multiple redshift accuracy b.

%\bi
%\item Catalog of... position, z, Luminosity, errors, Mstar, etc...
%\ei

%\begin{figure}[!t]
%\centering
%\includegraphics[width=88mm]{../Plots/nominal_v7_capability_simulation_histo.pdf}
%\caption[]{Recoverd galaxy density for multiple redshift accuracy requirement.}
%\label{fig:ng_density}
%\end{figure}
 
%========================
\section{Cosmological forecasts}
%========================
\label{sec:cosmo}

In this section we introduce the methodology used to forecast the
cosmological constraints obtainable with the SPHEREx galaxy catalog.

\subsection{General Methodology}
\label{subsec:method}

We will rely on the Fisher matrix formalism (see, e.g.,
\cite{TegTayHeav97}). As an
input, we will consider the galaxy catalog produced with the
simulation pipeline described above. We divide the output catalog into $N_z$ redshift
bins with bounds $\{z_{i}^{\rm min},z_{i}^{\rm
  max}\}_{i=1}^{N_z}$. We use a large number of redshift bins and have tested that we are in the regime
  where the forecasted constraints have become insensitive to the exact binning. We subsequently divide the galaxy catalogs into
$N_{\sigma_z}$ redshift accuracy bins with bounds  $\{\tilde{\sigma}_{z,j}^{\rm min},\tilde{\sigma}_{z,j}^{\rm
  max}\}_{j=1}^{N_{\sigma_z}}$,
where $\tilde{\sigma}_z \equiv \sigma_z/(1 + z)$ and $\sigma_z$ the estimated redshift uncertainty
of a galaxy.  
Ultimately, our input catalog leads to a
collection of mean densities $\bar n^g_{j}(z_i)$.
For each redshift bin $z_i$, the forecasts rely on not only the number densities of the subsamples,
but also the values of their mean redshift scatter, $(1 + z_i) \tilde{\sigma}_{z,j}$, and of their galaxy bias, $b_{j}(k,z_i)$.
The former is estimated directly from the catalog by computing the scatter in $(\hat{z} - z)/(1 + z)$
(where $\hat{z}$ is the redshift estimate in the catalog, and $z$ is the true/input redshift)
of all galaxies in the $j$-th $\tilde{\sigma}_z$-bin, and generally the resulting $\tilde{\sigma}_{z,j}$ lies somewhere
in between the defining bounds, $\tilde{\sigma}_{z,j}^{\rm min} - \tilde{\sigma}_{z,j}^{\rm
max}$.
The galaxy bias is estimated using abundance matching, as described in more detail in section \ref{subsec:bias}.
Dividing the sample into different $\tilde{\sigma}_z$-bins ensures that we have subsamples with different galaxy bias,
which enables us to use the multi-tracer technique to optimize cosmological constraints, especially
those on primordial non-Gaussianity \cite{seljak09,hamausetal12,RdPdore14}.

Once the above ingredients, i.e.,~number density, redshift scatter and galaxy bias for each redshift and redshift
uncertainty bin, are computed, we use standard Fisher matrix forecasting methods to calculate expected parameter constraints from
the galaxy power spectrum (Section \ref{subsec:pk}) and bispectrum (Section \ref{subsec:bispec}).
While the effects of cosmological parameters on large-scale structure and the expansion history of the universe
are well known, we wish to highlight the effect of the local
primordial non-Gaussianity parameter, \fnlloc,
as this parameter will be particularly well constrained by SPHEREx.
In the presence of primordial non-Gaussianity,
the linear halo bias receives a scale-dependent correction
\cite{dalaletal08,matver08,slosaretal08,desjselj10},
\be
\label{eq:fnl bias}
b_j(k,z) = b_{G,j}(z) + 2 \, f_{\rm NL}^{\rm loc} \, (b_{G,j}(z) - 1) \, \delta_c \, \frac{3 \Omega_m H_0^2}{2 k^2 T(k) D(z)}.
\ee
%\Eli{Added k-dependence on the left hand side, needed for BS section - hope that's not a problem for PS section}
Here $b_{G,j}(z)$ is the Eulerian, Gaussian halo bias in the $j$-th subsample at redshift $z$,
and $\delta_c$ is the critical overdensity for
halo collapse, here taken to be the critical density for spherical collapse,
$\delta_c = 1.686$. Furthermore, $\Omega_m$ is the matter density at $z=0$ relative to
the critical density, $H_0$ the Hubble constant ($z=0$), $T(k)$ is the transfer function
of matter perturbations, normalized to $1$ at low $k$, and $D(z)$ is the linear
growth function, normalized such that $D(z) = 1/(1 + z)$ during matter domination.
A key feature of this bias correction that will be important later is that
the effect is proportional to $k^{-2}/T(k)$ and therefore most important on
large scales. The bias correction is of order $f_{\rm NL}^{\rm loc}$
at the horizon scale. Moreover, the bias correction is proportional to $b_G - 1$,
so that there is no scale dependence for an unbiased tracer.

%Each sample
%allows us to map the three-dimensional galaxy clustering information
%in each redshift slice with a varying level of accuracy but as will be
%discussed below, their joint analysis will turn out to be very
%beneficial. 
%
%To perform this joint analysis, we will fully include their auto- and
%cross-correlation inside the Fisher matrix, otherwise the usual Fisher
%matrix calculated for the 3D power spectrum in  redshift space
%(including redshift space distortions, Alcock-Paczynski effect, etc.), 

\subsection{Galaxy bias prescription}
\label{subsec:bias}

The Gaussian galaxy bias $b_{G,j}(z)$ (i.e.,~the bias in the absence of primordial non-Gaussianity)
is a crucial input to the Fisher forecasts, with large bias
generally being very beneficial for $f_{\rm NL}^{\rm loc}$ studies.
The catalog directly gives us the mean value of $\tilde{\sigma}_z$ for each redshift accuracy subsample
and the number density as a function of redshift.

We obtain an estimate of the galaxy bias for each subsample (and each redshift bin)
using the abundance matching technique.
Specifically, at each redshift, we match the galaxies with the lowest redshift uncertainty
to the host halos with the largest total mass.
In practice, at a redshift $z$, we first find the minimum halo mass $M_{h,{\rm min}}$ of the $j$-th redshift accuracy subsample
by equating
\be
\bar{n}_h(M_h > M_{h,{\rm min}}; z) = f_{\rm cen} \, \bar{n}^g(\tilde{\sigma}_z < \tilde{\sigma}_{z,j}^{\rm max}; z).
\ee
The central galaxy fraction factor, $f_{\rm cen}$, accounts for the fact that if satellite galaxies
were included, there would not be a one-on-one mapping between galaxies and halos. We take its value to
be $f_{\rm sat} = 0.8$, consistent with recent halo occupation distribution studies (e.g.,~\cite{L12}).
For the halo number density, $\bar{n}_h$, we use the halo mass function fitting formula from \cite{tinkeretal08}.
Specifically, we use their universal mass function, calibrated on N-body simulations,
and apply the empirical redshift scaling $(1 + z)^{-0.26}$.
The value of $\bar{n}^g$ is obtained directly from the catalog.
Once $M_{h,{\rm min}}$ is found, we simply set
the galaxy bias to the halo bias of halos with mass $M_{h,{\rm min}}$,
\be
b_{G,j}(z) = b_h(M_{h,{\rm min}};z).
\label{eq:b_abundance}
\ee
The halo bias on the right hand side above is computed from the fitting formula provided in
\cite{tinkeretal10}, which is based on N-body simulations.

Our method is crude, but gives a reasonable enough estimate of the galaxy bias for the sake of our Fisher forecasts.
The abundance matching assumption is likely violated to some extent, which would lower the bias.
In this sense, our approach is optimistic, as large bias is beneficial for (especially) $f_{\rm NL}^{\rm loc}$.
On the other hand, we are making a conservative choice by using the halo bias {\it at} the halo mass $M_{h,{\rm min}}$
matched to galaxies at $\tilde{\sigma}_{z,j}^{\rm max}$, i.e.,~to the galaxy in the subsample with the largest
redshift scatter and, accordingly, the lowest halo mass and halo bias in the abundance matching picture. In this sense, our approach is conservative.

\subsection{Power spectrum forecasts}
\label{subsec:pk}
%---------------------------------------------

For the power spectrum forecast, we to a large extent follow \cite{seoeis07,pfsreport12}.
We use the information in the full shape of the three-dimensional, redshift-space, galaxy
power spectrum $P_g(k,\mu)$, where $k$ is the wave vector of a galaxy overdensity mode in Fourier space,
and $\mu$ is the cosine of the angle between the mode and the line-of-sight direction (we assume
the plane-parallel approximation throughout).
Since we restrict our analysis to linear and mildy non-linear scales, the starting point
for our power spectrum description is the linear Kaiser model \cite{Kaiser:1987qv}. For a given subsample,
\be
P_{g,j}(k,\mu; z) = b_j^2(k,z) \left[1 + \frac{f(z)}{b_j(k,z)} \mu^2 \right]^2 \, P_m^L(k; z).
\ee
%\Eli{Should this, and following equations in this section be $b_j(k,z)$ or $b_{G,j}(z)$?}
%For now, we drop redshift and subsample dependence of the galaxy bias, but note that for a given redshift ($z_i$) and redshift accuracy bin ($j$),
%the galaxy bias $b$ appearing here and in subsequent expressions
%is really $b_{j}(z_i)$ and is related to the Gaussian bias $b_{G,j}(z_i)$ by Eq.~(\ref{eq:fnl bias}).
Note that $b_{j}(k,z)$ is related to the Gaussian bias $b_{G,j}(z)$ by Eq.~(\ref{eq:fnl bias}).
In the above, $f$ is the linear growth rate of matter perturbations, which enters through redshift space distortions,
and $P_m^L$ is the linear matter power spectrum.

The power spectrum estimated from data is based on the three-dimensional coordinates of
galaxies, which are obtained from the observed angular positions and redshifts by assuming
a reference cosmology. If there is a mismatch between the reference cosmology and the true one,
the scales appearing in the estimated power spectrum will be off from the truth. Specifically,
wave vectors in the reference cosmology are related to the true wave vectors by
\begin{equation}
\label{eq:kref}
k_{\perp,{\rm ref}}= \frac{D_{A}(z)}{D_{A, {\rm
 ref}}(z)}
k_{\perp},
\hspace{1em}
k_{\parallel, {\rm ref}}=\frac{H_{\rm ref}(z)}{H(z)}k_\parallel,
\end{equation}
where $D_A(z)$ is the angular diameter distance, and $H(z)$ the Hubble parameter.
Moreover, the change in volume due to using the reference cosmology causes a vertical shift in
the estimated power spectrum.
The effects described above encode the geometrical information in the galaxy clustering
measurement, including information from the BAO scale and the Alcock-Paczynski effect.
We therefore take these geometric shifts into account in our model for the observed galaxy power spectrum.

Another effect that needs to be included is that, when a sample of galaxies has non-negligible redshift scatter, $\sigma_z = (1+z) \tilde{\sigma}_{z,j}$,
this causes a suppression of the observed power spectrum in the radial direction (see e.g., \cite{asoreyetal12}).
%Moreover, there is a shot noise contribution due to the finite number density of galaxies used to sample the underlying
%density modes. We assume this shot noise factor to be Poissonian, given by $1/\bar{n}^g$, where $\bar{n}^g$ is the galaxy number
%density.
Putting it all together (except for a non-linear correction that we will add at the level of the
Fisher matrix itself), we end up with the following model for the observed galaxy power spectrum (after shot noise subtraction),
\begin{eqnarray}
\label{eq:pk model}
&&P_{g,j}(k_{\perp,{\rm ref}},k_{\parallel,{\rm ref}}; z)= e^{-((1+z) \tilde{\sigma}_{z,j})^2/H^2_{\rm ref}(z) \, k_{\parallel, \rm ref}^2} \, \frac{D_{A,{\rm
 ref}}(z)^2H(z)}{H_{\rm ref}(z)D_A(z)^2} \, b_j^2(z) \,
\left[1+ \frac{f(z)}{b_j(k,z)} \, \frac{k_{\parallel}^2}{k^2}\right]^2 \, P^L_m(k;z) \, ,
\label{eq:Pg}
\end{eqnarray}
where the physical wave vectors $k_{\parallel}, k$, etc.~are related to the wave vectors in the reference cosmology by
Eq.~(\ref{eq:kref}).

Finally, we implement a prescription for the damping of the power spectrum and smearing of the BAO feature due to
the non-linear effect of bulk flows (e.g.,~\cite{seoeis07}). The damping factor in the power spectrum is
\be
\label{eq:nl damping}
\exp\left[-\frac{1}{2} k^2\Sigma_\perp^2 - \frac{1}{2} k^2\mu^2(\Sigma_\parallel^2-\Sigma_\perp^2)\right],
\ee
where the Lagrangian
displacement fields $\Sigma_\parallel$ and $\Sigma$ to model the
smearing effect are given as
\begin{eqnarray}
\Sigma_{\perp}(z)&\equiv & c_{\rm rec}D(z)\Sigma_0, \\
\Sigma_{\parallel}(z)&\equiv & c_{\rm rec}D(z)(1+f(z))\Sigma_0,
\label{eq:sigma}
\end{eqnarray}
with $D(z)$ the linear growth normalized to unity at $z=0$.
The present-day Lagrangian displacement field is
$\Sigma_0=11h^{-1}{\rm Mpc}$ for $\sigma_8=0.8$ \citep[][]{eisetal07}.
The parameter
$c_{\rm rec}$, which we set to $c_{\rm rec} = 0.5$, models the effectiveness of the reconstruction method of the
BAO peak \cite{eisetal07, padmanabhanetal12}. 
We take the exponential factor of the smearing
effect outside of the derivatives of the galaxy power spectrum in the Fisher matrix (see below). This is equivalent
to marginalizing over uncertainties in $\Sigma_\parallel$ and
$\Sigma_\perp$.

In the above, we have given the modeling of the power spectrum of a single sample. With this model established,
it is an easy step to the Fisher matrix for multiple samples. As explained in Section \ref{subsec:method},
our SPHEREx forecasts divide the simulated galaxy catalog into redshift bins, and the sample at each redshift into
subsamples with varying galaxy bias, based on redshift accuracy. To compute the Fisher matrix, we make the standard
assumption of a Gaussian galaxy density field
and assume the power spectra in different redshift bins are uncorrelated. Within each redshift bin,
we use the information in all possible auto- and cross-power spectra of the different subsamples.
The Fisher matrix element for parameters $p_\alpha$ and $p_\beta$ thus becomes
\be
F_{\alpha \beta} = \sum_{z_i} V(z_i) \, \int_{k_{\rm min}}^{k_{\rm max}} \int_{-1}^{1} \, \frac{k^2 dk d\mu}{2 (2\pi)^2} \,
{\rm Tr}\left[{\bf C}^{-1}(k,\mu;z_i) \frac{\partial{\bf C}}{\partial p_\alpha}(k,\mu;z_i) \, {\bf C}^{-1}(k,\mu;z_i) \frac{\partial{\bf C}}{\partial p_\alpha}(k,\mu;z_i) \right] \, \exp\left[- k^2\Sigma_\perp^2 - k^2\mu^2(\Sigma_\parallel^2-\Sigma_\perp^2)\right]
\ee
(we set the fiducial cosmology equal to the reference cosmology).
Here, $V(z_i)$ is the comoving volume in the $z_i$ redshift bin. The $N_{\sigma_z} \times N_{\sigma_z}$ matrix ${\bf C}(k,\mu;z_i)$
is the covariance matrix of the observed galaxy overdensity in the various
redshift uncertainty ($\tilde{\sigma}_z$) bins, including shot noise. In other words, the element corresponding
to redshift uncertainty bins $j_1$ and $j_2$, and redshift bin $z_i$, is given by
\be
{\bf C}_{j_1 j_2}(k,\mu;z_i) = \sqrt{P_{g,j_1}(k,\mu;z_i) \, P_{g,j_2}(k,\mu;z_i)} + \delta^K_{j_1 j_2} \, \frac{1}{\bar{n}^g_{j_1}(z_i)},
\ee
where the power spectrum terms on the right hand side are to be taken from Eq.~(\ref{eq:pk model}) applied to
the galaxy bias and redshift scatter of subsamples $j_1$ and $j_2$,
and the shot noise is assumed to be Poissonian.

We compute this Fisher matrix for the following cosmological parameters,
\begin{eqnarray}
p_{\alpha}&=&\{\omega_b, \omega_c, \omega_\nu, \Omega_{\rm DE}, \Omega_K, n_s, \alpha_s, \sigma_8, w_0, w_a, f_{\rm NL}^{\rm loc}  \}. 
\label{eq:parameters}
\end{eqnarray}
In addition, while the prescription of Section \ref{subsec:bias} gives us an estimate of the galaxy bias (in the absence of non-Gaussianity) of each
subsample, $b_{G,j}(z_i)$, we treat this galaxy bias as a free parameter,
with the value based on Section \ref{subsec:bias} its fiducial value.
This adds $N_z \times N_{\sigma_z}$ bias parameters, $b_{G,j}(z_i)$,
to the Fisher matrix,
which we marginalize over when calculating parameter uncertainties.
The effects of cosmological parameters on the matter power spectrum, which is the main ingredient
to the galaxy power spectrum, is obtained using CAMB \cite{LewChalLas00}.
For the range of scales included in the forecast, we use $k_{\rm min} = 0.001 h/$Mpc and $k_{\rm max} = 0.2 h/$Mpc.
Finally, we always include a Planck CMB prior to break parameter degeneracies.
This prior encodes the information in the CMB angular power spectra, and does {\it not} include
bispectrum information. Therefore, it does not provide any direct information on $f_{\rm NL}^{\rm loc}$.

\subsection{Bispectrum methodology}
\label{subsec:bispec}
%----------------------------------

\begin{figure}[!t]
\centering
\includegraphics[width=0.9\textwidth,height=0.33\textwidth]{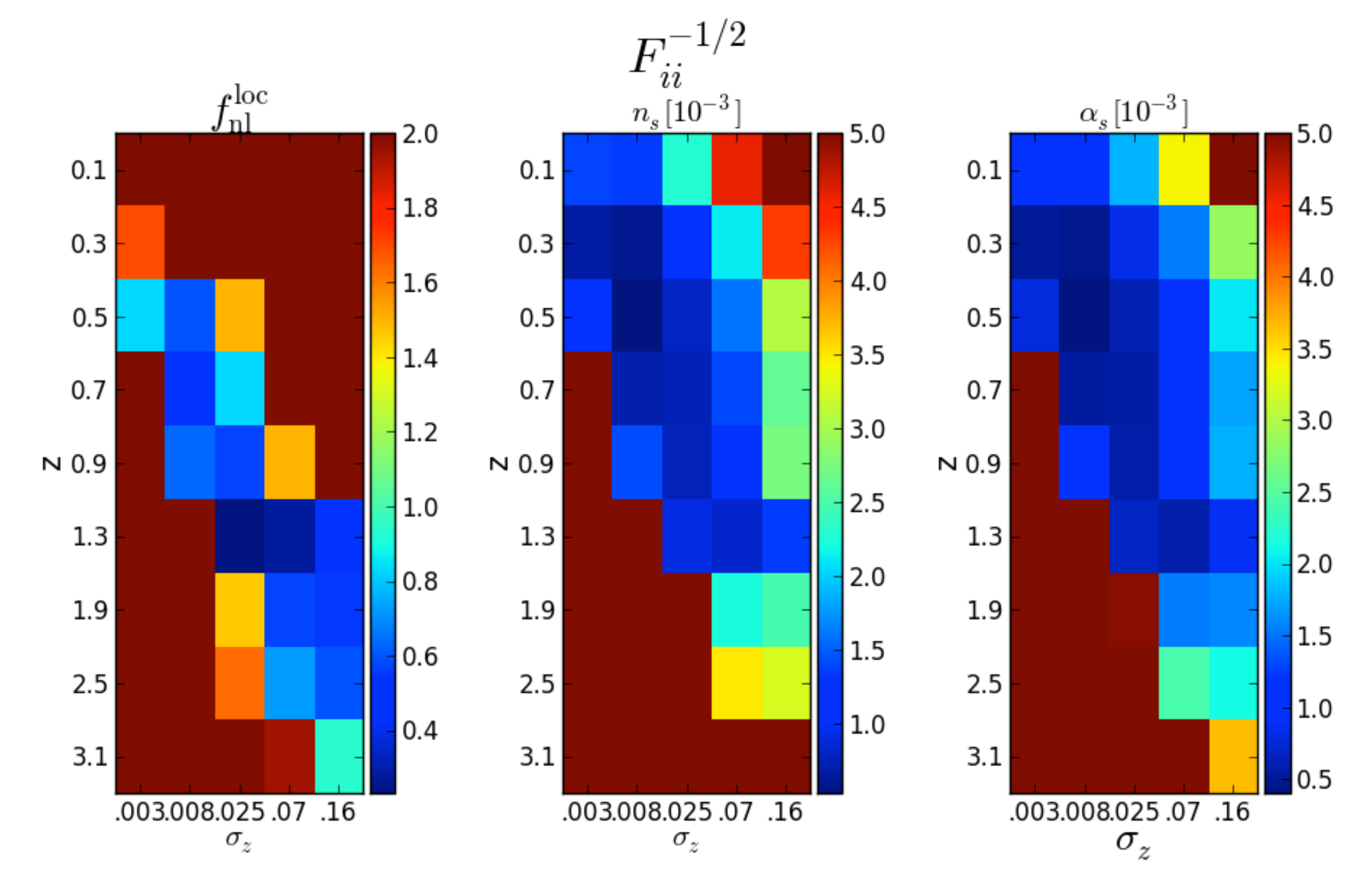}
\caption[]{Unmarginalized constraining power from the galaxy bispectrum in different bins in redshift and redshift accuracy.
The horizontal axis shows the mean redshift scatter $\tilde{\sigma}_z = \sigma_z/(1 + z)$ of each galaxy subsample.
From left to right, constraining power is shown on \fnlloc, $n_s$ and $\alpha_s$.}
\label{fig:BS_Fii}
\end{figure}

In addition to the power spectrum described in the previous
section, structure formation also imprints cosmological
information in the higher order moments of the density
field. Gravitational instability amplifies the initial perturbations
non-linearly on small scales so that the three-point correlation
function and its counterpart in Fourier space, the bispectrum, become
non-zero. These three-point statistics are sensitive test of the
gravitational instability paradigm but also of the galaxy
biasing. They can be used to break the degeneracy between the linear bias
and the matter density parameter present in power spectrum
measurements. In addition to the memory of the infationary bispectrum
present in the initial conditions, their sensitivity to the bias make
them very sensitivite probe of the primodial non-Gaussinianity, in a
complementary manner to the power spectrum.  Most recently, the
bispectrum has been measured  using the SDSS DR11 data and lead to
insightful constraints on the linear growth of structures
\cite{Gil-Marin:2014baa,Gil-Marin:2014sta}. 

We will focus here on third-order statistics in form of the $3D$ galaxy
bispectrum 
\begin{equation}
 \hat{B_{\rm g}}(\mathbf k_1, \mathbf k_2, \mathbf k_3) = (2 \pi)^3
 \delta_{\mathrm D}(\mathbf k_1 +\mathbf k_2+\mathbf k_3)
 \hat{\delta_{\rm g}}(\mathbf k_{1})\hat{\delta_{\rm g}}(\mathbf k_{2
 })\hat{\delta_{\rm g}}(\mathbf k_{3})\,,
\end{equation}
which has the potential to probe interactions in
the early Universe through measurements of the primordial bispectrum, and has considerable constraining power due
to the enormous number of fundamental triangles contained in the survey volume.

We model the late-time galaxy spectrum as the sum of the primordial bispectrum and 
the tree-level bispectrum, which is caused by non-linear structure
formation, modulated by non-Gaussianity induced scale-dependent
galaxy bias. For local primordial non-Gaussianity, this can be written as
\begin{align}
\nonumber B_{\mathrm g,j} \left(\mathbf{k}_1, \mathbf{k}_2;z_i \right) = &
\prod_{l =1}^{3}b_{j}(k_l,z_i) %\left(b_i+ f_{\rm NL}^{\rm loc}  \Delta b_i(k_j)\right)
\left\{
\left[
\frac{ \Omega_{\mathrm m} H_0^2 k_3^2 T(k_3)}{k_1^2 T(k_1) k_2^2 T(k_2) D(z_i)} + 
2 F_2^{(\mathrm s)}(\bm{k}_1,\bm{k}_2)
\right] P_{\mathrm m}(k_1;z_i)P_{\mathrm m}(k_2;z_i) +2\;\mathrm{perm.}\right\}\\
&
+b_{2,j}(z_i)b_j(k_1,z_i)b_G(k_2,z_i)
P_{\mathrm m}(k_1;z_i)P_{\mathrm m}(k_2;z_i) +2\;\mathrm{perm.}
\,,
\end{align}
with $b_{2,j}$ the quadratic galaxy bias parameter of galaxies in redshift bin $z_i$ and redshift scatter $\tilde \sigma_{z,j}$, which is determined analogously to Eq.~\ref{eq:b_abundance}, and where we have already imposed the triangle condition $\mathbf k_3 = -\mathbf k_1-\mathbf k_2$. In this expression, the two terms in curly brackets describe the inflationary matter bispectrum propagated to redshift $z$ and the tree-level matter bispectrum, the last term denotes the contribution from non-linear galaxy clustering with $b_{2}$ the quadratic galaxy bias.
As in the power spectrum forecast, we calculate the covariance in the
Gaussian approximation, so that different triangles configurations are uncorrelated, and the variance of each configuration is given by \citep{Jeong:2010phd}
\begin{align}
\mathrm{Var}\left(B_{\mathrm g,j}( \mathbf k_1,\mathbf
k_2;z_i)\right) = V_{\rm
 survey}(z_i) \left(P_{\mathrm g,j}(k_1;z_i)+\frac{1}{\bar{n}^{\mathrm g}_j(z_i)}\right)
\left(P_{\mathrm g,j}(k_2;z_i)+\frac{1}{\bar{n}^{\mathrm g}_j(z_i)}\right)
\left(P_{g,j}(k_3;z_i)+\frac{1}{\bar{n}^{\mathrm g}_j(z_i)}\right)\,.
\end{align}

The impact of redshift errors on the bispectrum is more
complicated than in the power spectrum case, as each triangle side now has as a
component along the line of sight. To simplify this calculation, we
restrict the bispectrum Fisher matrix to triangle configurations with
an an-isotropic upper limit $\tilde{ \mathbf k}_{\mathrm{max}} =
(k_\mathrm{max},k_\mathrm{max}, k(\tilde\sigma_{z,j};z_i))$ on each Fourier mode in the triangle; here the limit
perpendicular to the line of sight is given by the usual
$k_\mathrm{max}$ determined by the scale at which tree-level modeling
becomes inaccurate, and $k(\tilde\sigma_{z,j};z_i) = 2 \pi H(z_i)/(c \sigma_{z,j})$  
\begin{align}
F_{\alpha \beta} = \sum_{z_i} \sum_{\mathbf k_1} \sum_{\mathbf k_2}\, \Theta(k_\mathrm{max} - k_3)\, \Theta(k(\tilde\sigma_{z,j};z_i)) - k_{3,z})\,
\frac{\partial B_{\mathrm g,j}(\mathbf k_1,\mathbf k_2;z_i)}{\partial p_\alpha}
 \frac{1}{\mathrm{Var}\left(B_{\mathrm g,j}( \mathbf k_1,\mathbf
k_2;z_i)\right)}
\frac{\partial B_{\mathrm g,j}(\mathbf k_1,\mathbf k_2;z_i)}{\partial p_\beta} \,, 
\end{align}
where the binned sum over triangle sides is defined as
\be
 \sum_{\mathbf k} =  \sum_{k_{x} = k_{\mathrm{min}}}^{k_\mathrm{max}}\frac{\Delta k_x}{k_F}\sum_{k_{y} = k_{\mathrm{min}}}^{k_\mathrm{max}}\frac{\Delta k_y}{k_F}\sum_{k_{z} = k_{\mathrm{min}}}^{k(\sigma_{z,i})}  \frac{\Delta k_z}{k_F}\, \Theta(k_\mathrm{max} - k)\,,
\ee
with $\Delta \mathbf k$ the bin width, which we choose as $\Delta
\mathbf k = \tilde{ \mathbf k}_{\mathrm{max}}/10$ for computational
speed, and where the step function $\Theta$ restricts the analysis to $k <k_\mathrm{max}$. 

For galaxy samples with non-negligible redshift uncertainties, a multi-tracer
bispectrum analysis is computationally prohibitively expensive. Hence our Fisher matrix analysis includes one galaxy population per redshift bin, and we choose the $\tilde\sigma_j(z_i)$ as follows: In the cosmic
variance limit, galaxy samples with $k(\sigma_z;z)\sim k_\mathrm{max}$
have the most constraining power due the rapid increase of the number
of accessible triangles with $\tilde{ \mathbf
  k}_{\mathrm{max}}$. However, for the SPHEREx galaxy selection
function, the low redshift uncertainty galaxy samples become shot
noise dominated near $k_\mathrm{max}$ around $z \sim 1$, and the
constraining power is higher for  galaxy samples with lower shot noise
(and higher redshift uncertainty), as illustrated in
Figure~\ref{fig:BS_Fii}.

%\subsection{Modeling assumptions}
%%-------------------------------
%
%Abundance matching and sensitivity to various prescriptions.
%\bi
%\item Describe handling of multiple resolution and multiple bias.
%\item
%\ei

\subsection{Results}
%-----------------

\begin{figure}[!t]
\centering
\includegraphics[width=0.48\textwidth]{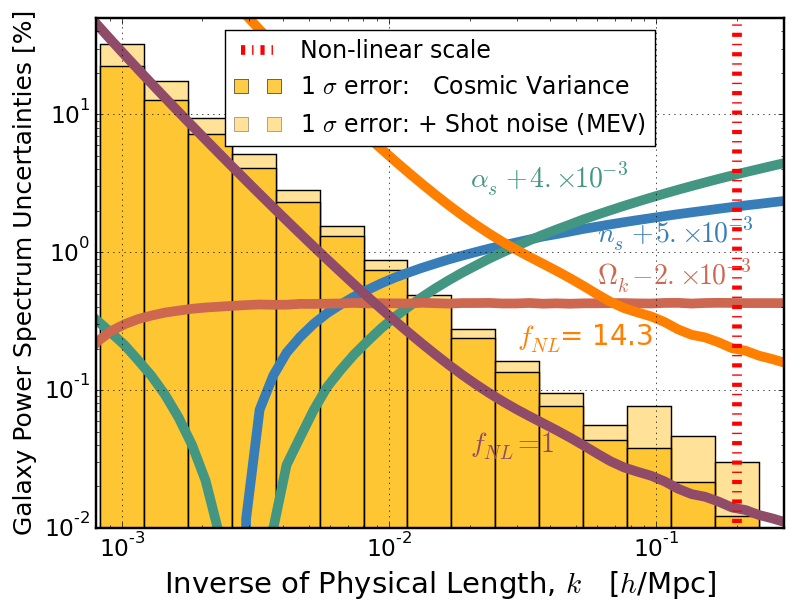}
\caption[]{SPHEREx relative errors for the galaxy 3-D power-spectrum as a function of wavenumber. The relative statistical and cosmic
  variance error bars correspond at the weighted  combination of all
  redshift. The solid curves correspond to the theoretical model
  relative variation with regards to the fiducial one. The theoretical
  models are computed at $z=1$.}
\label{fig:k_pk_relative}
\end{figure}

% \begin{figure}[!t]
% \centering
% \includegraphics[width=0.9\textwidth]{../Plots/plot_z_ngal_v3.pdf}
% \caption[]{} 
% \label{fig:nbar vs z}
% \end{figure}

In practice, we divide the simulated catalog into five redshift accuracy bins,
given by
\be
\tilde{\sigma}_z \equiv \frac{\sigma_z}{1 + z} = 0 - 0.003 - 0.01 - 0.03 - 0.1 - 0.2.
\ee
This leads to the cumulative comoving number densities shown in Figure \ref{fig:nbar_vs_z}
(the inset illustrates the distribution of the estimated redshift relative to the true redshift
for each subsample).
The number density in each subsample is of course given by the difference between subsequent curves.
We now turn to the cosmological constraints that can be obtained with these samples, using the methods
extensively discussed in the previous sections.
It is useful to separate the discussion of the constraints on primordial non-Gaussianity from those
on other parameters. The $f_{\rm NL}^{\rm loc}$ power spectrum constraints are driven by the largest scales available to the survey
and, as a consequence, do not require high redshift accuracy. The suppression of information
due to redshift scatter is such that almost the full power of the first four samples, i.e.,~up to $\tilde{\sigma}_z = 0.1$,
can be used. While the suppression becomes large around $\tilde{\sigma}_z = 0.1$, even the $\tilde{\sigma}_z = 0.1 - 0.2$
bin still contributes to the final constraint. Thus, for $f_{\rm NL}^{\rm loc}$, we have a very deep sample at our disposal.
The other parameters, on the other hand, rely strongly on the smallest scales accessible to the survey and to our modeling
capabilities. Since the main focus of the SPHEREx cosmology program is inflation, we are here particularly interested in
the parameters $f_{\rm NL}$, $n_s$, $\alpha_s$ and $\Omega_k$. In addition, we will be able to place tight bounds on the dark energy
parameters $w_0$ and $w_a$, and on the sum of neutrino masses, $\Sigma m_\nu$. For all these parameters, the
most important sample is the highest redshift accuracy one, $\tilde{\sigma}_z < 0.003$ (i.e.,~nearly spectroscopic redshift
quality for cosmology purposes), although the $\tilde{\sigma}_z = 0.003 - 0.01$ also contributes significantly,
despite its relatively large suppression of information due to redshift scatter.

Starting with $f_{\rm NL}^{\rm loc}$, Figure \ref{fig:k_pk_relative} illustrates the amplitude of the scale-dependent bias
signal for $f_{\rm NL}^{\rm loc}$ values at the level targeted by SPHEREx. While in practice, we use many redshift and
redshift accuracy bins, the plot condenses down the information to a single effective redshift for illustration purposes.
The error bars show the sensitivity of SPHEREx, with and without cosmic variance.
The plot shows that the SPHEREx power spectrum measurement should be sensitive to $|f_{\rm NL}^{\rm loc}| \sim 1$.
The results of the detailed Fisher forecast are shown in the first row of Table \ref{tab:forecasts}.
Indeed, we find that SPHEREx can reach $\sigma(f_{\rm NL}^{\rm loc}) = 0.87$.
The second column shows that the galaxy bispectrum can reach significantly higher precision, with the
best possible SPHEREx bound coming from the combination of the two (third column).
The EUCLID column shows the constraint expected from the EUCLID spectroscopic galaxy clustering
analysis, computed with the same power spectrum forecast method as used for SPHEREx.
We do not show EUCLID bispectrum forecasts as they do not exist yet.
Finally, the last column shows the current $f_{\rm NL}^{\rm loc}$ constraint from the Planck CMB
bispectrum \cite{plancknongauss14}.
Based on these forecasts, even with the galaxy power spectrum only,
we thus expect an \fnl~bound a factor $\sim 6$ better than the expected constraint
from the EUCLID spectroscopic sample, and a similar factor with respect to
the current CMB-based bound. When taking into account the bispectrum,
the improvement over EUCLID is substantially more.

\begin{table*}[hbt!]
\begin{center}
\small
\begin{tabular}{c|ccccc}
\hline\hline
$1\sigma$ errors &  PS & Bispec & PS + Bispec & EUCLID & Current \\
\hline\hline
$f_{\rm NL}^{\rm loc}$ & 0.87 & 0.23 & 0.20 & 5.59 & 5.8 \\
Tilt $n_s$ ($\times 10^{-3}$) & 2.7 & 2.3 & 2.2 & 2.6 & 5.4 \\
Running $\alpha_s$ ($\times 10^{-3}$) & 1.3 & 1.2 & 0.65 & 1.1 & 17 \\
Curvature $\Omega_K$ ($\times 10^{-4}$) & 9.8 & NC & 6.6 & 7.0 & 66 \\
Dark Energy FoM $= 1/\sqrt{{\rm Det Cov}}$ & 202 & NC & NC & 309 & 25 \\
\hline\hline
\end{tabular}
\caption{Forecasted constraints from SPHEREx on inflationary parameters and on the dark energy figure of merit (FoM). See text for details.
Estimates of current constraints and future constraints from the EUCLID spectroscopic galaxy survey are shown for comparison.
NC stands for ``Not Calculated''.
}
\label{tab:forecasts}
\end{center}
\end{table*}

\begin{figure}[!t]
\centering
\includegraphics[width=0.7\textwidth]{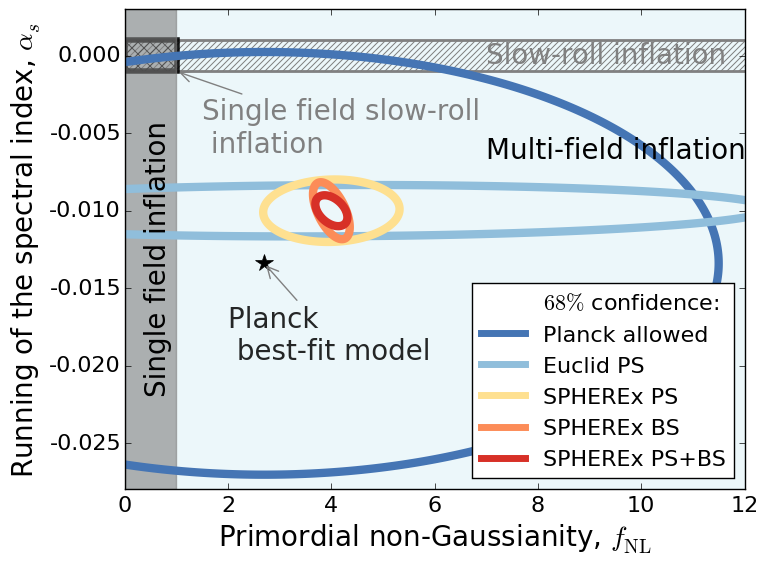}
\caption[]{68\% confidence level contours in the \fnl-$\alpha_s$ plane from the SPHEREx power spectrum and bispectrum analyses.
We show the constraints from current data and those expected from the EUCLID spectroscopic survey
for comparison. Various areas
  of interest for inflation model classes are presented as well.}
\label{fig:al_fnl_contour}
\end{figure}

Moving on to cosmology beyond primordial non-Gaussianity, the next rows in Table \ref{tab:forecasts}
show the expected constraints on the tilt $n_s$ and running $\alpha_s$ of
the primordial power spectrum, the spatial curvature of the Universe $\Omega_K$,
and the dark energy figure of merit (we here take FoM$=1/\sigma(w_p) \sigma(w_a)$,
where $w_p$ is the dark energy equation of state at the pivot redshift).
In all cases (including $f_{\rm NL}^{\rm loc}$), we marginalize over all parameters of the
vanilla $\Lambda$CDM model, $\omega_b, \omega_c, \Omega_\Lambda, n_s, \sigma_8, \tau$
(we remind the reader that we always include a CMB power spectrum prior to
break parameter degeneracies) and over galaxy bias.
For the prospected uncertainties on the inflation parameters $n_s$, $\alpha_s$ and $\Omega_K$,
we additionally marginalize over the full inflationary parameter space, including
$n_s$, $\alpha_s$, $\Omega_K$ and $f_{\rm NL}^{\rm loc}$.
The estimates of current constraints beyond $f_{\rm NL}^{\rm loc}$
come from Planck CMB data for $n_s$ (Table 5 of \cite{planckcosmoparam}),
from Planck CMB plus a compilation of BAO measurements for $\alpha_s$ and $\Omega_K$ (Table 10 of \cite{planckcosmoparam}),
and we base the dark energy FoM on the Planck + BAO $w_0 - w_a$ constraints in Table 15 of \cite{andersonetal14}.

For parameters beyond $f_{\rm NL}^{\rm loc}$,
we thus see that
SPHEREx will deliver
constraints comparable to those expected from the EUCLID spectroscopic survey. Since the SPHEREx constraints
are driven by the low-$\tilde{\sigma}_z$ sample of SPHEREx, which is restricted to $z < 0.9$, SPHEREx
provides information independent of and complementary to the EUCLID galaxies, which are at $z > 0.9$.

Figure \ref{fig:al_fnl_contour} provides an illustration of the ability of SPHEREx to constrain the inflationary parameter space.
SPHEREx will srongly narrow down the allowed parameter space in the $f_{\rm NL}^{\rm loc} - \alpha_s$ plane.
Specifically, while single-field models predict negligible \fnlloc (in the squeezed limit) \cite{maldacena03,cremzal04},
multi-field models generically
generate $|f_{\rm NL}^{\rm loc}| \gtrsim 1$ \cite{Alvarez14,lythetal03,zal04}.
Moreover, various non-primordial effects, such as ``GR effects'' and non-linear evolution,
produce interesting cosmological signals equivalent to
those of primordial non-Gaussianity of order $|f_{\rm NL}^{\rm loc}| \sim 1$.
SPHEREx, with its sensitivity $\sigma(f_{\rm NL}^{\rm loc}) = 0.87$ from the power spectrum,
and $\sigma(f_{\rm NL}^{\rm loc}) = 0.23$ from the bispectrum is thus in an excellent position to
detect the non-primordial signal, and to distinguish between single- and multi-field models.
Crucially, SPHEREx has two more or less {\it independent} measurements, since the bispectrum largely gets
its information from different scales than the power spectrum,
so that the new insights it will provide into the nature of inflation will be very robust.
Finally, Figure \ref{fig:al_fnl_contour} also shows that, by measuring $\alpha_s$,
SPHEREx will enable a stringent test of the class of slow-roll inflation models.

While {\it local} non-Gaussianity has a particularly striking effect on galaxy bias,
because it appears on the largest scales and is weakly degenerate with other parameters (with regards to
the galaxy power spectrum), other types of non-Gaussianity also leave a scale-dependent signature in the galaxy bias.
While more detailed studies remain to be performed, we have carried out preliminary forecasts of how well
SPHEREx will constrain non-Gaussianity of the equilateral type, parameterized by $f_{\rm NL}^{\rm eq}$.
In this case, the scale-dependence of the bias correction is given by $\Delta b \propto 1/T(k)$, e.g.,~\cite{schmidtkam10,desjetal11},
where $T(k)$ is the linear transfer function of matter perturbations normalized to unity at low $k$ (large scales).
Thus, the effect of equilateral non-Gaussianity is a bias correction at small scales, rather than large (since
the $k$-independent contribution to the bias at low $k$ is not observable).
In the large-$k$ limit, it scales like $\Delta b \propto k^2 \ln(k)$.
Making the same assumptions as for the $f_{\rm NL}^{\rm local}$
calculations, our forecasts show that the SPHEREx galaxy power spectrum
can reach a constraint \fnlequil$\approx 7$.
Our bispectrum forecasts suggest that the bispectrum can do at least as well on \fnlequil as the power spectrum.
However, we note that because of the scale
dependence in the equilateral case, \fnlequil has a much stronger degeneracy with other parameters,
and perhaps more worryingly, with non-linear effects. A more robust forecast would take this into account
by using a more detailed model of non-linear bias, non-linear evolution, and non-linear redshift space distortions.

%===========================
\section{Systematic effects analysis}
\label{sec:sys}
%===========================

\begin{figure}[!t]
\centering
\includegraphics[width=0.9\textwidth]{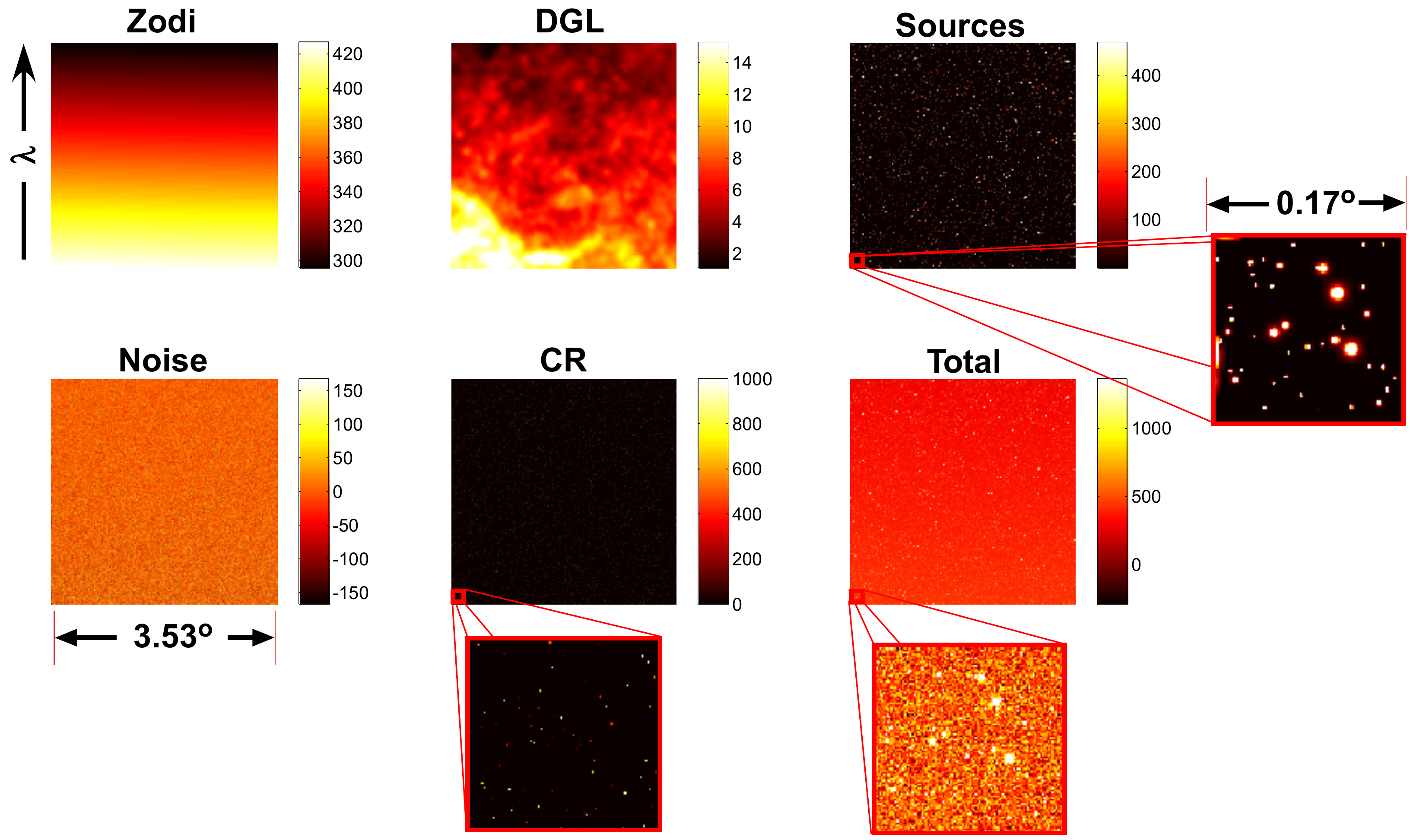}
\caption[]{Components of a single simulated SPHEREx exposure for the
  short wavelength band towards the region $[\alpha,\delta] =
  [16^{h}11^{m}36^{s}, +54^{d}23^{m}17^{s}]$.  Color stretch indicates
  $\lambda I_{\lambda}$ and is given in units of
  nW~m$^{-2}$~sr$^{-1}$.  Counter-clockwise from the upper right, the
  components show: a map of the unresolved sources, the Diffuse
  Galactic light, Zodiacal light, Noise (including contributions from
  read and photon statistics), cosmic rays, and the sum of all
  components.  Insets show a subsection of the same images.} 
\label{fig:imagesims}
\end{figure}

While the simulation suite described in Sec.~\ref{sec:gal_catalog}
provides a clear assessment of the statistical uncertainties in the
\spherex all-sky galaxy clustering survey, it does not encompass any
systematic effects associated with the unique scan strategy and spectral sampling techniques employed.
In this section, we discuss and quantify the various systematic
effects, both instrumental and astrophysical and the methods of
mitigation.  Of particular interest are the sources of error coherent
on large angular scales ($\sim10^\circ$), which introduce artificial
correlations in the 3-D source catalogs.  We estimate all sources of
error must be controlled to 0.2\% (rms per dex in wavenumber) to
measure \fnl=1. Table~\ref{tab:syst} provides a concise summary of the effects described below.  The images in
Figure~\ref{fig:imagesims} shows the various astrophysical components
which can contribute systematics as they appear on the array. 

\bi

\item {\bf Galactic Extinction} Variable atmospheric extinction has
  been a problem for past \fnlloc studies (e.g., \cite{Pullen:2012rd})
  SPHEREx avoids this by observing from space. Spatially variable
  galactic dust extinction leads to variations in depth of galaxy
  surveys if not corrected. It is an order of magnitude smaller at 3.2
  $\mu$m (the 1.6 $\mu$m bump at $z=1$) than in the $i$-band. Standard
  extinction  corrections and template projections have sufficient
  accuracy to remove this modest  variation.

\begin{figure}[!t]
\centering
\includegraphics[width=0.45\textwidth,height=0.3\textwidth]{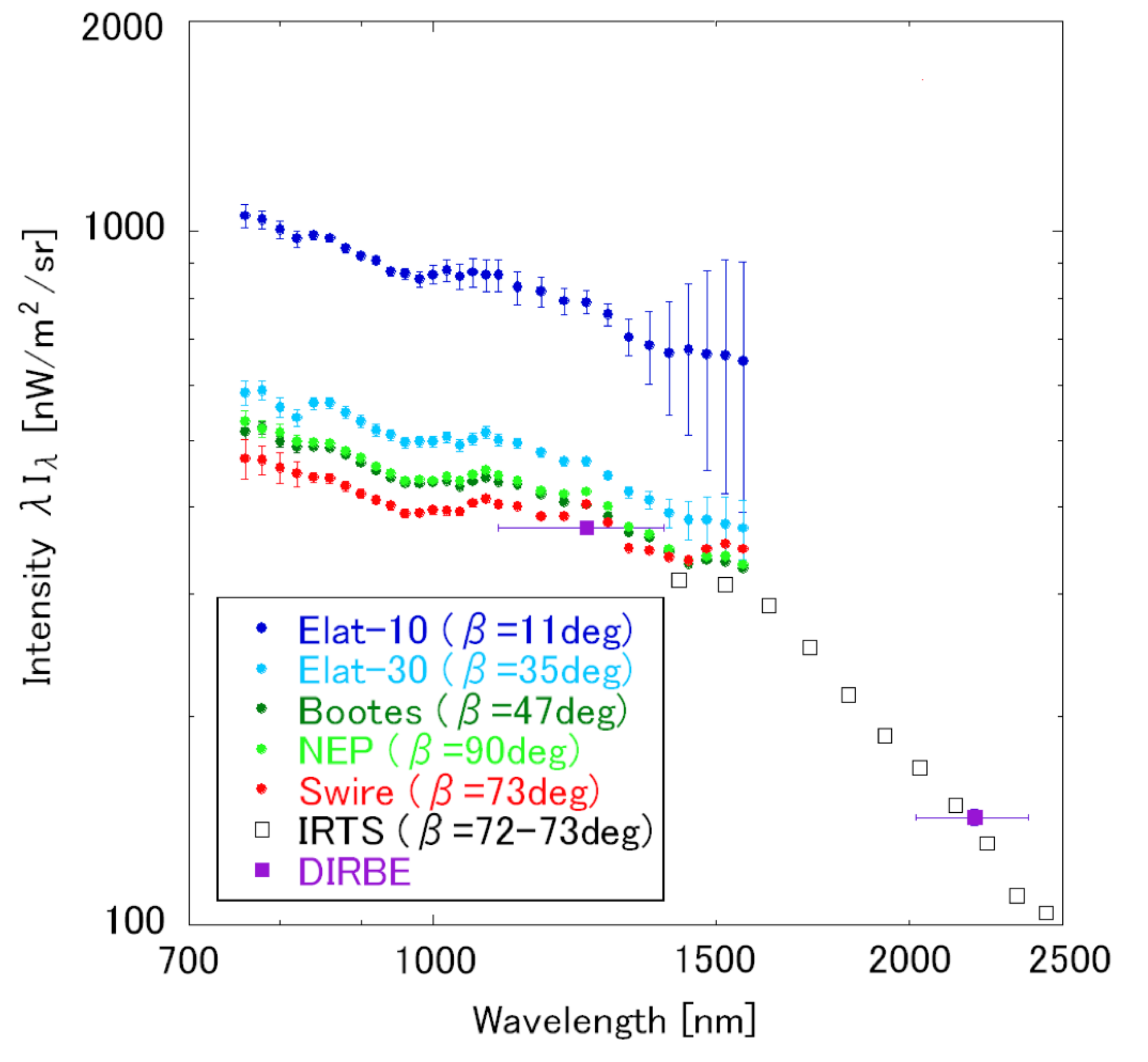}
\includegraphics[width=0.45\textwidth,height=0.3\textwidth]{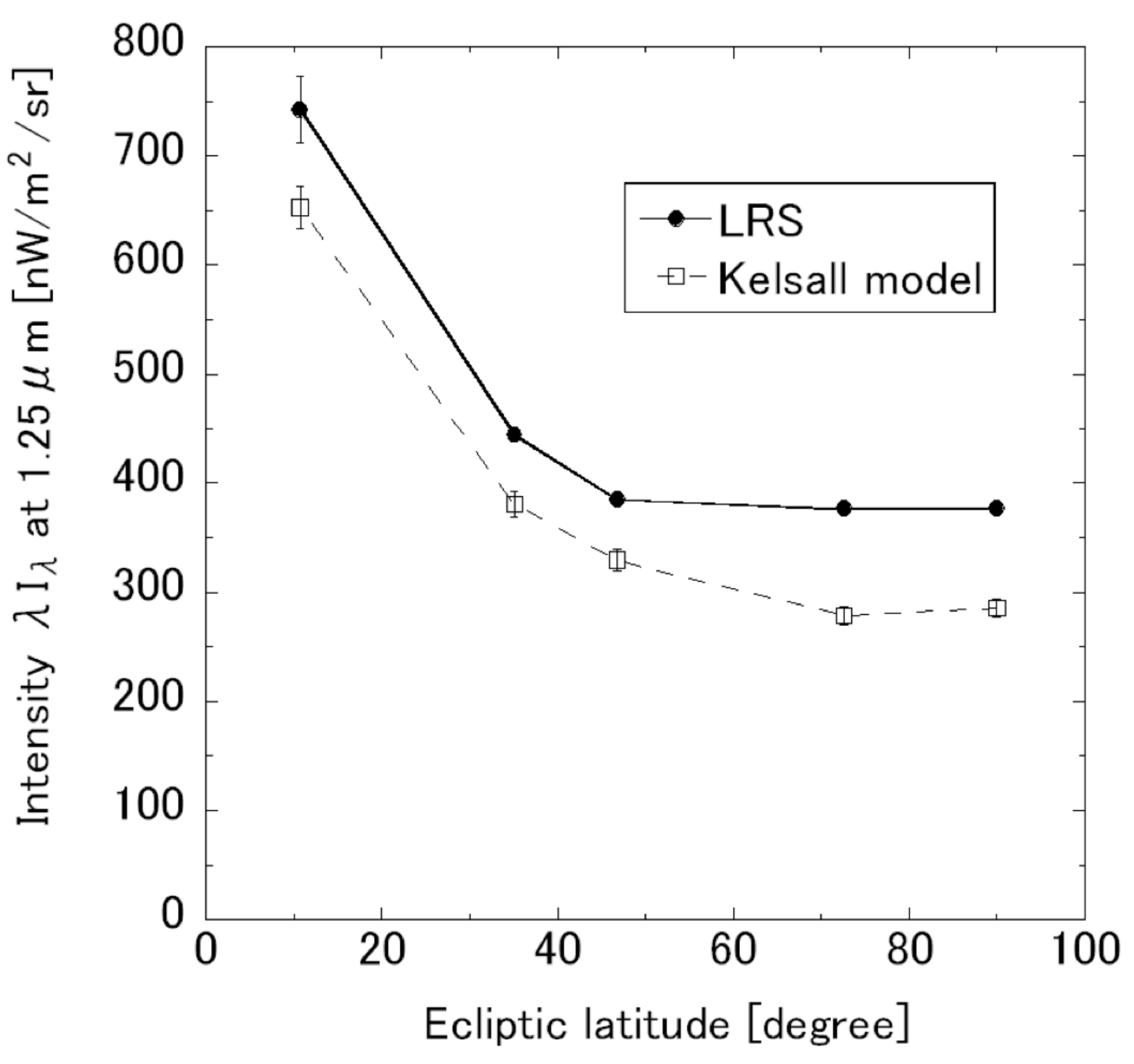}
\caption[]{Left: Measured spectra of the near-infrared sky
  brightness. Colored circles indicate the CIBER/LRS data used in
  \cite{Tsumura:2010jv}, open squares indicate IRTS data averaged at
  72 deg.-73 deg. ecliptic   latitude from Matsumoto et al. (1996), and purple squares indicate
  darkest DIRBE data from Hauser et al. (1998). Right: Ecliptic
  latitude dependence of the DIRBE all-sky zodiacal dust model
  (Kelsall et al. 1998) (open squares) and the LRS sky brightness
  (circles) after partial removal of the integrated Galactic starlight
  based on detected stars at $\simeq$ 1.25 $\mu$m. There is still a
  residual from undetected stars that explains the observed
  difference. The impact from the difference of the solar
  elongation or  ecliptic longitude should be small, because the range of them in our
  observation fields is narrow ($<$30 deg.◦; see Table 1) to see the ecliptic
  latitude dependence of the zodiacal light. The assumed ZL foreground
  in the SPHEREx sensitivity model is 575 nW/m$^2$ sr at 1.25 $\mu$m.
  Note that $\sim$2/3 of the sky area will have lower brightness than
  this assumed level, with somewhat better sensitivity than projected,
  while $\sim$1/3 of the sky area will have higher brightness.} 
\label{fig:zodi}
\end{figure}

\item {\bf Zodical light (ZL)} in the NIR arises from scattered sunlight by interplanetary dust in the
optical and near-infrared (and thermal emission from the same dust in
the mid- and far-infrared). For our baseline sensitivity we assumes a
sky brightness of 575 nW/m$^2$.sr (0.24 MJy/sr) at 1.25 $\mu$m.  According
to the DIRBE/Kelsall ZL model, this brightness corresponds to an
ecliptic latitude of 20 degrees at a solar elongation of 90 degrees.
This is empirically borne out by measurements of the total sky
brightness in Figs.~\ref{fig:zodi} borrowed from \cite{Tsumura:2010jv}.
The discrepancy noted in right panel of Figs.~\ref{fig:zodi} is
largely due to residual star light which is not fully removed in the
CIBER-LRS 1'$\times$1' pixels. On a source by source basis, we find that the
effect of ZL (and DGL) on optimal photometry (and thus redshift estimation) is negligible as
both components appear as a slow gradient across the full image. Each
images can provide a simple background gradient estimate to accurately
remove the contribution from any such diffuse emission.  The variation
in ZL brightness (and DGL to a lesser extent) produces a variation in
sensitivity in these background-limited images that will vary the
selection function. We will handle it as discussed below.

\item {\bf Diffuse Galactic Light (DGL)} in the NIR arises from
the Galactic radiation field scattering off interstellar dust.  While its absolute
intensity is much less than the ZL for most of the sky, the spatial
variation is on smaller angular scales more relevant to compact source
extraction.  To estimate the DGL in a SPHEREx FOV, we rely on the
scaling relation between DGL intensity and 100 $\mu$m emission
%\citep{ienakea,tsumuradgl} including line emission from PAH features.
\citep{tsumuradgl} including line emission from PAH features.
Using this scaling relation as an input we simulate an expected field,
as shown in the top center of Fig.~\ref{fig:imagesims} and combine
it with estimates of all other effects discussed in this section,
including simulated photon noise from all components.  We then mask
resolved sources and reconstruct the input DGL scaling relation.  This
is done for simulated maps of the sky as seen in 24 pointing centers
scanned across the region as is done in the SPHEREx scan strategy.  We
find that using this technique, we were able to reconstruct the DGL to
an RMS accuracy of 0.5~nW~m$^{-2}$ sr$^{-1}$, subdominant to the
statistical noise by an order of magnitude. But to a large extent, this
reconstruction is \emph{not} necessary. Just as for ZL, the variation
in DGL brightness will produce a small variation in sensitivity in
these background-limited images that will vary the selection function
and will be modeled as discussed below. 

\item {\bf Selection Non-Uniformity} Spatial variations in the
  selection function are introduced by changes in stellar density, sky
  brightness and noise level in the different fields (e.g.,
  \cite{Huff:2011aa}). This effect will be characterized by
  simulations where artificial sources are injected into the real
  data pipeline to characterize the selection function. The same
  technique can be applied to understand the wavelength dependence of
  this selection function which can lead to spectral $z$
  non-uniformity. The understanding of this selection function will
  result in the definition of a non-unfirorm 3D weighting in the
  clustering  analysis, just as is performed in 2D for full-sky survey
  with inhomogeneous noise such as WMAP or Planck.

\begin{figure}[!t]
\centering
\includegraphics[width=0.55\textwidth,height=0.3\textwidth]{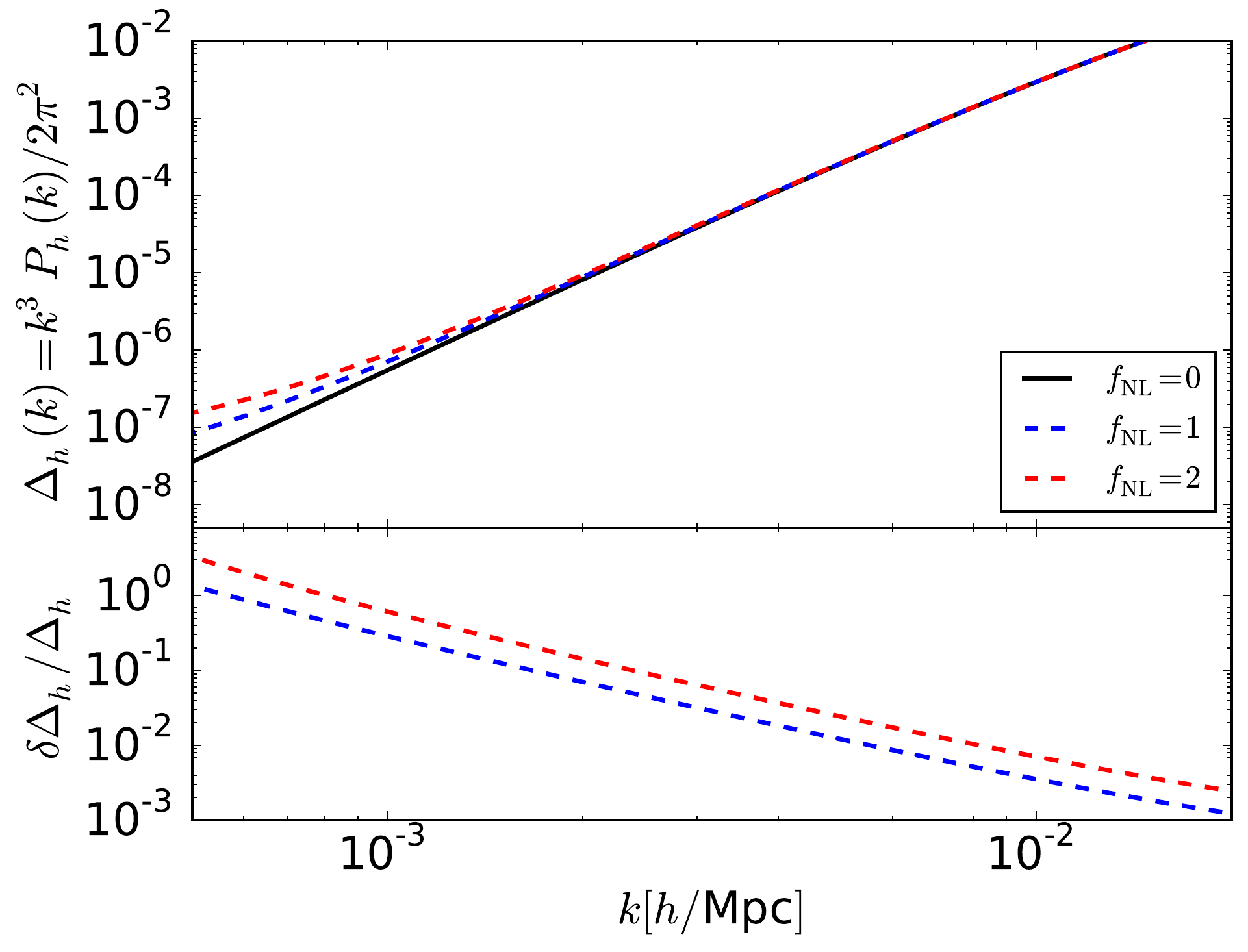}
\caption[]{Absolute (top) and fractional (bottom) power spectrum
  variation for the galaxy power spectrum at $z=1$ for various \fnlloc.
  To illustrate the required level of precision and systematics control for measuring \fnlloc $\sim 1$,
  we plot the dimensionless power spectrum in the top panel, i.e.~the variance in the galaxy overdensity
  per $\ln k$ interval.}
\label{fig:dhoh_requirement}
\end{figure}

\item {\bf Spectral Gain Errors} The gain between
  spectral channels is obtained by measuring the response to diffuse
  sky to inter-calibrate pixels within a spectral channel, and the 93
  spectral standard stars observed over the sky (STSCI
  database). SPHEREx provides an excellent data set for measuring
  variations in gain response across the array as it is highly
  redundant and dithered at the celestial poles. 

\item {\bf Flat Field}. The SPHEREx scan strategy provides an
  excellent data-set for measuring variations in gain response across
  the array (flat-field) as it is highly redundant and dithered at the
  celestial poles (deep regions). The self-calibration algorithm of
  Fixsen, Moseley and Arendt 2000 will be applied to reconstruct the
  Gain matrices using the ZL as the flat-field source.  Each day of
  Science operation, both celestial poles will be sampled ~50 times
  for the deep survey.  Assuming the \citet{Kelsall:1998bq} model for ZL
  intensity, the reconstruction of the flat-field will be accurate to
  1.4\% using data from a single day and accurate to 0.25\% using one
  month of data. 

\item {\bf Source Blending} We identify redshift targets from prior
  catalogs, and screen out blended galaxies using the
  higher-resolution WISE and Pan- STARRS/DES images. This method has
  been implemented successfully in previous selections of samples for
  spectroscopic follow-up by e.g., the Sloan Digital Sky Survey
  \citep{Strauss:2002dj}.

\item{{\bf PSF Errors}}. Flux extraction of unresolved sources using optimal
  photometry requires an accurate understanding of the PSF. For each
  wavelength in each exposure, we will generate a PSF template by
  stacking all stellar sources in the 2MASS catalog.  Each
  spatial/spectral sample in SPHEREx spans a solid angle of 0.53 deg2,
  for moderately high Galactic latitude fields, this will contain ~350
  stars with J-band magnitudes brighter than 14th.  Simulations of
  this measurement have shown a PSF reconstruction accurate to a level
  of 1\% on the FWHM of ~1\%.  This technique is well established and
  has been previously demonstrated on science data \cite{Bock:2012fw}.
  We have simulated SPHEREx reconstructed PSFs with a systematic error
  of 1\% on the FWHM, which produces an error of 0.1\% on the inferred
  flux of a point source. 

\item {\bf Absolute Pointing Reconstruction} To assess the worst case
  impact an imperfect pointing solution biases photometry result, we
  simulate the measurement of a $M_{AB}$(I band) =17 galaxy and rely
  on the centroiding of the object from SPHEREx data alone.  Using all
  90 independent measurements of the object, its location was fit for
  to a 0.035 rms level.  Using a centroid incorrect by this amount
  for optimal photometry produces a 0.04\% random bias. 

\item {\bf Saturation} H2RG arrays have a well depth of approximately
  $1\times10^{5}$ e-. For bright sources, which will generate more
  electrons than this in a single exposure, the inferred flux will be
  extracted by line-fitting only the early frames of the exposure
  \cite{Bock:2012fw}.  This technique will be implemented in firmware
  on board the spacecraft.   A pixel will therefore only be truly
  saturated if this threshold is reached in fewer than 3 frames,
  accounting for the Neff of the SPHEREx PSF, this corresponds to an
  I-band AB mag of 5.3.  For regions of moderate Galactic latitude, we
  expect $\sim$20 objects this bright in each FOV. 

\item {\bf Cosmic Rays} Investigations of the WPFC3 cosmic ray rate
  (using a similar detector at a similar altitude to SPHEREx) give a
  cosmic ray hit rate that is $\leq$0.1\% of pixels per minute of
  integration \citep{Barker2010,Russell2009}. Estimates for the JWST H2RG
  \citep{Robberto2011}\footnote{http://www.stsci.edu/jwst/instruments/nircam/docarchive/JWST-STScI-001650.pdf}
  yield a cosmic ray rate where $\leq$1\% of pixels
  are  affected in a 200s integration, which due to the different
  radiation  environment at L2 we regard as conservative. Affected pixels are
  flagged and removed from later analysis.

\item {\bf Image Persistence} Previous measurements of persistence curves in H2RGs have
    demonstrated reproducible results at 0.02\% accuracy over extended
    periods of time \cite{Smith2008}. For the relevant timescales
    associated with the SPHEREx scan cadence and the expected bright
    star counts, we estimate roughly 20 pixels will need to be
    discarded in each exposure due to excessive image persistence from
    the previous exposure.

\item {\bf Dark Current Correction}. The H2RG detectors implemented by
  SPHEREx have demonstrated excellent dark current (DC) performance
  with a mean level of 0.02e$^-$/s and a one sigma spread of
  0.005e$^-$/s. This dispersion is negligible compared with the
  $\simeq$0.070 e$^-$/s statistical noise per pixel in a given
  image. Based on previous measurements of H2RG thermal stability
  (Finger et al. 2008), we conservatively estimate a maximum 10\%
  variation in DC mean level. Additionally, as was implemented for the
  CIBER-LRS (Tsumura et al. 2013), where small regions of each array
  were masked to maintain a constant monitor of DC drifts, SPHEREx
  will use dark reference pixels provided in the H2RG design. 

%  As previously mentioned
% The H2RG detectors implemented by SPHEREx have been demonstrated to
% possess excellent dark current (DC) performance with a mean level of
% $\sim$ 0.01e-/s and a one sigma spread of 0.005e-/s. The individual
% arrays procured for SPHEREx will undergo an extensive testing prior to
% launch to characterize the array wide DC morphology. With these data
% we will generate template images to be subtracted from individual
% exposures. To maintain stability in orbit, the array temperature will
% be actively regulated, exploiting the heritage of the thermal control
% system implemented on CIBER (Zemcov et al 2013.). Based on previous
% measurements of H2RG thermal DC stability (Finger et al. 2008), we
% conservatively estimate a maximum 10\% variation in DC mean level.
% This corresponds to a maximum systematic residual of 1.5
% nW. Additionally, small regions of each array will be masked to
% maintain a constant monitor of DC drifts should they be present to
% further reduce this conservative estimate.

The formalism used to convert sky variations in the redshift estimator to galaxy
overdensities, $\delta n/n$, is given in
Appendix~\ref{sec:bias_photoz}. The fractional variation of
the galaxy/halo density as a function of \fnl is shown in
Figure~\ref{fig:dhoh_requirement}. 

To conclude, to constrain primordial non-Gaussianity at the $f_{\rm
  NL}\simeq 1$ level requires to accurately
measure the largest scales on the sky, which in turn requires  an
exquisite control of systematic effects. SPHEREx design uniquely
enables this control. The high-redundancy, the stability enabled by
space will allow abundant cross-checks. The ability to infer redshifts
from 97 bands from the same instrument in space, i.e., without the
large scale fluctuations created by the atmosphere is unique amongst
current or planned surveys.

\begin{table}[!t]
\centering
%\begin{tabular}{|p{3.0cm}|C{3.0cm}|C{3.0cm}|C{3.3cm}|C{3.3cm}|C{2.5cm}|}
 %       \hline & {LSST} & DESI & Euclid & SPHEREx & CHIME \\
  %  \hline
\resizebox{0.9\textwidth}{!}{
\begin{tabular}{|L{2.0cm}|L{4.0cm}|L{3.0cm}|C{2.0cm}|L{2.0cm}|C{2.0cm}|C{2.0cm}|}
 \hline
 \hline
  Systematic & Mitigation & Amplitude & Conversion to  $\delta n/n$ & Technique & Coherent on large scales? & $\delta
n/n$ \% rms/dex\\
\hline
\hline
Galactic extinction & Observe in NIR, template projection & 0.007 mag
rms before mitigation & 0.92/mag & e.g., Pullen \& Hirata 2013 & Yes & 0.064\\
\hline
Noise selection non-uniformity & Inject simulated objects into real
data & Template projection 0.2 mag rms (before mitigation) & 
1.8$\times 10^{-3}$/mag &  e.g., Huff et al. 2014 & Yes & 0.036\\
\hline
Noise spectral $z$ non-uniformity & Inject simulated objects into real
data & Template projection 0.2 mag rms (before mitigation) & 0.46/mag
& e.g., Huff et al. 2014 & Yes & 0.092\\
\hline
Spectral gain errors& Measure flat field, calibrate on spectral
standards & $\leq$ 0.25 \% pixel-pixel gain &NA & Fixsen et al. 2000 &
No & NA\\
\hline
Source blending & High resolution Pan-STARRS/ DES/ WISE catalog &
Negligible for bright sources & NA & Jouvel et al 2009 & No & NA\\
\hline
PSF and Astrometry Error & Stack on 2mass catalogs &  $\leq$ 0.1\%
flux &  1 & Zemcov et al. 2013 & No & 0.10 \\
\hline
Cosmic Rays & Flag contaminated pixels & $\leq$ 1\% pixels
lost/exposure & NA & Russell et al. 2009 & No & NA\\
\hline
Bright Sources & Mask persistent pixels & $\leq$ 2\% pixels
lost/exposure & 1 & Smith et al. 2008 & Yes & 0.04\\
\hline
Dark Current & Thermal stability & $\leq 10$\% of statistical error &
NA & Zemcov et al. 2013 & No & NA\\
\hline
\hline
\end{tabular}}
\caption{Main systematic effects in the SPHEREx inflationary science
  data analysis, their mitigation method with heritage and their
  impact on the galaxy over-density ($\delta n/n$) before and after
  mitigation. Adding the residuals errors of the last column in
  quadrature we obtain a 0.160\% rms per dex which gives us a margin
  of 0.121 over our 0.2\% rms per dex goal.}
\label{tab:syst}
\end{table}

\ei

\section{SPHEREx synergies with Euclid and WFIRST}
%----------------------------------------------------

%The ESA/NASA Euclid and NASA WFIRST satellite missions overlap in
%their cosmological science case, however the mission design differs
%substantially in instrumentation and survey strategy, which makes SPHEREx highly complementary to both.
%We will first consider SPHEREx-Euclid synergies, which are extremely strong, then
%briefly elaborate on SPHEREx-WFIRST. 

SPHEREx will enable new scientific opportunities in conjunction with
Euclid \cite{Euclid2011} and WFIRST \cite{wfirst} datasets, largely as
a result of the complementary redshift coverage (see
Figure~\ref{fig:fnl_effect}). In the following we highlight the most
prominent examples.

\subsection{SPHEREx - Euclid}

Euclid is an ESA led satellite mission aiming to constrain dark
energy through a combination of wide-field imaging and
spectroscopy. The imaging component of Euclid encompasses 15,000 deg$^2$
to a nominal depth of 24.5 mag in one broad filter band (550-900nm)
aiming at precision shape measurements for weak gravitational lensing
(WL). Spectroscopic information, i.e., the galaxy clustering
(GC)/baryonic acoustic oscillations (BAO) survey, will only be
available for 1/30 of the WL galaxies \citep{Euclid2011}. 
% For a
% comparison of the spectroscopic survey component of Euclid to 
% SPHEREx see CH. \ref{sec:comparison}; here we focus on the Euclid WL
% science.   

\begin{figure}[!t]
\centering
\includegraphics[width=0.9\textwidth]{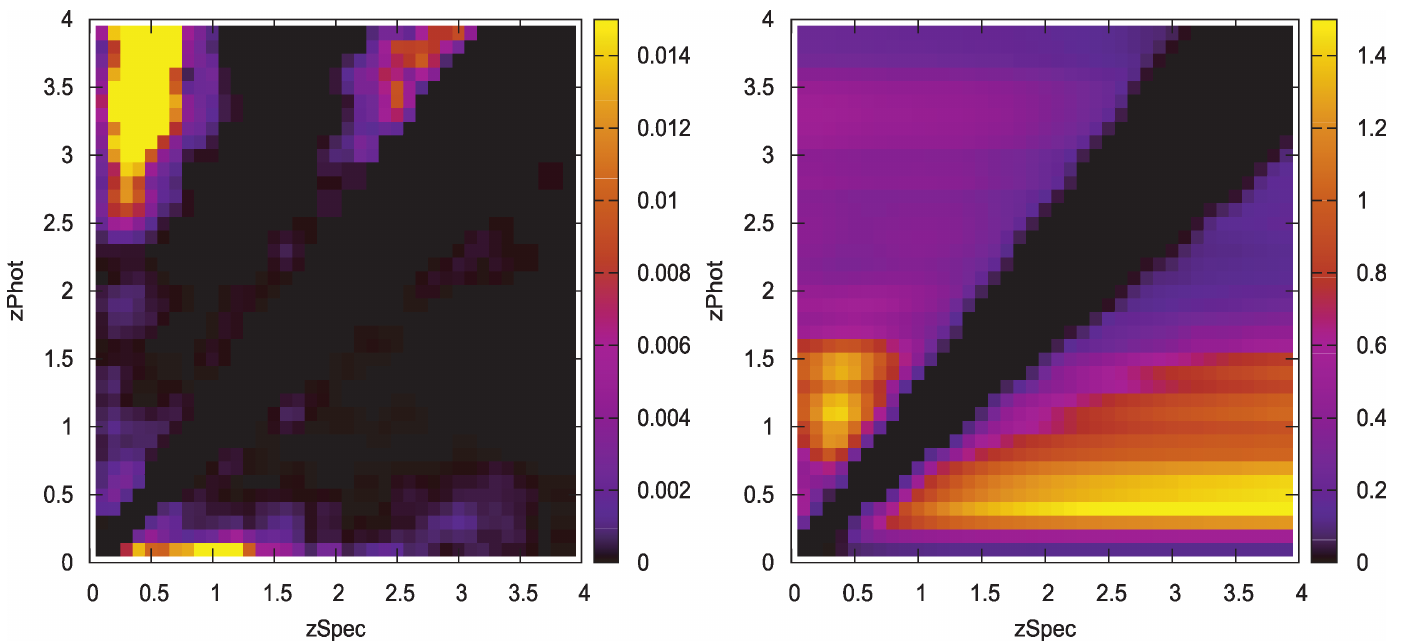}
\caption[]{\textit{Left:} Contamination fraction of catastrophic
  photo-$z$ per unit redshift in spectro-$z$ (considered a good proxy for the
  true redshift of a galaxy). For a perfect photo-$z$ survey, this
  plane should be empty. However, we see that some prominent contamination is
  happening for $z_{phot}>2.5$ and $z_{phot}<0.2$. The highest 
  contaminations are at $z_{phot}>2.5$ , $z_{spec} < 0.6$ which are
  likely caused by confusion of high-$z$ Lyman breaks with low-$z$ 0.4
  $\mu$m  breaks. Note that a small leakage from shallow $z \sim 0.5$ galaxies
  can have a large impact on true galaxies at $z > 2.5$ since the
  latter are relatively rare.  \textit{Right:} The subsequent bias on dark energy
  parameters $w_0$, $w_a$ from catastrophic photo-$z$ errors (bright
  regions indicate severe bias). The plot indicates that the outlier
  contamination rates must be known to one to three parts per thousand
  over all photometric redshift bins, most sensitively at $0.3 <
  z_{phot} < 1.5$ (\textit{Credit: \cite{Bernstein2010}}). SPHEREx
  high-precision redshift measurements at $z<1$ will resolve the
  confusion of high and low redshift galaxies.}
\label{fig:photoz_tim}
\end{figure}

The combination of sophisticated imaging from Euclid and
high-precision redshift information from SPHEREx will greatly enhance
the cosmological constraining power of Euclid WL by substantially
improving on photo-$z$ uncertainties and catastrophic photo-$z$
errors. Removing these uncertainties is critical to Euclid's
cosmological constraining power. Since Euclid's spectroscopic 
component will only cover 1/30 of its WL galaxies and the broad-band
imaging prohibits an internal photo-$z$ estimation, Euclid relies on
ground based coverage in multiple bands to the same depth and over the
full survey in order to determine its galaxy redshift. Although SPHEREx is more shallow than Euclid, it will 
improve Euclid's overall photo-z estimation through cross-calibrating
photo-$z$'s using clustering information
\cite{newman08,phzpaper,menardetal13}. More importantly it will
resolve potential catastrophic photo-$z$ outliers at low redshifts,
where their occurrence is most frequent and their impact most severe
(see Figure~\ref{fig:photoz_tim}). The improved photo-$z$ estimation
affects all of Euclid WL science; in addition to the increase in
cosmological constraining power from cosmic shear, it also improves
the cluster weak lensing signal and removes all confusion about
cluster member galaxies and background galaxies from clusters below
$z=1$.

% FIG. 2: Left: The contamination rate csp/∆zs of the photo-z bin per unit redshift in spectro-z is plotted for our example case. Note that the contamination is highest at zp > 2.5 and zP < 0.2, where there are relatively few source galaxies and hence a small number of outliers can become a large fractional contamination. Right: The quantity hsp which specifies the significance of the w0 − wa bias caused by a contamination rate of 0.001 δzp across a range δzp of photo-z. This plot indicates that the outlier contamination rates must be known to 1–3 parts per thousand over all photometric redshift bins, most sensitively at 0.3 < zp < 1.5.

Most importantly perhaps the SPHEREx-Euclid data set will also enable
new science via high precision galaxy-galaxy lensing
measurements. Galaxy-galaxy lensing, i.e., the (stacked) shear signal
of background galaxies (from high-precision Euclid WL) around the
position of foreground galaxies (with high-precision redshift
information from SPHEREx), allows for a direct measurement of the dark
matter environment surrounding the foreground galaxies. This technique
unlocks a number of interesting science studies that are not (or only
to a much lesser precision) accessible when using the individual data
sets. 
\begin{itemize}
\item \textbf{Study the Galaxy-Halo Connection} as a function of
  galaxy luminosity, type, and environment. Galaxy-galaxy lensing
  measurements allow for constraints on the luminous (stellar) mass
  with respect to the surrounding dark matter halo of the
  galaxy. These studies are extremely valuable for galaxy formation
  and evolution theories.  
\item \textbf{Intrinsic alignment of source galaxies.} A major astrophysical uncertainty for Euclid WL is the intrinsic alignment
(IA) of sources, which mimics a WL signal and can substantially bias the inferred cosmological information. Overlapping imaging (from Euclid) and (near-) spectroscopic information (from SPHEREx) allows for measuring
and removing the IA contamination from the WL signal
(\cite{Joachimi2011}, \cite{Mandelbaum2011}).  This again will be
limited to low redshift galaxies ($z<1$), which however covers the
redshift range where the IA contamination is damaging.     
\item \textbf{Baryonic physics inside Halos} The role of baryons in
  structure formation and the exact physical mechanisms associated
  with baryons is a fascinating topic in astrophysics and an extremely
  worrisome uncertainty in cosmological parameter estimation (e.g.,
  \cite{Zentner2013}, \cite{Eifler2014}, and references therein). The
  impact of Baryonic Cooling, AGN, and SN feedback on the halo
  concentration parameter and the matter power spectrum can be
  constrained with galaxy-galaxy lensing.   
\item \textbf{Constrain Galaxy Bias} The combination of galaxy
  clustering and galaxy-galaxy lensing allows for constraints on the
  galaxy bias and thereby removes the largest uncertainty from galaxy
  clustering as a cosmological probe when inferring the matter power
  spectrum. This joint signal can be used as a cross-check for the
  matter power spectrum inferred from cosmic shear, or if in agreement
  the combination of all three probes can be used to substantially
  increase cosmological constraining power (e.g., \cite{Joachimi2010},
  \cite{Eifler2014b}, and references therein). 
\item \textbf{Modified Gravity Studies} \cite{Reyes2010} use the combination of galaxy-galaxy lensing and galaxy clustering to build an estimator that constrains modified gravity theories on very large scales. Their analysis comprised $\sim 70,000$ large red galaxies only; the combination of Euclid and SPHEREx will allow for a substantial improvement in these constraints. 
\item \textbf{Higher-order moments of the density field}
  Galaxy-galaxy-galaxy lensing (\cite{Simon2012}) is a powerful tool
  to investigate higher-order correlations of the density field, in
  particular whether this correlation varies as a function of galaxy
  type and luminosity (\cite{Simon2013}). 
\end{itemize}
The precision of these studies strongly depends on the accuracy to
which the lens redshift is known, hence the high-precision low-z
information from SPHEREx combined with the Euclid shapes of background
sources is an ideal data set for these science cases. 

\subsection{SPHEREx-WFIRST}
%----------------------------

\begin{figure}[!t]
\centering
\includegraphics[width=0.7\textwidth]{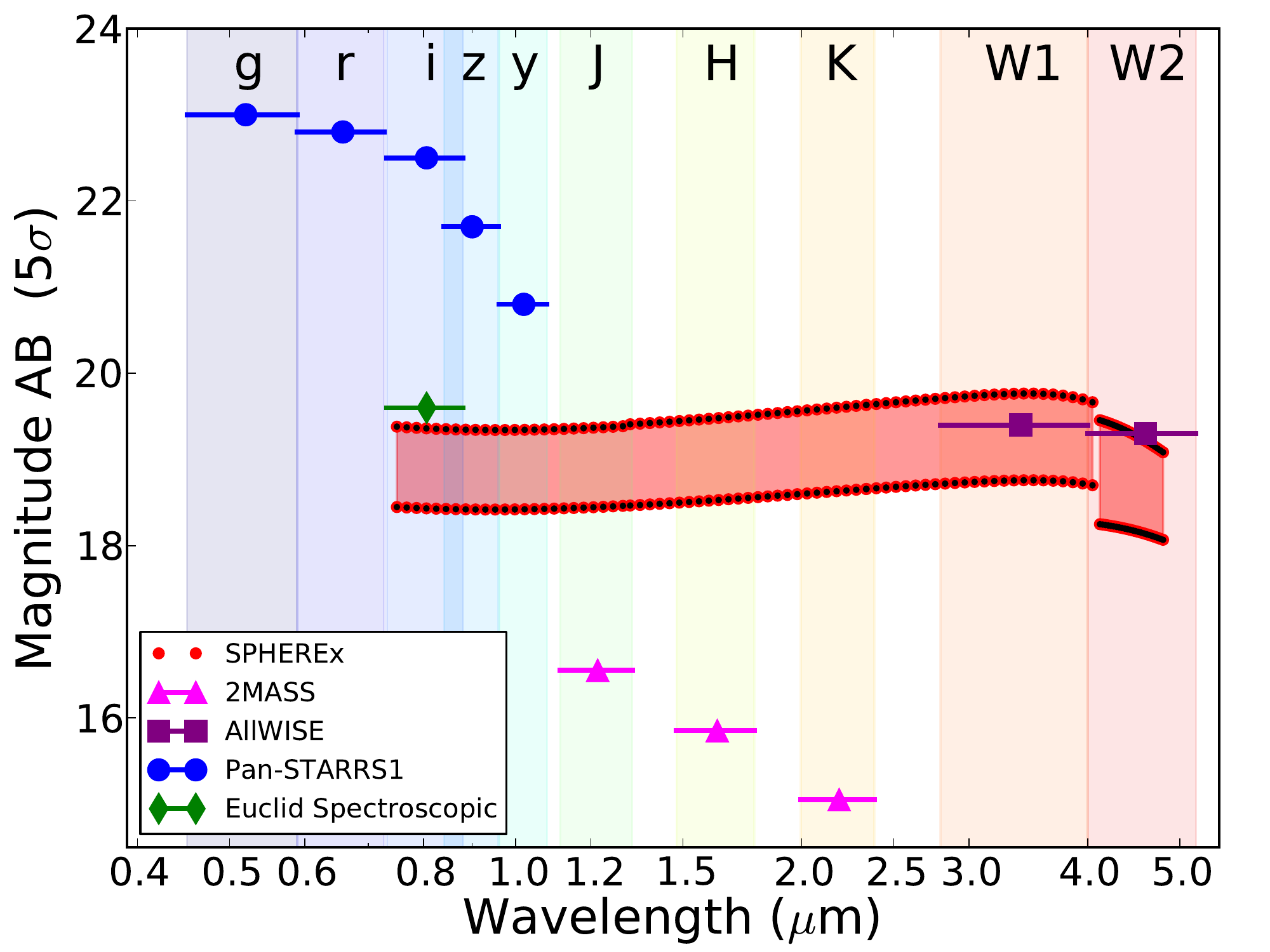}
\caption[]{SPHEREx all-sky survey depth as a function of wavelength
  compared to other large scale cosmological survey. Each red dot
  corresponds to one of SPHEREx 97 wavelength element. The top red
  curve corresponds to the current best estimate sensitivity while
  the bottom curve corresponds to the instrument sensitivity based on
  specifications that each sub-system can meet with contingency.}
\label{fig:lam_mag}
\end{figure}

The exact details of the WFIRST multi-band and grism instrumentation
are still being determined, however most of the synergies outlined in
the previous section apply very similarly to WFIRST. In the following
we want to point out additional properties of a joint WFIRST-SPHEREx
data set: 
\begin{itemize}
\item  A novel science case will be enabled by the combination of the
  unprecedented high-redshift and low-redshift cosmological volumes
  mapped by WFIRST and SPHEREx respectively. Jointly analysing the two
  samples will enable new cosmological measurements, probing directly
  the longest line-of-sight modes \cite{Raccanelli:2013gja}.  Comparing
  these line-of-sigth modes with the modes perpendicular to the
  line-of-sights will enable potent new tests of general relativity on 
  cosmological scales and test the isotropy hypothesis
  \cite{Raccanelli:2013dza}.  
  % cosmological volume. The joint analysis f , 
  % very deep (but covers only a limited area on the sky, i.e., 2,500
  % $\rm{deg}^2$), whereas SPHEREx is a full sky survey (but
  % shallow). The combination of these data sets will for the first time
  % enable a test of the isotropy of the Universe (radial vs angular
  % scales) using the galaxy clustering signal.   
\item The depth of the WFIRST mission causes it to be more susceptible
  to catastrophic photo-$z$ confusion between high and low redshift
  (see Figure~\ref{fig:photoz_tim}). Although WFIRST's instrumentation
  requirements will certainly take this into account, the mission will
  benefit from SPHEREx high-precision redshift measurements. 
\item The high number density of background galaxies enables a higher
  signal-to-noise ratio for the galaxy-galaxy lensing and cluster
  lensing measurements from the joint WFIRST-SPHEREx data set compared
  to the Euclid-SPHEREx case. 
\end{itemize}

\section{SPHEREx All-sky Spectral Data-base Legacy Science}
\label{sec:legacy}

All-sky surveys (IRAS, WMAP, Planck, WISE) have played a major role in
advancing modern astrophysics, enabling ground- breaking science and
producing versatile legacy archive that will prove valuable for
decades. SPHEREx will contribute to this proud heritage by carrying
out the first all-sky spectroscopic survey at near-infrared
wavelengths. 

SPHEREx will measure the spectrum of every object in the 2MASS PSC
(1.2 $\mu$m, 1.6 $\mu$m, 2.2 $\mu$m) catalog to at least (20$\sigma$,
40$\sigma$, 120$\sigma$) per spectral channel (see Figure~\ref{fig:lam_mag}). Most objects in the WISE
catalog will also be detected by SPHEREx spectroscopically, the
faintest detected at $\simeq 3\sigma$ in each spectral channel. Thus the SPHEREx
data will contain high quality spectra of hundreds of millions of WISE
and 2MASS sources, as well as a significant population of unique
SPHEREx source detections at 1-3 $\mu$m. Such a rich archival spectral
database will support numerous scientific investigations (e.g., Table
\ref{tab:legacy}) of great interest to the wider scientific
community. We outline below a few examples.

SPHEREx enables new studies of galaxy formation. The entire catalog
will contain ~120 million galaxies with $\sigma (z)/(1+z) <$ 0.03,
sufficient for statistical galaxy evolution studies
\cite{Ilbert:2008hz}, and enabling clustering studies that link galaxy
properties to their underlying dark matter haloes
\cite{Coupon:2011mz}. Importantly, SPHEREx will provide the first
statistically significant spectroscopic dataset in the important 1-4 $\mu$m wavelength
range. In local galaxies, SPHEREx will detect infrared diagnostic
lines such as CO, H2O, SII, and PAH emission that probe the interplay
of stellar populations and AGNs with the density and excitation of the
ISM. At higher redshifts, SPHEREx probes the H$\alpha$, H$\beta$, O[III], and O[II]
lines. In individual bright galaxies these spectra will find objects
with unusually strong emission lines, due to AGNs or very low
metallicities, and identify rare interesting objects such as quasars
or heavily obscured sources. In fainter objects SPHEREx spectra allow
ensemble studies of metallicity, obscuration and star formation over a wide range of redshifts and galaxy types.

As another example, a spectral catalog of main sequence stars over the
entire sky opens numerous opportunities, e.g., a search for strong
near- infrared excesses due to warm circumstellar dust radiating
shortward of 5 $\mu$m. Large amounts of such dust signal catastrophic
transient processes in an exoplanetary system orbiting the
star. Kennedy \& Wyatt (2013) estimate that such excesses should be
seen around $\simeq 1\%$ of young stars and around 0.01\% of stars
older than 1 Gyr. Confirming these predictions and studying the nature
of systems with recently injected dust would increase our
understanding of planetary system evolution and tie exoplanet
observations to events, such as the collision thought to have scooped
the moon out of the Earth, which occurred in our Solar System.

\begin{table}[!h]
\centering
%\begin{tabular}{|p{3.0cm}|C{3.0cm}|C{3.0cm}|C{3.3cm}|C{3.3cm}|C{2.5cm}|}
 %       \hline & {LSST} & DESI & Euclid & SPHEREx & CHIME \\
  %  \hline
\resizebox{0.9\textwidth}{!}{
\begin{tabular}{|L{4.0cm}|L{2.0cm}|L{4.0cm}|C{4.0cm}|}
 \hline
 \hline
 Object & \# Sources & Legacy Science & Reference \\
\hline
\hline
Detected galaxies & 1.4 billion & Properties of distant and heavily
obscured galaxies & Simulation based on COSMOS and Pan- STARRS\\
\hline
Galaxies with $\sigma (z)/(1+z)<0.1$ & 301 million & Study large scale
clustering of galaxies & Simulation based on COSMOS and Pan- STARRS\\
\hline
Galaxies with $\sigma (z)/(1+z)<0.03$ & 120 million & Study
(H$\alpha$, H$\beta$, CO, OII, OIII, SII,
H$_2$O) line and PAH emission by galaxy type. Explore galaxy and AGN life
cycle & Simulation based on COSMOS and Pan- STARRS\\
\hline
Galaxies with $\sigma(z)/(1+z)<0.003$ & 9.8 million & Cross check of
Euclid photo-$z$. Measure dynamics of groups and map
filaments. Cosmological galaxy clustering, BAO, RSD. &
Simulation based on COSMOS and Pan- STARRS\\
\hline
QSOs & $>$ 1.5 million & Understand QSO lifecycle, environment, and
taxonomy & \citet{Ross:2012dt} plus simulations\\
\hline
QSOs at $z > 7$ & 0-300 & Determine if early QSOs exist. Follow-up
spectro- scopy probes EOR through Ly$\alpha$ forest  &
\citet{Ross:2012dt} plus simulations\\
\hline
Clusters with $\geq$ 5 members & 25,000 & Redshifts for all eRosita
clusters. Viral masses and merger dynamics & \citet{Geach:2011gu} \\
\hline
Main sequence stars & $>100$ million & Test uniformity of stellar mass
function within our Galaxy as input to extragalactic studies & 2MASS
catalogs\\
\hline
Mass-losing, dust forming stars &	Over 10,000 of all types & Spectra of M supergiants, OH/IR stars, Carbon stars. Stellar atmospheres, dust return rates, and composition of dust &	Astro-physical Quantities, 4th edition [ed. A.Cox] p. 527\\
\hline
Brown dwarfs & $>$400, incl. $>$40 of types T and Y & Atmospheric
structure and composition; search for hazes. Informs studies of giant
exoplanets & {\texttt dwarfarchives.org} and J.D. Kirkpatrick,
priv. comm.\\
\hline
Stars with hot dust & $>$1000 & Discover rare dust clouds produced by
cataclysmic events like the collision which produced the Earth’s moon
& Kennedy \& Wyatt (2013)\\
\hline
Diffuse ISM &	Map of the Galactic plane & Study diffuse emission
from interstellar clouds and nebulae; hydro-carbon emission in the 3$\mu$m
region &	GLIMPSE survey (Churchwell et al. 2009) \\
\hline
\hline
\end{tabular}}
\caption{SPHEREx spectral database populations}
\label{tab:legacy}
\end{table}

%=================================
\section{The \spherex Galactic Ice Investigation}
\label{sec:ice_investigation}
%=================================

\spherex will carry out a groundbreaking survey of water and 
other biogenic ices in the Milky Way.  By exploiting an optimal 
spectral region for ice spectroscopy, and increasing by 100-fold 
or more the number of ice spectra toward molecular clouds, 
young stellar objects and protoplanetary disks, the SPHEREx
Galactic Ice Investigation will resolve long-standing questions about 
the amount and evolution of such biogenic molecules as H$_2$O, CO, 
CO$_2$, and CH$_3$OH, through all phases of star and planet formation.
In this Section we describe the \spherex Ice Investigation technique, 
targets, data products, and science.

\subsection{Introduction}

\begin{figure}[t]
%\begin{minipage}[b]{0.48\linewidth} % A minipage that covers half the page
\centering
\includegraphics[scale=0.38]{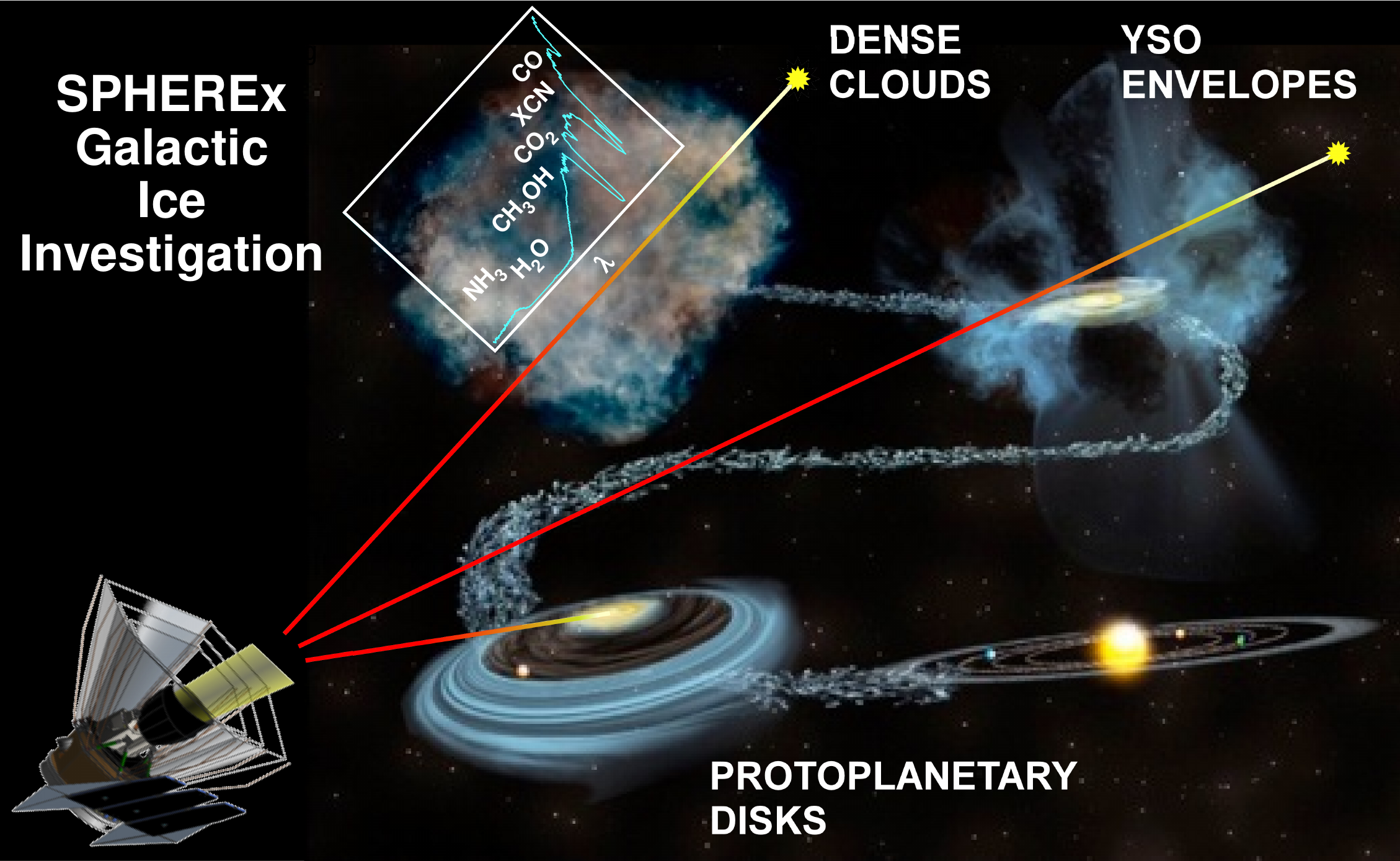}
%\renewcommand{\baselinestretch}{0.95}
%\end{minipage}
%\hspace{0.3cm} % To get a little bit of space between the figures
%\begin{minipage}[b]{0.48\linewidth}
%\centering
\includegraphics[scale=0.52]{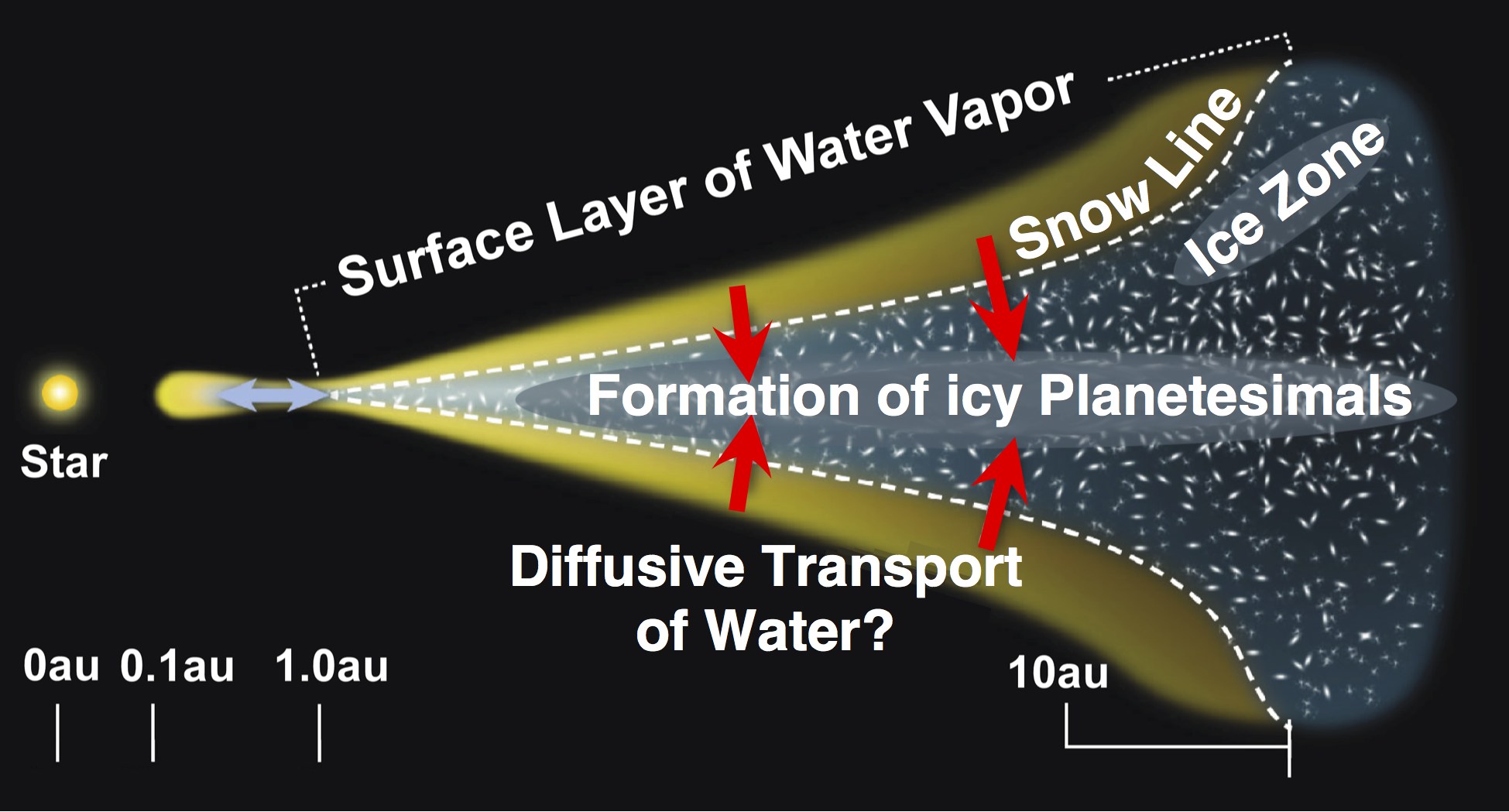}
%\renewcommand{\baselinestretch}{0.95}
%\vspace{-3.7mm}
\caption{Left: Evolutionary stages of star and planet formation. SPHEREx will obtain absorption spectra
against more than 20,000, and as many as 10$^6$, background stars and hundreds of protostars to reveal 
the ice content of these regions. Right: Schematic of a young protoplanetary disk showing the model-predicted zones of gas-phase
water and water ice. Most of the water ice is expected to lie in the disk mid-plane beyond the snow line.
SPHEREx will observe hundreds of protoplanetary disks with inclinations low enough to permit absorption
spectroscopy against the central star.}
\label{iceintrofig}
\label{protodisk}
%\end{minipage}
\end{figure}

Based on the fewer than 250 existing near- and mid-infrared absorption spectra toward Galactic 
molecular clouds, circumstellar envelopes, and protoplanetary disks (see Fig.~\ref{iceintrofig})
there is clear evidence that: 
(1) ices are common in dense (i.e., n(H$_2) \ge 10^3$\,cm$^{-3}$), well-shielded regions; 
(2) they accumulate in measureable quantities on the surface of dust grains; and, 
(3) for some important biogenic molecules, such as water (H$_2$O), 
carbon dioxide (CO$_2$), and methanol (CH$_3$OH), there is evidence that 
the amount of these species locked in ice, while variable from source to 
source, far exceeds that in the gas phase \citep{Oberg:2011kk, Caselli:2012xa,Sakai:2008}.

Why is this important?  It has long been known that star formation occurs 
within the cores of dense interstellar clouds and that the gas and dust within 
these cores are the reservoirs from which stars and planets subsequently 
assemble. In young protoplanetary disks, both models and the limited amount 
of data presently available suggest that most of the water and, perhaps, 
other biogenic molecules, are locked in ice toward the disk mid-plane and 
beyond the snow line \citep[Fig.~\ref{protodisk};][]{Hogerheijde:2011pq}.
%(Fig.~\ref{protodisk}; \citealt{Hogerheijde11}).  
A recent 
theoretical study \citep[e.g.,][]{Cleeves:2014} goes further and finds 
that much of the solar system's water was likely inherited directly from 
interstellar ices, with little further processing in the pre-solar disk.

It is not known, however, whether the link between interstellar water ice and water 
in our own solar system is correct and typical of most planet-forming 
disks, or whether the ice composition in clouds and disks varies broadly.  
This ignorance is largely the result of the limited data currently 
available. To date, the {\sl Infrared Space Observatory (ISO)}, 
{\sl Spitzer}, and {\sl AKARI} have obtained spectra in 10s of lines 
of sight toward a few dense clouds while an additional $\sim$100 lines 
of sight have been probed toward protostars using {\sl Spitzer} and 
{\sl ISO}, revealing a highly variable ice composition 
\citep{Oberg:2011kk}.  The number of ice observations toward 
protoplanetary disks is much lower \citep[e.g.,][]{Pontoppidan:2004dv, Aikawa:2012av}
because of the low disk inclination required for ice absorption studies.

\subsection{Hundreds of Thousands of Ice Absorption Spectra}

The vast majority of objects with rich ice absorption 
spectra are concentrated towards the galactic plane, so the \spherex
Ices Survey nicely exploits the portions of the orbit that are 
inappropriate for extragalactic work.  However, to conduct a survey 
based on absorption spectroscopy, strong background point sources are 
required that are well separated and display evidence for extinction 
due to intervening material.  Here we describe the \spherex strategy 
for identifying promising spectroscopy targets for the Ices Survey.

Our strategy relies on a combination of all-sky, near-infrared catalogs
from the 2 Micron All-Sky Survey \citep[\sl 2MASS;][]{Skrutskie:2006wh} and 
the Wide-field Infrared Survey Explorer \citep[\sl WISE;][]{Wright:2010qw}.
As a starting point, we begin with the {\sl WISE} catalog of 
$>10^7$ 3.4\,$\mu$m (W1) and 4.6\,$\mu$m (W2) point sources within 
$\pm5^\circ$ of the Galactic plane as a starting point.  We assess
the suitability of each {\sl WISE} source for our science with the
following four criteria:

First, the source fluxes measured in the {\sl WISE} W1 and W2 bands must 
be greater than 100 times the projected \spherex (1$\sigma$) sensitivity per 
resolution element at both 3.4 and 4.6\,$\mu$m, allowing for high 
SNR spectra. 
Second, the source must carry $>99$\% of the total flux detected in the
{\sl WISE} W1 and W2 bands in a local box of 2$\times$2 \spherex 
pixels of 6.2\,$^{\prime\prime}$.  Third, there must be evidence that the source lies behind an intervening 
molecular cloud and is heavily

%\begin{wrapfigure}{l}{0.45\textwidth}
\begin{figure}[t]
%\vspace{-0.25in}
\begin{center}
\includegraphics[width=0.6\textwidth,height=0.4\textwidth]{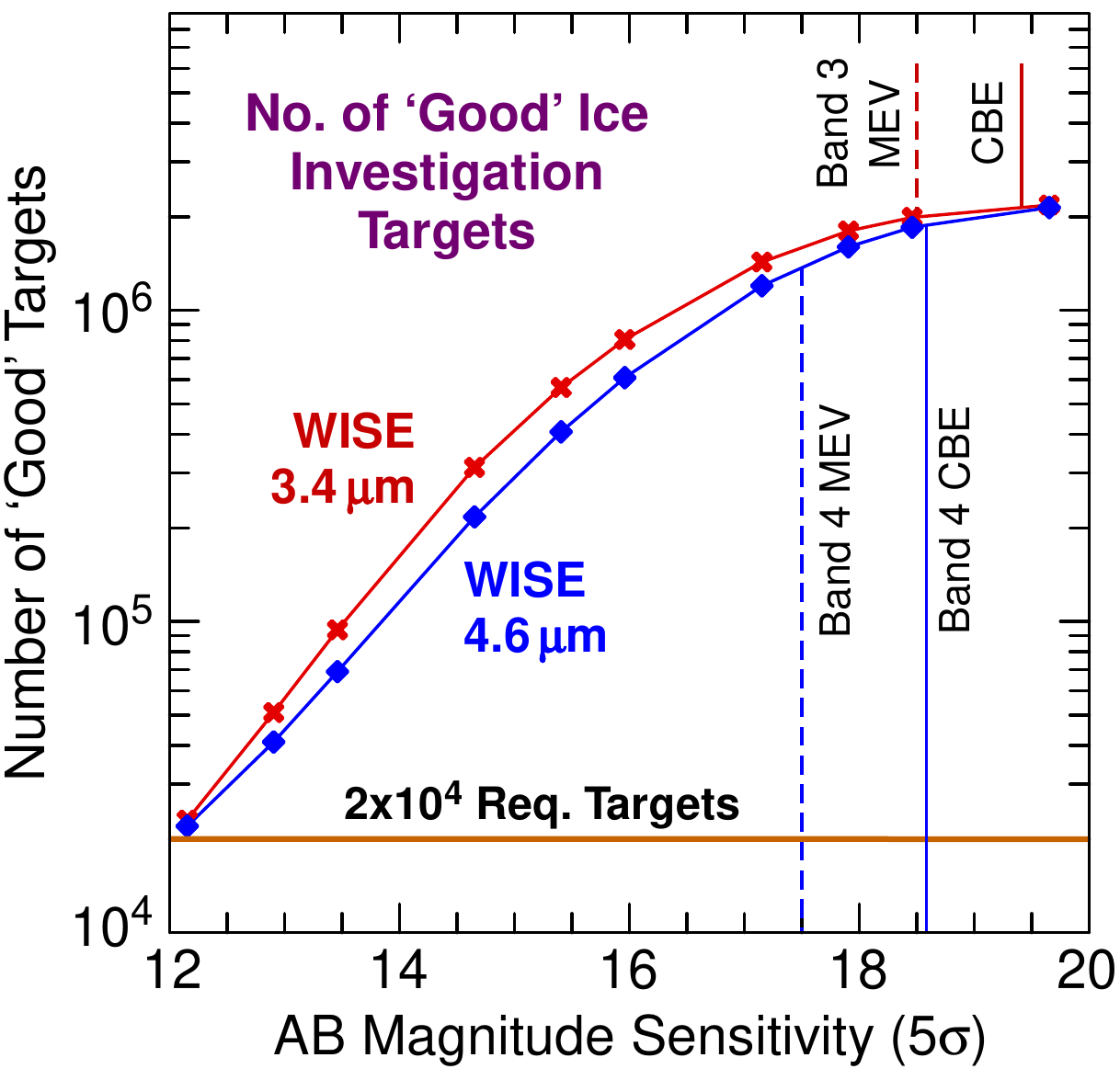}
\end{center}
%\vspace{-5.3mm}
\caption{Number of 3.4\,$\mu$m and 4.6\,$\mu$m point sources detected by
WISE within $\pm$\,1$^{\rm o}$ of the Galactic plane that satisfy the criteria described
in the text. The required 20,000 high-quality absorption
spectra are realized with a Band 4 sensitivity of 12 AB mag (5$\sigma$) per
spectral channel. The MEV performance set by the Inflation
investigation at $\lambda <\:$4.1\,$\mu$m and the Ice investigation at $\lambda >\:$4.1\,$\mu$m, will
result in the detection of more than 10$^6$ targets.}
%\vspace{-0.3in}
\label{numsources}
\end{figure}
%\end{wrapfigure}

\noindent  extinguished.  For the identification of 
such anomalous reddening, we used the Rayleigh-Jeans color excess 
method \citep{Majewski:2011md} and only retained stars with an infrared 
color of $H-$[W2]$>1.5$.  
Fourth, we applied a color cut based on $K-$[W2] to reject the few 
unobscured stars that pass the $H-$[W2] cut, i.e., very
evolved (unobscured) stars which are red in $H-$[W2] but not in $K-$[W2].

These four criteria correctly identified 22 of the 24 stars behind 
quiescent molecular clouds previously observed by {\sl AKARI} 
\citep{Noble:2013tta} and {\sl ISO} \citep{Gibb:2004}.  As expected, 
the spatial distribution of these sources is well aligned with regions 
of high CO emission measured by {\sl Planck}, concentrating near the 
galactic plane.  Applied to the entire 2MASS+{\sl WISE} database, 
these criteria identify over one million ``good sources" within 1\,degree
of the galactic plane, providing a strong and robust lower limit on 
the number of \spherex lines of sight sampling dense interstellar clouds.

Large numbers of protoplanetary disk targets are important for this investigation because 
the presence of a disk leads to an orientation-dependent spectrum; 
only a fraction ($\le10$\%) may be sufficiently edge on to show an optimum 
absorption spectrum against the embedded central star.  
Protoplanetary surveys using both {\sl Spitzer} and {\sl WISE} 
find over 3000 young and forming stars, of which about 750 are in the 
early collapse and protostellar phase and another 1500 in the 
protoplanetary disk phase.  Many of these objects are bright 
enough to enable \spherex spectra with SNR $>100$ in every spectroscopic
pixel; again, this is 
a lower limit to the total number of such objects to be sampled by 
\spherex's unbiased survey, because the existing surveys are not 
unbiased.  Thus there will be ample spectra to trace the evolution 
of ices from dense interstellar clouds through the stages of the 
star and planet formation.

With an orbit that covers the entire Galactic plane every 6 months, 
high sensitivity, and good spectral resolving power, \spherex will 
increase the sample of high-quality Galactic ice absorption spectra by 
100-fold or more in an unbiased survey that includes a rich diversity 
of interstellar clouds, young stellar object (YSO) envelopes, and 
protoplanetary disks.  Thus, the \spherex catalog of $>2\times10^4$ 
ice spectra can be used to evaluate the relation between ice content 
and cloud evolutionary stage through to the protoplanetary phase 
in ways not possible with the limited sample of spectra available 
today.  Moreover, the breadth and depth of the groundbreaking \spherex ice 
spectra catalog is sure to reveal new and exciting sources that will 
merit follow-up by JWST.

\begin{table}[t]
\begin{center}
\caption{~Solid-State Ice Features:~~2.96\,$\mu$m$\;\leq\,\lambda\,\leq\;$4.8\,$\mu$m}
\vspace{1mm}
\begin{tabular}{lcclc}  \hline  \\*[-4.6mm]  \hline
\rule{0mm}{5.6mm}  &  $\lambda$  &  $\Delta\nu$  &   &  Band Strength  \\*[-0.1mm]
Molecule\hspace{15mm}  &  ~~($\mu$m)~~  &  ~~~(cm$^{-1}$)~~~  &  Vibration Mode   &  ~~(10$^{-17}$ cm molecule$^{-1}$)~~  \\*[1.6mm] \hline
\rule{0mm}{5.6mm}NH$_3$ \dotfill  &  2.96  &  45  & --N--H stretch  &  1.1  \\*[1mm]
H$_2$O \dotfill  &  3.05  &  335~~  & O--H stretch  &  20~~~~  \\*[1mm]
--CH$_2$--, --CH$_3$ \dotfill  &  3.47  &  $\sim\,$10~~~~  &  C--H stretch  &  ~\hspace{3.6mm}$\sim\,$0.1\,--\,0.4  \\*[1mm]
CH$_3$OH \dotfill  &  3.53  &  30  &  C--H stretch  &  ~~0.76  \\*[1mm]
CH$_3$OH \dotfill  &  3.95  &  115.3  &  C--H stretch  &  ~~0.51  \\*[1mm]
H$_2$S \dotfill  &  3.95  &  45  &  S--H stretch  &  2.9  \\*[1mm]
CO$_2$ \dotfill  &  4.27  &  18  &  C--O stretch  &  7.6  \\*[1mm]
$^{13}$CO$_2$ \dotfill  &  4.38  &  ~\hspace{2.2mm}12.9  &  $^{13}$C--O stretch  &  7.8  \\*[1mm]
H$_2$O \dotfill  &  4.5~~  &  700~~  &  3$\nu_L$ and/or $\nu_2 + \nu_L$  &  1.0  \\*[1mm]
``XCN" \dotfill  & 4.62  & ~\hspace{3mm}29.1  &  CN stretch  &  $\sim\,$5~~~~~~  \\*[1mm]
CO \dotfill  &  4.67  &  ~~~~~~~9.71  &  $^{12}$CO stretch &  1.1  \\*[1mm]
$^{13}$CO \dotfill  &  4.78  &    &  $^{13}$CO stretch &  1.3  \\*[1.5mm] \hline
%OCS \dotfill  &  4.91  &  ~\hspace{1.1mm}19.6  &  C--S stretch  &  17~~~~  \\*[1.5mm]  \hline
\end{tabular}
\end{center}
\vspace{-5.9mm}
\phantom{0}\hspace{-2.0in}After Gibb et\,al.~2004, {\em ApJ Suppl. Series}, 151, 35.
\label{icetable}
\end{table}%

\begin{figure}[b]
\centering
%\vspace{-0.8mm}
\includegraphics[scale=0.85]{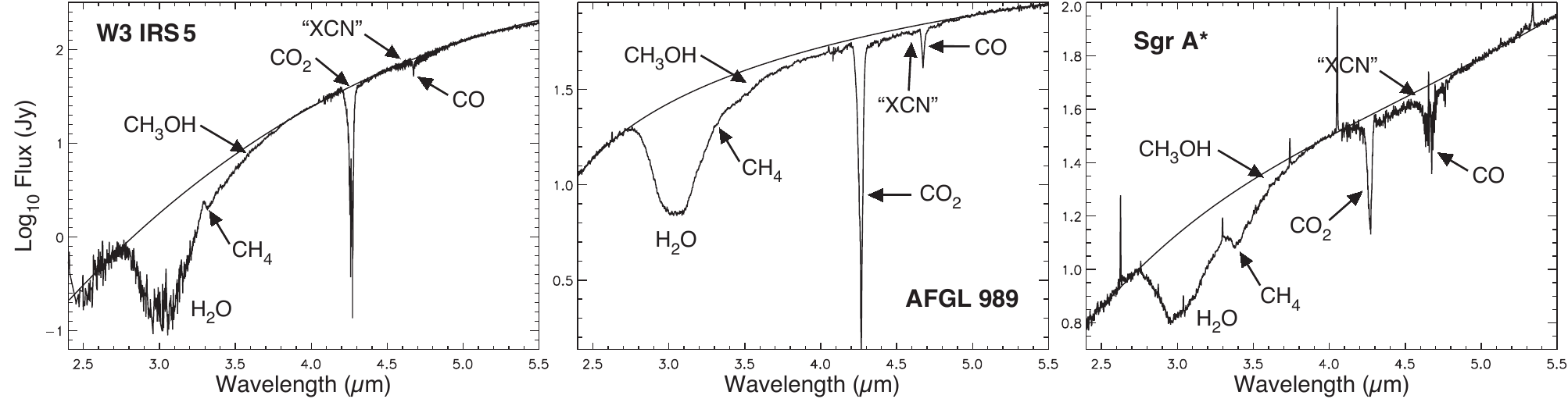}
%\renewcommand{\baselinestretch}{0.99}
%\vspace{-2.7mm}
\caption{Spectra toward a sample of Galactic star forming regions showing
absorption features due to ices.  These spectra were obtained using the Short Wavelength Spectrometer
aboard the {\em Infrared Space Observatory} (ISO)
with a spectral resolving power, $R = \lambda/\Delta\lambda$, of $\sim\,$750
(Gibb et$\:$al.~2004, {\em ApJ Suppl. Series}, 151, 35).}
\label{icespectra}
\end{figure}

\begin{figure}[t]
\centering
%\vspace{-0.8mm}
\includegraphics[scale=0.58]{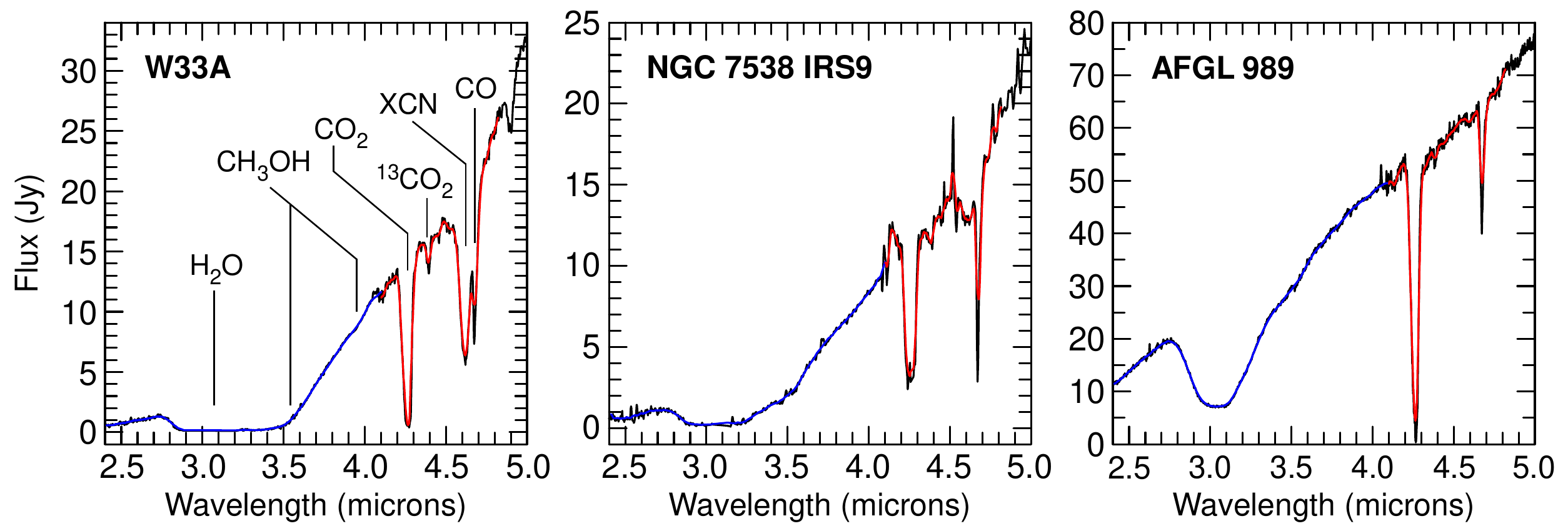}
%\renewcommand{\baselinestretch}{0.99}
%\vspace{-2.0mm}
\caption{Near-infrared spectra of two massive young stellar objects (YSO; W33A and NGC 7538 IRS9) and one
intermediate mass YSO (AFGL 989) obtained by {\em ISO} with $R =\:$750 (black line).  The spectral region
shortward of 4.1$\:\mu$m has been smoothed to $R =\:$40 (blue line) and between 
of 4.1$\:\mu$m and 4.8$\:\mu$m to $R =\:$150 (red line).  Because the solid-state ice features are inherently
broad, little information is lost when sampling these spectra at $R$ of 40 and 150 at $\lambda <\:$4.1$\:\mu$m
and $\lambda >\:$4.1$\:\mu$m, respectively.  (Note: Unlike Fig.~\ref{icespectra}, 
these spectra are plotted on a linear scale to highlight the weaker features.)}
\label{specres}
\end{figure}

Finally, based on laboratory ice spectroscopy and existing observations 
of interstellar ices, the 2.5\,--\,5\,$\mu$m region is the richest part of 
the electromagnetic spectrum for the study of ices.  This wavelength 
region includes strong absorption features for the main ice constituents 
H$_2$O, CO, and CO$_2$ as well as chemically important minor constituents 
NH$_3$, CH$_3$OH, X-CN, $^{13}$CO and $^{13}$CO$_2$ (see Table~\ref{icetable} and Fig.~\ref{icespectra}).
The SPHEREx spectral resolving powers were selected in order to ensure that these
key features can be unambiguously distinguished from one another, i.e., no line
blending that would make the assignment of a unique optical depth to each ice feature
uncertain.  This is achieved with $R = \lambda/\Delta\lambda =\:$40 at $\lambda \leq\:$4.1$\:\mu$m,
which is dominated by the very broad H$_2$O-ice feature, and $R =\:$150 at $\lambda >\:$4.1$\:\mu$m.,
where $R$ is set by the need to spectrally separate the ``XCN'' feature at 4.62$\:\mu$m and 
CO feature at 4.67$\:\mu$m.  This is illustrated in Fig.~\ref{specres} in which
the spectra obtainable with SPHEREx with $R =\:$40 and $R =\:$150 are shown superposed
on the {\em ISO} spectra.

\subsection{Interstellar, Circumstellar, and Protoplanetary Ices: Volatile Reservoirs and Chemical Factories}

Within the well-shielded interiors of molecular clouds, ices are the main 
repositories of many molecules, but their composition is line-of-sight 
specific \citep{Gibb:2004,Oberg:2011yd}.  However, these icy 
mantles are not merely passive volatile reservoirs, but are instead active 
sites of chemistry.  For example, hydrogenation and oxygenation reactions 
form H$_2$O from O, and CH$_3$OH from CO \citep{Tielens1982, Cuppen:2009dz}
The exposure of interstellar ices to radiation 
is expected to create reactive radicals in the ice, which recombine to 
form complex organic molecules \citep{Garrod:2008wy,Oberg:2009jz}.

Though disks are expected to inherit some pristine cloud material 
\citep{Visser:2009ws,Cleeves:2014}, they are also chemically 
active based on direct observations of the trace gases \citep{Oberg:2011yd,Hogerheijde:2011pq}.  
The limited data presently available 
prevent us from knowing whether the main reservoirs of volatiles during 
planet formation (i.e., grain ice mantles) also evolve \citep{Aikawa:2012av}.
However, the observed range of comet compositions in our Solar 
System indicates that ices do evolve \citep{Mumma:2011xj}.  Constraining 
these ice abundances is key to modeling planet formation as the snow 
lines for different species strongly affect the initial steps as well 
as the bulk and surface volatile and organic compositions of nascent 
planets forming at different disk locations 
\citep{Qi:2013qra,Drazkoska:2014,Oberg:2011dc}.

\subsection{The Impact of the \spherex Galactic Ice Investigation}

SPHEREx will be a game changer for the study of interstellar, 
circumstellar, and protoplanetary disk ices.  The tens of thousands of 
high-quality ice absorption spectra toward a wide variety of regions 
distributed throughout the Galaxy will reveal possible correlations 
between ice content and environment not possible with only 100s of 
spectra.  For example, the influence of cloud density, internal 
temperature, presence or absence of embedded sources, external UV and 
X-ray radiation, elemental abundances (such as C/O ratio), gas-phase 
molecular composition, cosmic-ray ionization rate, and cloud evolutionary 
stage, among other factors, can be correlated with the ice content 
in a statistically significant way.

Based on isotopic composition, the \citet{Cleeves:2014} finding 
that terrestrial and cometary H$_2$O ice in the Solar System formed in situ 
in the protostellar cloud and survived intact (rather than undergoing 
modification) through the star formation process exemplifies the type 
of science \spherex will enable.  \spherex spectra, when properly 
corrected for continuum extinction, can be analyzed quantitatively to 
determine the fractional abundance of water ice relative to the entire 
dust content along the line of sight.  This can be done for targets 
at all steps in the evolutionary path from interstellar cloud to
protoplanetary disk, and as a function of the numerous variables 
listed above.  It might be found that the relative abundance of H$_2$O 
ice is fairly constant in many environments, which would be strongly 
supportive evidence for the survival of the original interstellar 
material into the cold icy regions of the forming planetary system 
and suggest as well that water (essential for life as we know it) 
should be widespread throughout exoplanetary systems.

\subsection{ Ice Survey Follow Up Capabilities}

Within the \spherex team we have direct access to state-of-the-art 
laboratory ice spectroscopy and chemistry experiments at Harvard 
University (Co-I \"Oberg).  The ice spectroscopy experiments will be 
used to generate libraries of ice (mixture) spectra as needed to 
interpret \spherex data and also to develop new tools to analyze 
ice spectroscopic data, using, e.g., different decomposition methods.  
The ice chemistry experiments will be used to test new ice formation 
and destruction pathways and how they relate to the local radiation 
field, aiding in the chemical interpretation of observed ice abundance 
and composition patterns.  We also can use the Submillimeter Array 
at the CfA to follow up specific lines of sight using submillimeter 
observations of the gas-phase to obtain a more complete view of 
the volatile reservoir.  For other targets, we will propose follow up 
with either ALMA or JWST, for which the the catalogue of SPHEREx
ice spectra will be an invaluable resource.

%==============================================
\section{SPHEREx Galaxy Evolution and EOR Investigation}
\label{sec:eor_investigation}
%==============================================

SPHEREx's terminator orbit naturally accumulates observatory time at
the celestial poles. We exploit this feature to produce two deeply
mapped regions, enabling a unique large-scale measurement of spatial
fluctuations in the extragalactic background light (EBL). These maps
provide a novel probe of the origin and history of galaxy
formation. The deep regions are also ideal for monitoring long-term
instrument properties such as the flat-field response, gain, and noise
performance. 

\subsection{Explaining the Extra-galactic Background Fluctuations with broad bands}
%===============================================================

\begin{figure}[t]
\begin{center}
\includegraphics[width=0.49\textwidth,height=0.5\textwidth]{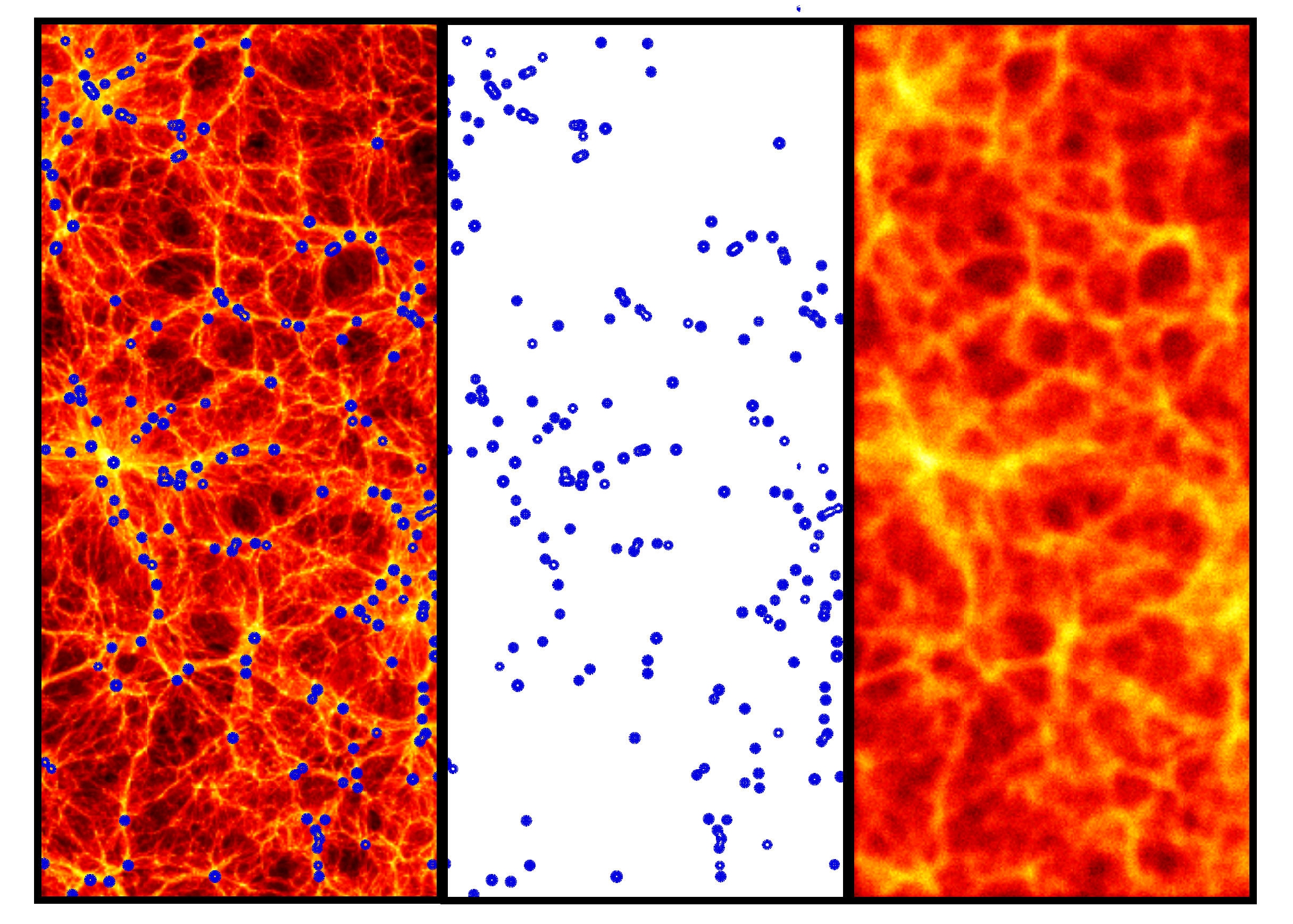}
\includegraphics[width=0.49\textwidth,height=0.5\textwidth]{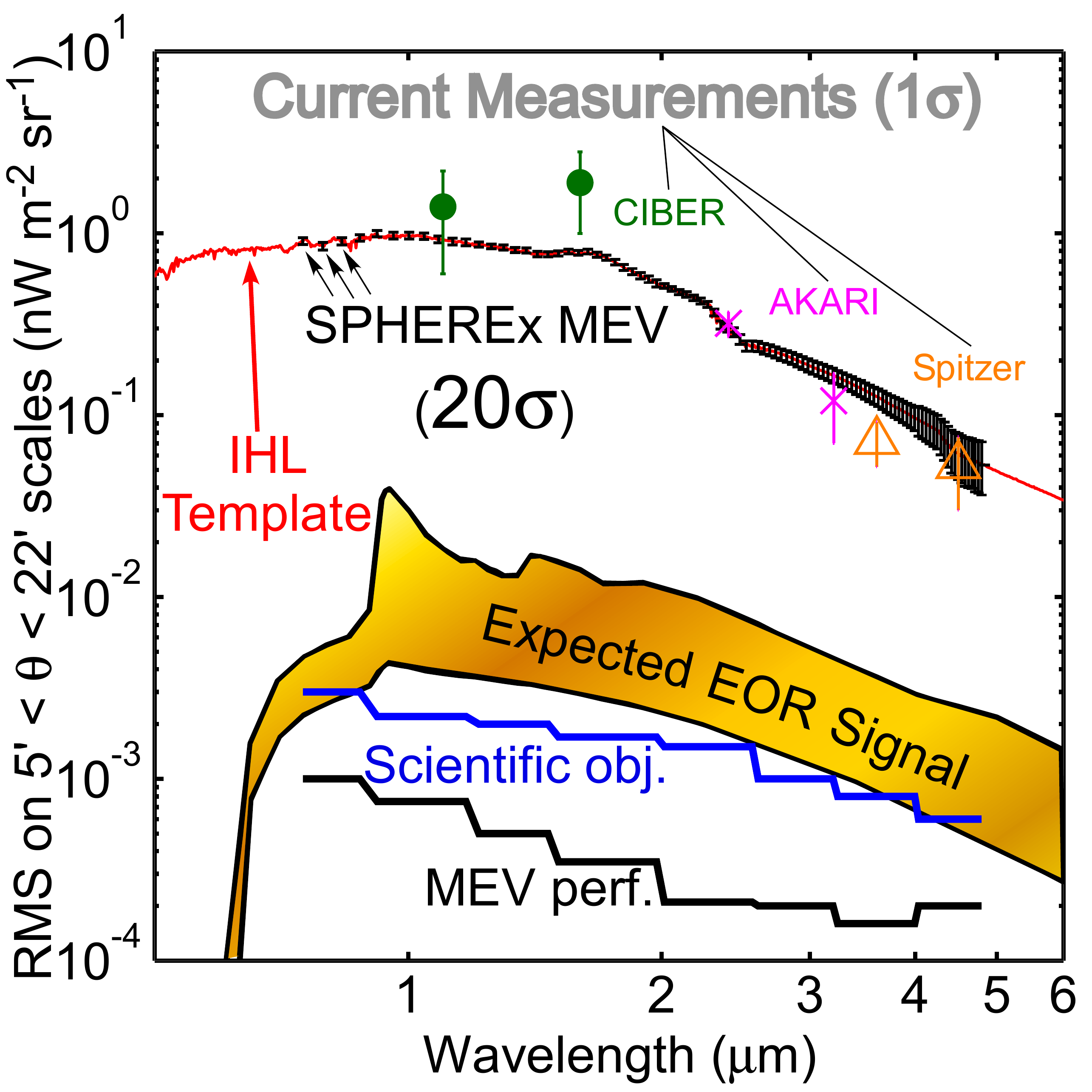}
\end{center}
\caption{Left: A large-scale mapping measurement like SPHEREx traces the
  total emission from diffuse components as well as the emission from
  individual galaxies. The left panel shows a numerical simulation of
  galaxies superposed with a diffuse emission component, such as IHL
  and early dwarf galaxies, that follows the structure of dark
  matter. A galaxy survey (middle) recovers the galaxies but misses
  the diffuse light component. A large-scale mapping measurement
  (right), traces the total emission from the diffuse component as
  well as the individual galaxies due to their clustering. Right:
  Amplitude of large-scale EBL fluctuations measured by CIBER,
  Spitzer, and AKARI, after removing the contribution from known
  galaxy populations. The solid lines show the expected IHL (red) and
  EOR (orange) signals. The bottom of the EOR range is the minimum
  signal that must be present given the existing $z > 7$ Lyman-break
  galaxy luminosity functions \cite{Bouwens:2013hxa}. The top of the
  EOR range allows for fainter galaxies below the detection level of
  deep HST surveys. We show the MEV instrumental performance as the
  RMS between 5 and 22 arcmin in eight bands between 0.75 and 4.8 $\mu$m by the black lines.} 
 \label{fig:ebl}
\label{fig:im}
\end{figure}

EBL fluctuations trace the underlying clustering of faint emission
sources, such as dwarf galaxies and intra-halo light (IHL), components
not readily detected in point source surveys (Figure~\ref{fig:im}, left). The
amplitude of the linear clustering signal is proportional to the total
photon emission and thus an important tracer of star formation
history.  Near-infrared EBL fluctuations have been measured in broad
continuum bands by a number of groups using AKARI
\cite{Matsumoto:2010pz}, Spitzer \cite{Kashlinsky:2005di,Kashlinsky:2005di,Cooray:2012dx} 
and CIBER \cite{Zemcov:2014eca}. The methods for these mapping
measurements are well established. The fluctuation
amplitude, robustly consistent across these experiments, exceeds that
expected from the large-scale clustering of known galaxy populations
\cite{Helgason:2012bs}. While the spectrum of this excess is best
described by low redshift emission such as IHL, other explanations
have been proposed \cite{Cappelluti:2012mj,Yue:2013hya}. The
contribution from the first galaxies during the epoch of reionization
(EOR) must be present in the background (Figure~\ref{fig:ebl}). SPHEREx has the
sensitivity to determine the origin and history of the light
production associated with EBL fluctuations, and search for an EOR
component to minimum levels.

%\subsubsection{Intra-halo light}

The measured near-infrared spectral energy distribution of EBL
fluctuations (Figure~\ref{fig:ebl}, right) is well explained by diffuse IHL emission
at low-redshifts ($z < 1$) arising from a population of stars no longer
associated with their parent galaxies. When individual galaxies
collide and merge, a fraction of the stars are stripped through
dynamical friction to form an extended stellar halo. Indeed images of
nearby galaxies show stars extending out to 300 kpc distances
\cite{Tal:2011vt}.   These stray stars lead to substantial IR 
fluctuations implying that IHL is a significant fraction of the
background light \cite{Cooray:2012xj,Zemcov:2014eca}. The exact
spectral shape, and the degree of correlation between spectral bands,
depends on stellar spectrum and the redshift history of IHL
production. A sensitive multi-band fluctuations measurement thus can
be used to probe the history of stars producing the IHL. 
% \subsubsection{Epoch of Reionization}
EBL fluctuations contain the imprints of the first stars that ended
the cosmic dark ages. Sometime between 200 Myr and 1 Gyr after the Big
Bang ($z =$ 6-20), the first collapsed objects formed and produced
energetic UV photons that reionized the surrounding hydrogen gas. This
EOR marks the end of the dark ages, and is the first chapter in the
history of galaxies and heavy elements.

% \begin{figure}[t]
% \begin{center}
% \includegraphics[width=0.6\textwidth,height=0.4\textwidth]{Clplot_IHL_EOR_v4.pdf}
% \end{center}
% \caption{Amplitude of large-scale EBL fluctuations measured by CIBER,
%   Spitzer, and AKARI, after removing the contribution from known
%   galaxy populations. The solid lines show the expected IHL (red) and
%   EOR (orange) signals. The bottom of the EOR range is the minimum
%   signal that must be present given the existing $z > 7$ Lyman-break
%   galaxy luminosity functions \cite{Bouwens:2013hxa}. The top of the
%   EOR range allows for fainter galaxies below the detection level of
%   deep HST surveys. We show the MEV instrumental performance as the
%   RMS between 5 and 22 arcmin in eight bands between 0.75 and 4.8
%   $\mu$m by the black lines.} 
% \label{fig:ebl}
% \end{figure}

\subsection{Line Intensity Mapping}

The SPHEREx deep spectro-imaging survey also produces the ideal data
set for full tomographic mapping of large-scale structure, a
statistical study based on galactic emission lines. These emission lines trace
linear large-scale galaxy clustering, but unlike 2D continuum
measurements outlined above, provide 3D redshift information.  

\begin{figure}[t]
\begin{center}
\includegraphics[width=0.49\textwidth]{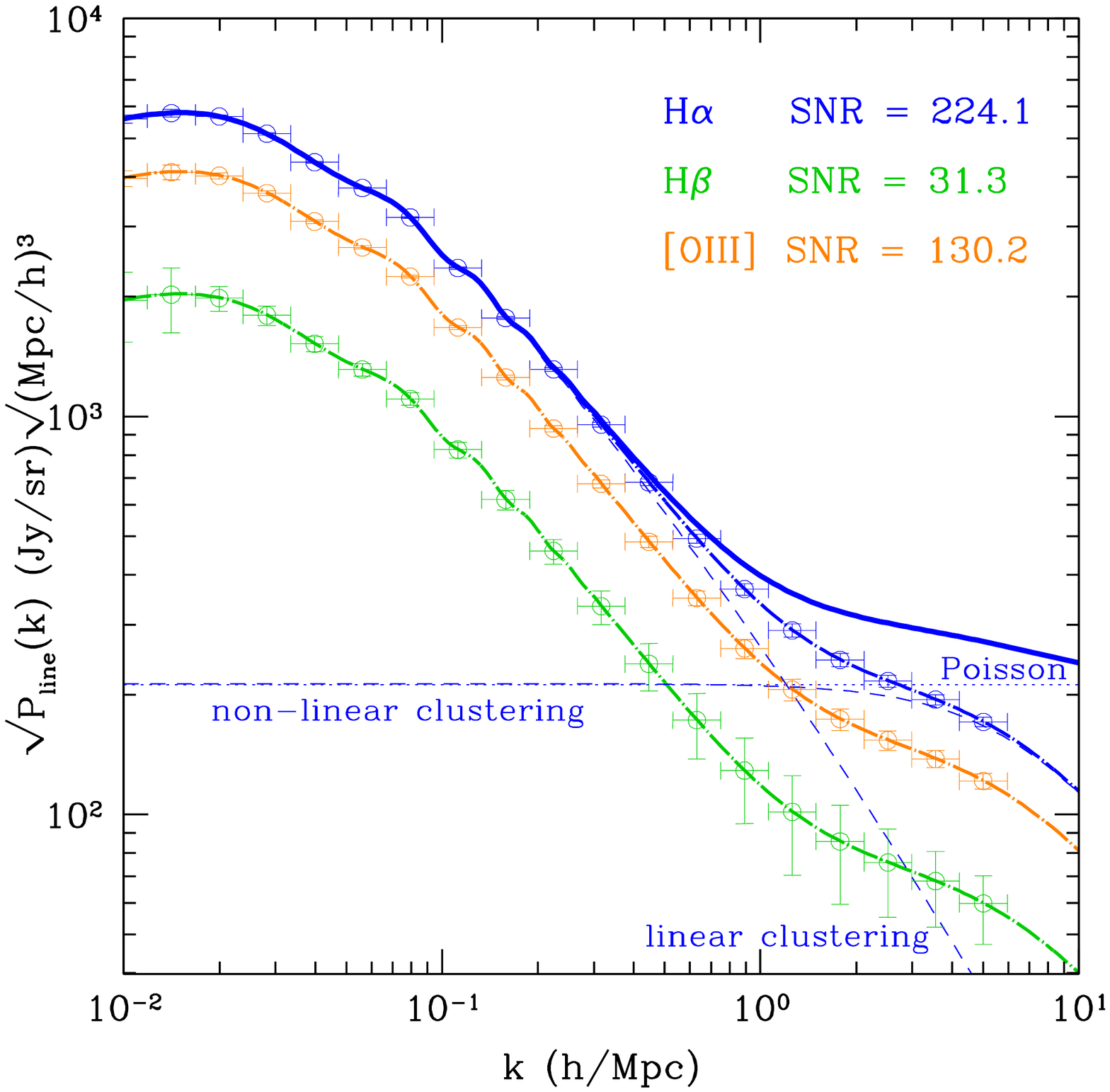}
\includegraphics[width=0.49\textwidth]{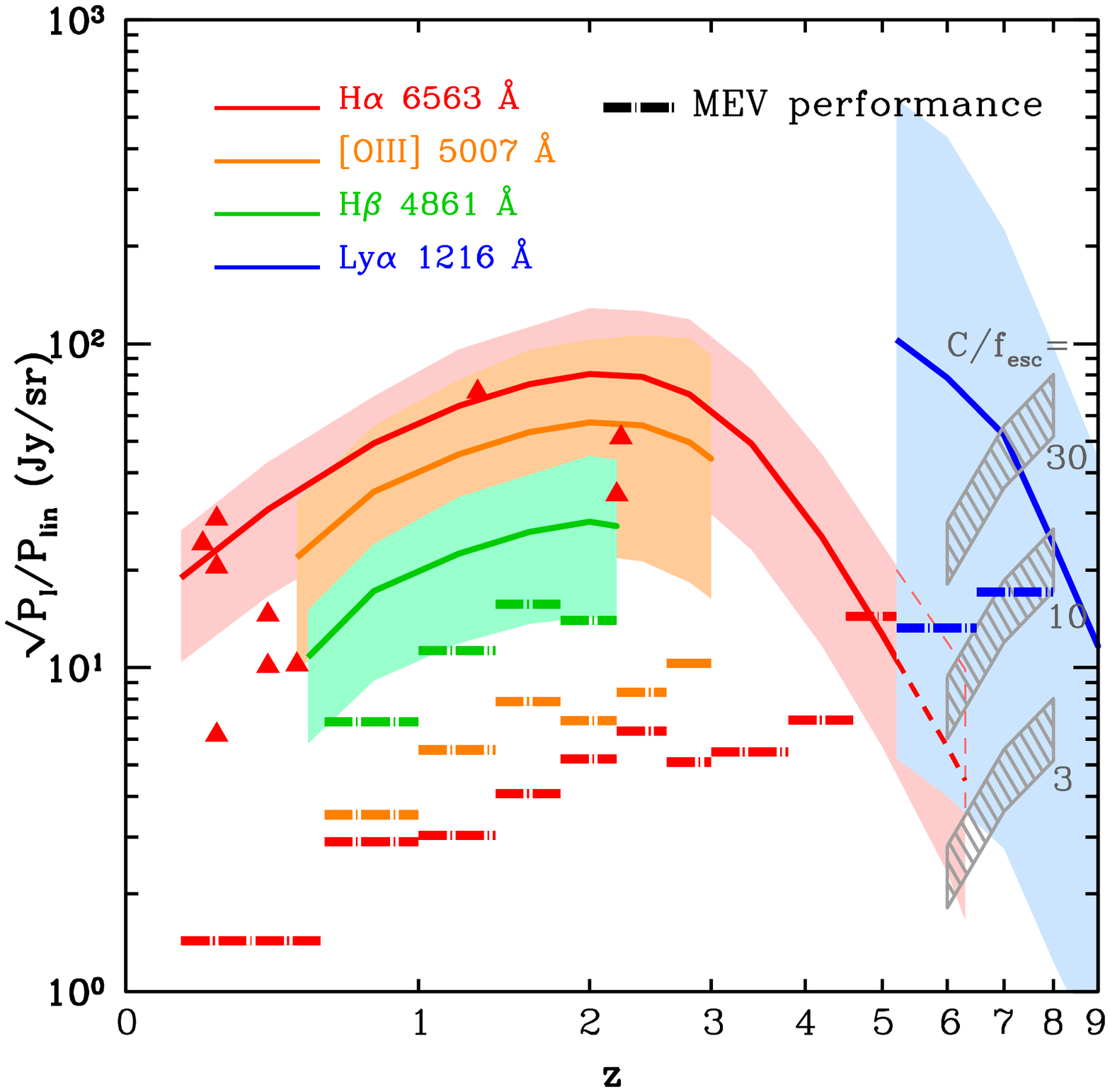}
\end{center}
\vspace{-2.cm}
\caption{Left: Three-dimensional power spectrum of the H$\alpha$ ,
  H$\beta$ and  [OIII] lines at redshift 2. The data points with
  errors show the  statistical uncertainty from the SPHEREx deep survey, which easily
  detects the linear clustering signal at $k<1 h$/Mpc, as well as
  non-linear clustering and Poisson variations at $k>$1$h$/Mpc, with
  high SNR on the linear clustering amplitude in all 3 lines. Right:
  SPHEREx will robustly measure the amplitude of linear galaxy 
clustering in multiple lines over a wide redshift range. The solid
curves show the estimated clustering amplitude, in emission lines
(H$\alpha$, H$\beta$, Ly$\alpha$ and OIII), with the shaded regions
showing present modelling uncertainties. At low redshifts 0.5$<z<$3,
SPHEREx will measure the H$\alpha$, H$\beta$ and [OIII] lines with
high signal to noise. The MEV performance sensitivity (1$\sigma$) is shown by
the colored long-dashed lines in redshift bins. SPHEREx will measure
H$\alpha$ fluctuations over a wide redshift range 0.2$<z<$ 6.3 with high
significance throughout. Furthermore, H$\alpha$ and Ly$\alpha$ overlap
at 5.2$<z<$ 6.3 and can be measured in cross- correlation. At high
redshifts, SPHEREx probes Ly$\alpha$ EOR fluctuations that are detectable for models
with clustering to escape fractions $C/f_{esc} >$ 10.}
\label{fig:im_snr}
\end{figure}

The spectral line intensity cubes from SPHEREx are an ideal tracer of
galaxy evolution. At low redshifts SPHEREx will detect multiple lines
with high signal to noise (Figure~\ref{fig:im_snr}), the dominant lines being H$\alpha$
for redshifts 0.1$<z<$5, H$\beta$ for redshifts 0.5$<z<$2, and [OIII]
for redshifts 0.5$<z<$3. At high redshifts 5.2$<z<$8, SPHEREx accesses
the Ly$\alpha$ line, providing a crucial probe of the formation and
evolution of EOR galaxies.  Traditionally H$\alpha$, after accounting
for dust extinction, has been used as a reliable measure of the cosmic
star-formation rate. In the deep SPHEREx region, we can measure the
H$\alpha$ power spectrum in 10 redshift intervals, with SNR $>$ 100 
(Figure~\ref{fig:im_snr}). The measurement of H$\alpha$ clustering thus traces bolometric
line emission, integrated over all galaxy luminosities and including
emission from any diffuse IHL component. Foreground line confusion
from lower redshift [OIII] and H$\beta$ lines can be robustly removed by
cross-correlating spectral lines in multiple bands. For example, z = 3
H$\alpha$ line fluctuations are detected in a band centered at 2.62 $\mu$m, while
at the same redshift [OIII] fluctuations are present in a band
centered at 2.00 $\mu$m. Cross-correlating two independent bands thus
traces the galaxies at z = 3 without masking, and naturally rejects
any line contaminants that may be present in one of the two bands.
Intensity mapping at high redshifts may provide an additional probe of
reionization \cite{Gong:2012iz,Gong:2011qf,Lidz:2011dx,Visbal:2010rz}. The EOR
epoch may also be probed by mapping the Ly$\alpha$ line
\cite{Silva:2012mtb,Pullen:2013dir}. With SPHEREx sensitivity,
Ly$\alpha$ fluctuations are detectable at S/N $\simeq$ 10 if the star formation rate
density maintains reionization at z $>$ 6 (Figure~\ref{fig:im_snr}). This measurement
provides a further consistency test on any EOR component detected in
the broad- band measurements described above. Furthermore the line
intensity mapping measurement is sensitive to the integrated emission
from dwarf galaxies and is thus complementary with JWST surveys that
individually resolve galaxies at higher luminosities. Line confusion
from lower redshifts requires optimized spectral masking of foreground
sources, although we note that the sensitivity in the deep region
appears to be sufficient to remove the dominant H$α$ contaminant for
Ly$\alpha$ EOR studies \cite{Pullen:2013dir}. We can develop multi-band
cross-correlations as an internal check on foreground removal.

%===============
\section{Conclusion}
%===============

In this paper, we introduced the SPHEREx mission  and its
approach to measuring the three dimensional structure of the
Universe. SPHEREx will be the first near-infrared all-sky spectral survey of its kind and will produce an
archive of lasting legacy value for the astronomy community. If
selected, it will launch in 2020, and nominally acquire data over a
two year lifespan, although there are no consumables that preclude a
longer lifetime. 

While we discussed a few cosmological applications focused on the
physics of inflation, it is clear that the 3-D mapping of such an
unprecedented cosmological volume would enabe numerous other novel cosmological
measurements.  In the galactic plane, it will exploit the coincidence
of its wavelength coverage and the optimal spectral region for ice
absorption spectroscopy to resolve long-standing questions about the
amount and evolution of such biogenic molecules as H$_2$O, CO, CO$_2$,
and CH$_3$OH through all phases of star and planet formation. The
SPHEREx near-infrared all-sky survey will also enable ground-breaking
observations of the extra-galactic background light (EBL). 
% These topics will be presented elsewhere. 

\appendix

\section{Biases in large scale structure maps due to redshift errors}
\label{sec:bias_photoz}

Redshift errors introduce two types of errors into a
large-scale structure survey \footnote{The same variations in photometry that lead to spatially varying redshift errors may also lead
to spurious clustering by pushing galaxies over the threshold defining the sample (e.g.~a signal-to-noise cut),
i.e.~by modulating the selection function. This well studied effect is not included
in this Appendix.}. The most familiar such error is the
smearing of the density perturbations along the line of sight: like
the familiar fingers of God due to true peculiar velocities of
galaxies \cite{1972MNRAS.156P...1J}, they reduce the observed power
spectrum for modes with large $k_\parallel$. There is also a bias in
the density perturbation map if the redshift error varies across the
sky, due e.g., to variations in sky background and hence in the noise
contribution to the observed photometry. 

To compute this, let us suppose that there is a true number density of galaxies meeting some threshold $N_{\rm g}(z,\hat{\bf n})$, where $\hat{\bf n}$ denotes a direction on the sky and $z$ the true redshift. This can be re-written as
\begin{equation}
N_{\rm g}(z,\hat{\bf n}) = \bar N_{\rm g}(z)\,\left[ 1 + \delta_{\rm g}(z,\hat{\bf n}) \right],
\end{equation}
where $\delta_{\rm g}(z,\hat{\bf n})$ is the density perturbation of the galaxies. Now we suppose that the galaxies are observed with a redshift $z_{\rm p}$, whose distribution depends on sky position, i.e., that there exists a $P(z_{\rm p}|z,\hat{\bf n})$. Then the observed number of galaxies is
\begin{equation}
N_{\rm g,obs}(z_{\rm p},\hat{\bf n}) = \int \bar N_{\rm g}(z)\,\left[ 1 + \delta_{\rm g}(z,\hat{\bf n}) \right] P(z_{\rm p}|z,\hat{\bf n})\,dz.
\label{eq:c:Ngobs}
\end{equation}
If we expand $P(z_{\rm p}|z,\hat{\bf n})$ itself into a sky average and a perturbation,
\begin{equation}
P(z_{\rm p}|z,\hat{\bf n}) = \bar P(z_{\rm p}|z) + \delta P(z_{\rm p}|z,\hat{\bf n}),
\end{equation}
then Eq.~(\ref{eq:c:Ngobs}) separates into 4 terms:
\begin{eqnarray}
N_{\rm g,obs}(z_{\rm p},\hat{\bf n}) &=& \int \bar N_{\rm g}(z)  \bar P(z_{\rm p}|z)\,dz
+ \int \bar N_{\rm g}(z) \delta_{\rm g}(z,\hat{\bf n}) \bar P(z_{\rm p}|z)\,dz
+ \int \bar N_{\rm g}(z)  \delta P(z_{\rm p}|z,\hat{\bf n})\,dz
\nonumber \\ &&
+ \int \bar N_{\rm g}(z) \delta_{\rm g}(z,\hat{\bf n}) \delta P(z_{\rm p}|z,\hat{\bf n})\,dz.
\end{eqnarray}
On large scales (our main interest here), we have $|\delta_{\rm g}|\ll 1$, and for a well-designed survey we should have $\delta P(z_{\rm p}|z,\hat{\bf n})$ small. Thus we will drop the last term, which contains the product of these two, and focus on the leading-order signal terms (the 2nd term) and non-uniformity-induced bias (the 3rd term); the 1st term becomes the survey mean. The observed density fluctuation is then
\begin{eqnarray}
\delta_{\rm g,obs}(z_{\rm p},\hat{\bf n}) &=& \frac{N_{\rm g,obs}(z_{\rm p},\hat{\bf n})}{\bar N_{\rm g,obs}(z_{\rm p})} - 1
\nonumber \\
&=& \frac{\int \bar N_{\rm g}(z) \delta_{\rm g}(z,\hat{\bf n}) \bar P(z_{\rm p}|z)\,dz}{\int \bar N_{\rm g}(z)  \bar P(z_{\rm p}|z)\,dz}
 + \frac{ \int \bar N_{\rm g}(z)  \delta P(z_{\rm p}|z,\hat{\bf n})\,dz}{\int \bar N_{\rm g}(z)  \bar P(z_{\rm p}|z)\,dz}
\nonumber \\
&=& \int \delta_{\rm g}(z,\hat{\bf n}) \bar P(z|z_{\rm p})\,dz
 + \frac{ \int \bar N_{\rm g}(z)  \delta P(z_{\rm p}|z,\hat{\bf n})\,dz}{\int \bar N_{\rm g}(z) \bar P(z_{\rm p}|z)\,dz}.
\label{eq:c:sp}
\end{eqnarray}
In the last line, the first term was simplified using Bayes's theorem. It represents the smoothed density field. The second term, which we will call $\delta_{\rm g,sp}(z_{\rm p},\hat{\bf n})$, is the spurious large-scale over density field induced by inhomogeneities in survey properties.

The spurious large-scale density perturbation field can be simplified in the case that $P(z_{\rm p}|z,\hat{\bf n})$ can be
described by a bias $\Delta_z(z,\hat{\bf n})$ and a scatter $\sigma_z(z,\hat{\bf n})$. Then we may use the derivative expansion of
the probability distribution. Taylor-expanding the expectation value of any function $f(z_{\rm p})$ (given $z$) as
\begin{equation}
\int f(z_{\rm p}) P(z_{\rm p}|z,\hat{\bf n})\,dz_{\rm p} =
\langle f(z_{\rm p})\rangle|_z = \sum_{j=0}^\infty \frac1{j!} \langle(z_{\rm p}-z)^j\rangle|_z f^{(j)}(z)
= \sum_{j=0}^\infty \frac1{j!} \langle(z_{\rm p}-z)^j\rangle|_z \int f^{(j)}(z_{\rm p}) \delta(z_{\rm p}-z)\,dz_{\rm p} ,
\end{equation}
where $\delta$ is the Dirac delta function, the superscript $^{(j)}$ indicates a $j$th derivative and the averages are taken over the probability distribution $P(z_{\rm p}|z,\hat{\bf n})$. Integration by parts $j$ times then implies that
\begin{equation}
P(z_{\rm p}|z,\hat{\bf n}) = \sum_{j=0}^\infty \frac{(-1)^j}{j!} \langle(z_{\rm p}-z)^j\rangle|_z  \delta^{(j)}(z_{\rm p}-z).
\end{equation}
(This should be thought of as an asymptotic expansion; in most cases it does not converge.)
Working to order $j=2$, and using that $\langle(z_{\rm p}-z)^j\rangle|_z = \Delta_z$ (for $j=1$) or $\sigma_z^2+\Delta_z^2$ (for $j=2$), we find
\begin{equation}
P(z_{\rm p}|z,\hat{\bf n}) \approx \delta(z_{\rm p}-z) - \Delta_z(z)\delta'(z_{\rm p}-z) + \frac{\Delta_z^2(z)+\sigma_z^2(z)}2\delta''(z_{\rm p}-z).
\end{equation}
Substituting this into the numerator of the second term in Eq.~(\ref{eq:c:sp}) -- and using the $j=0$ (trivial) expansion for the denominator -- gives
\begin{equation}
\delta_{\rm g,sp}(z_{\rm p},\hat{\bf n}) = \frac1{\bar N_{\rm g}(z_{\rm p})} \int \bar N_g(z) \left\{
-[\Delta_z(z,\hat{\bf n}) - \bar\Delta_z(z)] \delta'(z_{\rm p}-z)
+ \frac{\Delta_z^2(z,\hat{\bf n})+\sigma_z^2(z,\hat{\bf n}) - \overline{\Delta_z^2}(z) - \overline{\sigma_z^2}(z)}2\delta''(z_{\rm p}-z)
\right\}\,dz.
\end{equation}
Integrating by parts and simplifying gives
\begin{eqnarray}
\delta_{\rm g,sp}(z_{\rm p},\hat{\bf n}) &\approx& \frac1{\bar N_{\rm g}(z_{\rm p})}
\partial_{z_{\rm p}} \left\{
[\Delta_z(z_{\rm p},\hat{\bf n}) - \bar\Delta_z(z_{\rm p})] N_{\rm g}(z_{\rm p}) \right\}
\nonumber \\ &&
+ \frac1{2\bar N_{\rm g}(z_{\rm p})} \partial_{z_{\rm p}}^2 \left\{
[\Delta_z^2(z_{\rm p},\hat{\bf n})+\sigma_z^2(z_{\rm p},\hat{\bf n}) - \overline{\Delta_z^2}(z_{\rm p}) - \overline{\sigma_z^2}(z_{\rm p})] \bar N_{\rm g}(z_{\rm p}) \right\}.
\label{eq:c:sp2}
\end{eqnarray}

Equation~(\ref{eq:c:sp2}) gives the general formula for biases in the
large scale structure map due to variations in redshift estimation
performance. It can be thought of as a ``photo-$z$ Eddington bias,''
since it involves the change in the observed redshift distribution due
to redshift errors (direction-dependent in this case) and -- like the
original Eddington bias -- involves the second derivative of the
underlying distribution \cite{1913MNRAS..73..359E}. 

A case of special interest is when the redshift bias is small (or has
been calibrated out), so that the $\Delta_z$ terms may be
dropped. However the scatter varies with survey depth, i.e., 
\begin{equation}
\sigma_z(z,\hat{\bf n}) = \bar\sigma_z(z) [ 1 + \alpha(\hat{\bf n}) ],
\end{equation}
where $\alpha(\hat{\bf n})$ is the fractional variation in noise
amplitude (i.e., the photometric RMS noise in all bands is $\propto
1+\alpha$, with $\alpha=0$ at the survey mean depth, $\alpha>0$ in
shallower regions, and $\alpha<0$ in deeper regions). We further
suppose that the redshift distribution varies faster than the
redshift error. (This is a good first approximation for
SPHEREx, since the $N_{\rm g}$ is a rapidly falling function of
redshift, and the design of the experiment avoids sharp features in
redshift estimation performance as a function of $z$ such as the 4000 \AA\ break
redshifting out of the traditional $griz$ bands.) Under these
approximations, Eq.~(\ref{eq:c:sp2}) reduces to 
\begin{equation}
\delta_{\rm g,sp}(z_{\rm p},\hat{\bf n}) \approx
\frac{\bar N''_{\rm g}(z_{\rm p})}{\bar N_{\rm g}(z_{\rm p})} \alpha(\hat{\bf n}).
\label{eq:c:spurious-approx}
\end{equation}
In regions of $\bar N''_{\rm g}(z)>0$ (such as the falling tail of a
redshift distribution), this leads to a positive apparent density
fluctuation for RA and Dec with larger noise values, and a negative
apparent density fluctuation for RA and Dec with smaller noise
values. This dependence on the local noise is used in the systematics
budgeting in the main text. There $\alpha$ has been quoted in
magnitudes instead of fractional perturbations in the RMS noise:
e.g., if $\alpha=0.01$, then we say that the noise fluctuations are
increased by $0.01\times 1.086$ magnitudes. 

The spurious large-scale structure power implied by
Eq.~(\ref{eq:c:spurious-approx}) is larger than SPHEREx requirements
and must be mitigated. Fortunately, like all of the Eddington-like
biases, it can be calculated and corrected. In this case, the key is
to measure the variation over the sky of the redshift bias and
scatter, $\Delta_z(z_{\rm p},\hat{\bf n}) - \bar\Delta_z(z_{\rm p})$
and $\sigma_z^2(z_{\rm p},\hat{\bf n}) - \overline{\sigma_z^2}(z_{\rm
  p})$. The contribution of increased instrument noise in regions of
brighter zodiacal glow to these quantities can be measured by taking
low-noise images (near the ecliptic poles), adding artificial noise,
and measuring the shift and broadening of the redshift
distribution. Note that this is true independent of our knowledge of
the absolute bias and scatter $\bar\Delta_z(z)$ and
$\overline{\sigma_z^2}(z)$. 

\bibliography{refs}

\end{document}